\documentclass[11pt,twoside, spanish]{book}

\usepackage[spanish]{babel}
\usepackage[applemac]{inputenc}
\usepackage[Glenn]{fncychap}
\usepackage{boxedminipage}
\usepackage{amssymb}
\usepackage[dvips]{graphicx}
\usepackage{amsmath}
\usepackage{fancyhdr}
\usepackage[mathscr]{euscript}
\usepackage{flafter}
\usepackage{enumerate}
\usepackage{epsfig}
\usepackage{array}
\usepackage[usenames,dvipsnames,svgnames,table,array]{xcolor}
\usepackage{colortbl}
\usepackage{verbatim}
\usepackage{appendix}
\usepackage{bm}
\usepackage{float}
\usepackage[font=small,labelfont=it]{caption}
\usepackage{subcaption}
\usepackage{subfig}


\def\nnum{\nonumber}
\def\mb{\mathbf}
\def\bdet{\begin{vmatrix}}
\def\edet{\end{vmatrix}}
\def\bmat{\begin{pmatrix}}
\def\emat{\end{pmatrix}}

\def\barrt{\begin{array}{ccc}}
\def\earrt{\end{array}}
\def\barrf{\begin{array}{cccc}}
\def\earrf{\end{array}}

\def\balg{\begin{aligned}}
\def\ealg{\end{aligned}}
\def\beq{\begin{equation}}
\def\eeq{\end{equation}}
\def\ber{\begin{eqnarray}}
\def\eer{\end{eqnarray}}
\def\bse{\begin{subequations}}
\def\ese{\end{subequations}}
\def\bdm{\begin{displaymath}}
\def\edm{\end{displaymath}}
\numberwithin{equation}{section} 
\newcommand{\half}{\mbox{$\textstyle \frac{1}{2}$}}
\newcommand{\ket}[1]{\left| #1 \right\rangle}
\newcommand{\bra}[1]{\left\langle #1 \right|}

\newcommand{\braket}[2]{\langle #1 | #2\rangle}

\let\tmp\oddsidemargin
\let\oddsidemargin\evensidemargin
\let\evensidemargin\tmp
\reversemarginpar

\decimalpoint

\begin{document}

\renewcommand{\tablename}{Tabla}
\fancyhead{}
\fancyhead[CE]{\scriptsize \leftmark}
\fancyhead[CO]{\scriptsize \leftmark}
\fancyhead[LE,RO]{\thepage}
\fancyfoot{}

\frontmatter
\thispagestyle{empty}
\begin{center}
\includegraphics[scale=0.15]{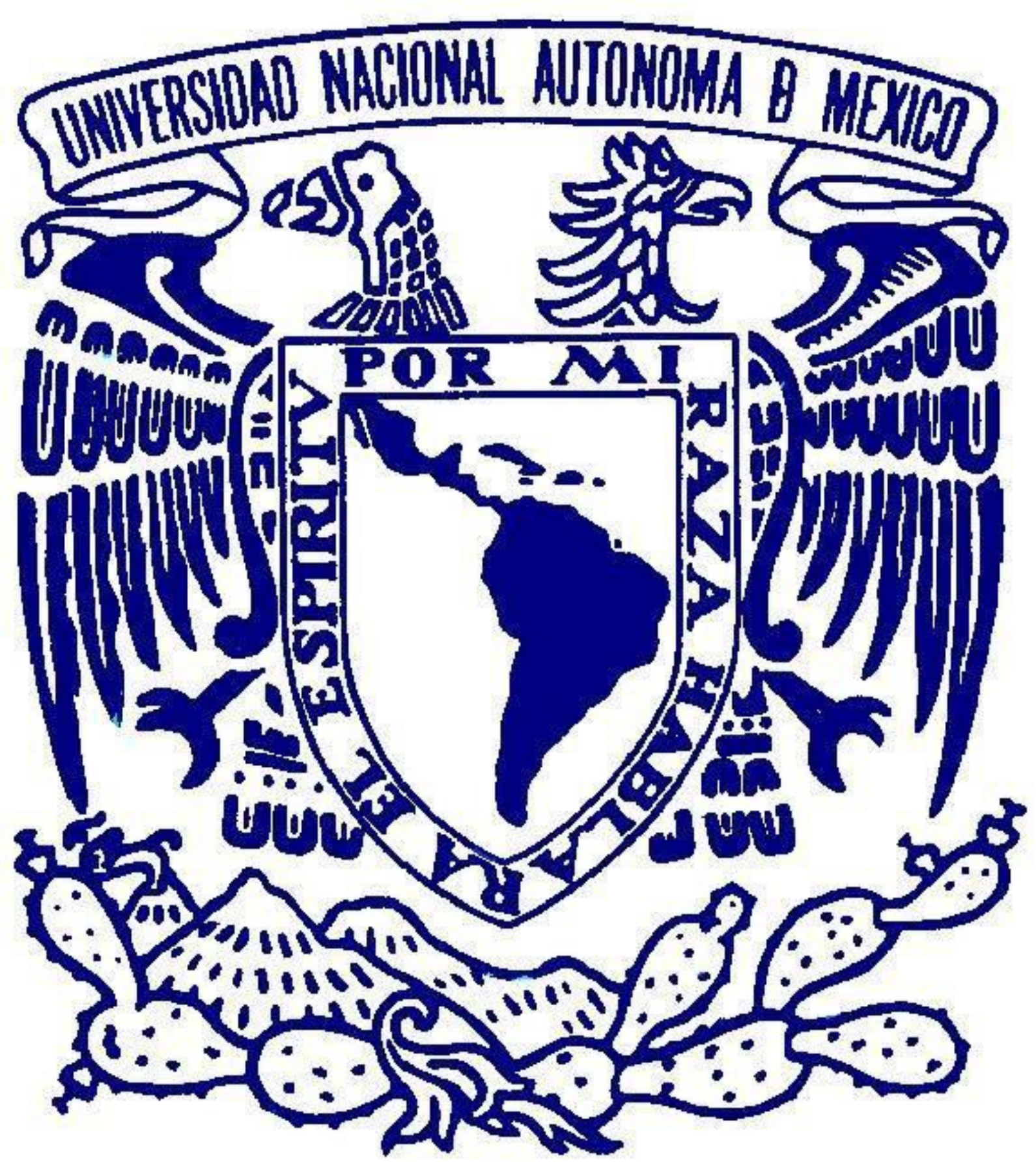}
\end{center}
\begin{center}
\includegraphics[scale=0.65]{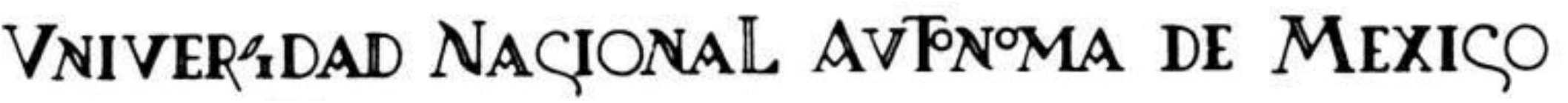}
\end{center}

\begin{center}
{ POSGRADO EN CIENCIAS FSICAS}\\
\vspace{1cm}
{ CONSECUENCIAS GEOMTRICAS Y DINçMICAS DE LA MòLTIPLE ADSORCIîN DE LITIO Y OTROS ALCALINOS EN POLIACENOS, POLI-PARA-FENILENOS Y HOJUELAS DE GRAFENO}\\
\vspace{1cm}
{ TESIS \\ QUE PARA OPTAR POR EL GRADO DE:\\ DOCTORA  EN CIENCIAS (F\'ISICA)}\\
\vspace{1cm}
PRESENTA:\\
YENNI PRISCILA ORTIZ ACERO \\
\vspace{1cm}

DIRECTOR DE TESIS: DR THOMAS H. SELIGMAN SCHURCH\\
INSTITUTO DE CIENCIAS FSICAS DE LA UNAM\\
\vspace{0.2cm}
MIEMBRO DEL COMIT TUTORAL: DR CARLOS BUNGE MOLINA\\
IINSTITUTO DE FSICA DE LA UNAM\\
\vspace{0.2cm}
MIEMBRO DEL COMIT TUTORAL: DR JOSE RECAMIER ANGELINI\\
INSTITUTO DE CIENCIAS FSICAS DE LA UNAM

\vfill
MEXICO, D.F., MARZO DE 2014
\end{center}

\newpage
\thispagestyle{empty}
\mbox{}


\newpage
\thispagestyle{empty}
 \begin{flushright}
 \hspace{5cm}\textsl{Gracias a mi madre Mirian Acero Gonzalez, pues todo lo que soy se lo debo a mi ella. Atribuyo todos mis xitos en esta vida a la enseanza moral, intelectual y fsica que recib de ella"  }
\end{flushright}

\newpage
\thispagestyle{empty}
\mbox{}


\newpage
\begin{center}
{\huge Agradecimientos}
\end{center}

Gracias por los apoyos financieros recibidos de  CONACyT , con la beca de doctorado y en el marco del proyecto No 79613; y de DGAPA-UNAM en el marco del proyecto PAPIIT No  IN 114310. 

Gracias a mi tutor, Thomas Seligman, por su orientacin en este y otros campos de conocimiento, y por ser el mejor ejemplo en el aspecto social y humano. 

Gracias a la Prof. Dr. Ute Kaiser y a su grupo de investigacin de la universidad de ULM, Alemania, por darme la oportunidad de tener un acercamiento experimental hacia el grafeno y por la colaboracin en estudios de adsorcin de Hidrogeno. 

Gracias al Prof. Dr. Matthias Troyer  y a al Dr. Iztok Pizorn, de la universidad ETH Zrich, Suiza, por la instruccin en el programa CP2K, el cual nos permiti comparar resultados en ste trabajo.

Gracias al Prof. Dr. Douglas Klein de la universidad A\&M en Galveston, USA, por sus aportes conceptuales.

Gracias a el Dr. Carlos Bunge, Dr. Francois, Dr. Jose Recamier, Dr. Ascencio, Dr. Hernan Larralde e Isabel por su amistad y gratas conversaciones.

Gracias a Natalia Carriazo por ser tan admirable ejemplo para m y por su apoyo incondicional.

Gracias a Elda Patricia Rueda por su amistad y cario, por su apoyo incondicional y por ser admirable ejemplo para m.

Gracias a mi madre Mirian Acero por su apoyo y amor incondicional, por ser la mujer que ms admiro.  A mis hermanos Hasbleidy y Anderson por su amor y apoyo. A mi  precioso sobrino Sebastian por sacar la mejor parte de mi. A mi madrina Lucero Garzn por sus invaluables consejos y por su confianza y apoyo.   A  mi adorable abuela Rita Gonzalez  y a la familia Acero por su confianza y amor, en especial a mi tio Orlando y su familia.

Gracias a mi novio Diego Espitia  por su amor y apoyo constante, por brindarme las mayores alegras.

Gracias a mis amigos y casi familia en Mexico: Paco, Jose  "Pieman", Gabriela Romero y familia, Luisa Muriel, Aaron en puebla, Silvia Mendez, Jose Soriano y familia,  Sara Rodriguez y sus hermosos hijos, a los Salsamaniacos, Christofer Hernandez, Juliio, Luisana.

En Colombia gracias a mis amigos constantes: Rosa Parra, Jacky, Edwin Roa, Jorge Moreno, Mabel Pareja, Iveth Rodriguez, Diana Rodriguez. Jeison Montao, Diego Mulato, David Palomino, Raul MAuricio.

Gracias a mis recientes amigos, Thomas Stegman, Adriana Fontalvo y Carlos Gonzalez.

\newpage
\thispagestyle{empty}
\mbox{}


\newpage
\begin{center}
{\huge Resumen}
\end{center}

En este trabajo estudiamos el rompimiento espontneo de simetra debido a la adsorcin de tomos de Litio y otros alcalinos en poliacenos y molculas aromticas consistentes de anillos de carbono cerrados en las orillas por enlaces con tomos de hidrgeno. Se realizaron clculos de Hartree Fock y DFT para poliacenos, poli-para-fenilenos y hojuelas de tamaos que muestran propiedades de grafeno. \\

Como resultado, en poliacenos encontramos que el rompimiento de simetra por la adsorcin mltiple de pares de Litios en lados opuestos  tiene que ver con la distorsin de Peierls \cite{peierls}, adems encontramos que no hay rompimiento de simetra debido a la adsorcin de otros alcalinos en poliacenos y que el nico caso que muestra una distorcin del poliaceno es cuando se adsorbe un par de tomos de Sodio. \\

Aunque existen estudios previos sobre la adsorcin de metales en hojas grandes de carbono \cite{js1}, nosotros quisimos simplificar el problema, reduciendo el sistema a poliacenos, lo que nos permiti obtener una mejor comprensin de las razones de la ruptura de la simetra y luego extendernos a estructuras complejas, en las que encontramos que la distorsin de Peierls es observable en hojuelas pequeas, con la simetra apropiada, pero no se puede generalizar a hojuelas de tamao mayor y menos a tiras de grafeno, ya que los estados de mnima energa para stos casos no corresponden a sta distorsin. \\

Como una aplicacin de los mtodos usados en el presente trabajo se propone un mecanismo para la descomposicin de las reacciones de clorometano, diclorometano y el formaldehdo de cloro sobre superficie de grafeno. Para esto se hicieron clculos sobre una hojuela de grafeno con la adsorcin de un Litio en el centro de la misma y del lado opuesto a las reacciones, en donde se encontr que se reduce la produccin de ciertos radicales intermedios. 

\newpage
\thispagestyle{empty}
\mbox{}


\newpage
\begin{center}
{\huge Abstract}
\end{center}
We study the spontaneous symmetry breaking due to adsorption of Lithium atoms on polyacenes and aromatic molecules consistent on carbon rings with edges closed by bond of hydrogen atoms. Hartree Fock and DFT calculations were made for polyacenes, poly-para-phenyls and carbon sheets of sizes that show properties of graphene.\\

As a result,  the spontaneous symmetry  breaking on polyacenes due to the adsorption of multiple pairs of Lithium atoms on opposite sides find explanation on the Peierls distortion \cite{peierls}. We also found that there are no spontaneous symmetry breaking due to adsorption of other alkalines in polyacenes and  the only case showing a distortion of polyacene is when a pair of Sodium atoms is adsorbed on it. \\

Although there were previous studies about  adsorption of metals on large sheet of carbon \cite{js1}, we initially wanted to simplify the problem so that allowed us to obtain a better understanding of the reasons for symmetry breaking and then extend it to complex structures, for which, we found that the Peierls distortion is observable in small flakes, with appropriate symmetry, but can not be generalized to larger flakes and neither graphene strips as the minimum energy states of these cases do not correspond to this distortion.\\

As an application on the methods we have used in the present work, a mechanism for the decomposition reactions of chloromethane, dichloromethane and formyl chloride on graphene surface is proposed. For this, we calculate the reactions on the graphene surface with a Lithium atom absorbed on the center of  it at the opposite side and we found intermediate production of radicals is reduced.

  \tableofcontents

\mainmatter


\chapter{Introduccin}

En los ltimos aos el grafeno \cite{geim} ha recibido gran atencin desde el descubrimiento de  los fulerenos (\textit{fullerenes}) \cite{fullerenes},  seguido por el surgimiento de los nanotubos \cite{nanotubes}, ms la atencin enfocada a tiras de grafeno \cite{ribbons1}.\\

El grafeno es una red cristalina hexagonal de carbonos que puede ser considerada  como una molcula aromtica indefinidamente larga \cite{aromatic, aromaticity}. Mientras que una pelcula delgada de un material tridimensional tiene un ancho alrededor 50 $nm$, para el caso del grafeno se ha mostrado que la estructura electrnica evoluciona rpidamente con el nmero de capas y se aproxima al grafito (tridimensional) con 10 capas \cite{capas}, alrededor de 3.35 $nm$. La carga en el grafeno es descrita por medio de la ecuacin de Dirac \cite{1947, novo} que permite la investigacin de fenmenos cunticos relativistas en experimentos de laboratorio \cite{gr_exp1}-\cite{gr_exp5}. Su carga exhibe gran movilidad  (mayor a 15000 $cm^2/V.s$) por lo que tiene densidades de corriente 6 veces mayores que el cobre\cite{mobil}.\\

Como el grafeno es un sistema de muchos cuerpos, para su estudio a nivel terico, son necesarios mtodos aproximados. Uno de los mtodos ms usados para obtener las propiedades de sistemas de muchos cuerpos es la teora del funcional de la densidad (DFT por sus siglas en ingls, \textit{Density Functional Theory}), que fue propuesta por Hohenberg y Kohn en 1964 \cite{dft}, aunque en 1927 Thomas y Fermi \cite{thomas, fermi} ya haban dado indicios del formalismo. La teora se basa en los teoremas de Hohenberg y Kohn \cite{dft}, los cuales son una herramienta que permite convertir la ecuacin Schrdinger para un sistema de muchos cuerpos en un conjunto de ecuaciones de un slo cuerpo, conocidas como las ecuaciones de Kohn-Sham \cite{ks} (KS) y que son ecuaciones exactas pero que contienen un trmino desconocido que debe aproximarse. ste trmino, llamado funcional de intercambio y correlacin, contiene parte de las interacciones electrn-electrn (intercambio dado por el principio de exclusin de Paui y efecto de correlacin dado por la repulsin coulombiana).\\

Por otro lado, en el ltimo siglo, gran parte de la fsica se ha construido estudiando las propiedades de simetra de los sistemas fsicos. Ejemplos de ello son la relatividad especial \cite{relativity} y el ferromagnetismo. Este ltimo en particular se describe a partir del rompimiento espontneo de simetra \cite{ssb} (SSB por sus siglas en ingls, \textit{Spontaneous Symmetry Breaking}) \cite{ssb}. El SSB ocurre cuando la simetra de las leyes que gobiernan un sistema no se manifiestan en el estado real del mismo, en otras palabras, cuando el hamiltoniano de un sistema tiene ciertas simetras y la solucin  que da la energa ms baja no exhibe tales simetras. Esto a diferencia del rompimiento de simetra explcito que se debe a la adicin de agentes  externos que no tengan la misma simetra del sistema.\\

El primer estudio que se hace en el presente trabajo es el referente al rompimiento de simetra \cite{ssb} debido a la adsorcin de tomos de Litio en hojuelas o laminillas pequeas de carbono tratadas como molculas aromticas  con hidrgenos en cada enlace externo, teniendo como antecedente el trabajo realizado en 2009 \cite{js1} donde se  estudi la transferencia de carga del Litio en laminillas de carbono. Para ello estudiamos la adsorcin de Litio particularmente en laminillas de carbono que geomtricamente tienen simetra de reflexin  \cite{article_flakes} y se observa una deformacin de la laminilla que rompe la simetra de reflexin si se adsorbe un solo tomo.  A primera vista, se podra esperar recuperar la simetra con la segunda adsorcin en el mismo anillo de lado opuesto, pero lo que sucede es que la deformacin se acenta en la misma direccin. La sospecha inicial fu que se trataba del efecto Jahn-Teller  \cite{jahn-teller} conocido tambin como distorsin de Jahn-Teller. Hermann Arthur Jahn y Edward Teller demostraron, utilizando teora de grupos \cite{group}, que las molculas no lineales con estados degenerados no pueden ser estables, es decir, cualquier molcula no lineal con un estado fundamental electrnico degenerado sufrir una distorsin geomtrica que elimina la degeneracin debido a que la distorsin disminuye la energa total de la molcula.  

\begin{figure}[h!]\centering
\includegraphics[height=0.27\textheight]{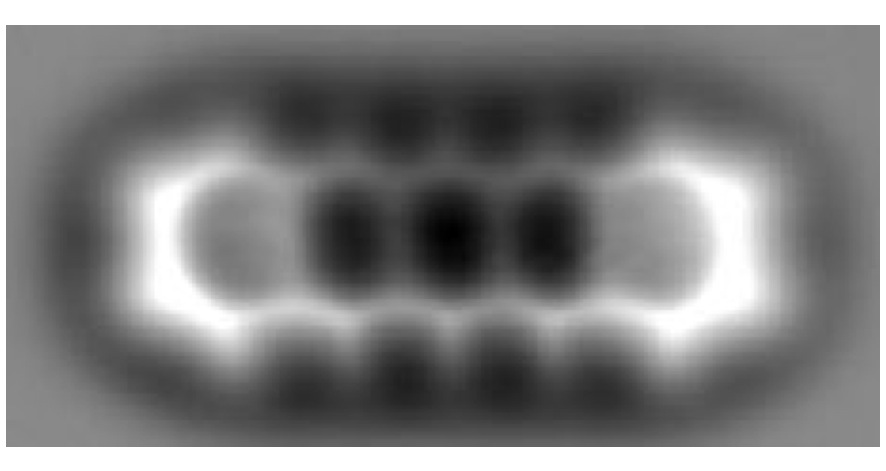}
\caption{Los investigadores de IBM han logrado esta imagen del pentaceno mediante un microscopio de fuerza atmica \cite{pentacene}.}
\label{penta}
\end{figure}

Con el fin de  simplificar el problema y obtener un mejor entendimiento del rompimiento de simetra, se redujo el sistema de hojuelas de grafeno a poliacenos  \cite{acenes} .  Los poliacenos son una clase particular de molculas aromticas compuestas de anillos de benceno dispuestos linealmente y son de inters por sus propiedades electrnicas, termodinmicas y pticas para el desarrollo de nuevos materiales\cite{poly1}-\cite{poly5}  y aplicaciones en diversas reas de ciencia de materiales \cite{ofet}-\cite{opv}. Cabe mencionar que el pentaceno se ha observado mediante un microscopio de fuerza atmica \cite{pentacene}, como se muestra en la figura \ref{penta}. \\

Para el estudio en poliacenos de la adsorcin de Litio y otros alcalinos se realizaron clculos computacionales usando los mtodos de Hartree-Fock (HF) \cite{hartree, fock} y la teora del funcional de la densidad \cite{dft} y encontramos que cuando se adsorben dos tomos de Litio dispuestos de lados opuestos en poliacenos de tamaos desde el naftaleno (2 anillos de benceno) hasta heptaceno (7 anillos de benceno) se obtiene un doblamiento del mismo \cite{articulo}.  Al aumentar el nmero de pares de Litio en anillos no adyacentes, encontramos que la configuracin de mnima energa es una configuracin tipo zigzag. Nuestros resultados en poliacenos se pueden explicar con la transicin de Peierls \cite{peierls}. Tambin encontramos que no hay rompimiento de simetra debido a la adsorcin de otros alcalinos en poliacenos y que el nico caso que muestra una distorcin del poliaceno es cuando se adsorbe un par de tomos de Sodio. \\

Buscando extender nuestros clculos a molculas aromticas ms grandes, especficamente a tiras de grafeno \cite{tiras1},  encontramos que algunas propiedades fsicas de estas dependen del borde que tengan. Por sta razn se estudi la adsorcin de Litios en poli-para-fenilenos \cite{polyph1}, que corresponderian al borde canasta de las tiras de grafeno.\\

 Se realizaron calculos de DFT  para la adsorcin de un par de Litios en lados opuestos un anillo desde bifenileno hasta pentafenileno y se encontr que los anillos cercanos al anillo en el que se adsorbe el par de litios se alinean con el mismo, esto teniendo en cuenta que sin la adsorcin de Litio los anillos adyacentes no son coplanares. \\
 
 De igual forma se extendieron nuestros estudios realizados en poliacenos a hojuelas ms grandes en forma de tiras de grafeno \cite{tiras1}-\cite{tiras3} y encontramos que la distorsin de Peierls es observable en hojuelas pequeas, con la simetra apropiada, pero no se puede generalizar a hojuelas de tamao mayor y menos a tiras de grafeno, ya que los estados de mnima energa para stos casos no corresponden a sta distorsin.
\\

Finalmente, como una aplicacin de los mtodos usados en el presente trabajo se propuso un mecanismo para la descomposicin de las reacciones de clorometano y diclorometano sobre superficie de grafeno, ste problema es interesante por el descubrimiento de aromticos grandes en la atmosfera \cite{atmos}. Para esto se hicieron clculos sobre una hojuela de grafeno con la adsorcin de un Litio en el centro de la misma del lado opuesto a las reacciones y se encontr que se reduce la produccin de radicales intermedios \cite{art_atm}. \\


\chapter{Marco Terico}

En este captulo se explican los diferentes mtodos que se usaron en el presente trabajo. Para fijar notacin haremos ciertas consideraciones.

Una molcula consiste de $K$ ncleos (con masas $M_k$ y cargas $Z_ke$) y N electrones (con masa $m$ y carga $-e$) y con la condicin $|Ne|=\sum_kZ_k$. El movimiento de dicha molcula en un estado con energa total $E$ es descrito por la ecuacin de Schrdinger independiente del tiempo: 

\beq \label{sch}\hat{H}\Psi=E\Psi\eeq 
donde el hamiltoniano $H$ es:

\beq \label{ham} \hat{H}=\hat T+ \hat V=-\frac{\hbar^2}{2m}
\sum^N_{i=1}\nabla_i^2-\frac{\hbar^2}{2}\sum_{k=1}^K\frac{1}{M_k}\nabla^2_k+V(\mathbf r,\mathbf R),\eeq
$\hat T$ es el operador de la energa cintica de los electrones y los ncleos y $\hat V$ es la energa potencial que explcitamente corresponde a:

\ber V(\mathbf r, \mathbf R) & = &V_{\text{nuc, nuc}}+ V_{\text{nuc, el}}+V_{\text{el, el}}\\
 & = &\frac{e^2}{4\pi \epsilon_0}\left[\sum_{k>k'}\sum_{k=1}^K\frac{Z_kZ_{k'}}{{R_{k,k'}}}- \sum_{i=1}^N\upsilon(r_i)+\sum_{i>i'}\sum_{i'=1}^N\frac{1}{r_{i,i'}}\right]\label{pot}
\eer
donde 
\beq \upsilon(\mb r_i)=\sum_{k=1}^K\frac{Z_k}{r_{i,k}}\label{v_nuc}\eeq
y

$$ R_{k,k'}=|\mathbf R_k-\mathbf R_{k'}|,\hspace{5pt} r_{i,k}=|\mathbf r_i-\mathbf R_k|,\hspace{5pt}r_{i,i'}=|\mathbf r_i-\mathbf r_{i'}|.$$
El primer trmino describe la interaccin coulombiana entre los ncleos, el segundo la atraccin entre ncleos y electrones y  el tercero la repulsin entre electrones. Aqu se han ignorado las interacciones relacionadas con el espn nuclear y electrnico ya que el corrimiento de la energa debido a estas interacciones es pequeo comparado con la energa total. 

Por simplicidad de ahora en adelante usaremos unidades atmicas que se obtienen definiendo:

$$m_e=1, \hspace{7pt}\hbar =1,\hspace{7pt}e=1, \hspace{7pt}\epsilon_0=1.$$

Vale la pena resaltar que usando unidades atmicas las ecuaciones resultantes no tienen las dimensiones en el sentido usual, a cambio se tienen la unidades atmicas de conversin:\\

La unidad atmica para la longitud es \textit{bohr} que es igual al radio $a_o$ de la menor rbita de Bohr en el tomo de hidrgeno:

\beq a_0= \frac{4\pi\epsilon_0\hbar^2}{me^2}\approx 0.05nm.\eeq

La unidad atmica para la energa  es \textit{hartree} definido como el doble de la energa de ionizacin del tomo de hidrgeno (-$E_{pot}$ para el electrn en la rbita ms baja):

\beq E_{pot}=-\frac{me^4}{(4\pi\epsilon_o)\hbar^2n^2}\approx 27 eV.\eeq

\section{Aproximacin de Born-Oppenheimer}

La aproximacin Born-Oppenheimer \cite{ap-b-o} (BO) consiste en que, debido a que el ncleo es ms pesado que los electrones,  a los ncleos se les asignan posiciones fijas, y entonces el movimiento de los electrones en el entorno nuclear fijo se determina a partir de la ecuacin de Schrdinger. La distribucin electrnica resultante origina un campo actuando sobre los ncleos, y stos se mueven gobernados por la energa potencial de interaccin de los ncleos y el campo de los electrones. En otras palabras, para cada $\mb R$ existe una  funcin de onda bien definida $\phi_n^{el}(\mb r,  \mb R)$ para el estado electrnico $n$, la cual depende de las variables de posicin de los electrones $\mb r$ y depende paramtricamente de las posiciones de los ncleos $\mb R$.\\

Aunque nosotros no estamos interesados en el estudio del movimiento de los ncleos, realizamos el tratamiento formal de dicha aproximacin con lo cual se dejan abiertos posibles estudios del movimiento nuclear.  Usando teora de perturbaciones \cite{sakurai} se ilustrar la conexin entre el movimiento nuclear y el movimiento de los electrones. Teniendo en cuenta que la energa cintica de los ncleos es pequea comparada con la energa cintica de los electrones la podremos considerar como una perturbacin al sistema cuyo ncleo es fijo ($\mb R=constante$). \\

Para el caso del estudio del movimiento de los electrones se usa el principio variacional \cite{sakurai} con el que se calcula el mnimo de la energa tomando como parmetro el vector $\mb R$. Con esto tambin se encuentra la configuracin de mnima energa. Es importante mencionar que este clculo es vlido siempre y cuando las superficies de energa para cada nivel $n$ no se crucen.   La ecuacin de Schrdinger no perturbada que describe una molcula en la cual los ncleos se consideran fijos en $\mb R$ \cite{demtro},  es:

\beq \left(\hat T_{el}+\hat V_{\text{el,el}}+\hat V_{\text{nuc, nuc}}+\hat V_{\text{el, nuc}}\right)\phi^{el}_n(\mb {r,R})=E_n^{(0)}(\mb R)\phi^{el}_n(\mb {r,R}),\eeq

\beq \hat H_0(\mb R)\phi^{el}_n(\mb {r,R})=E_n^{(0)}(\mb R)\phi^{el}_n(\mb {r,R})\label{schnp},\eeq\\

de tal manera  que $\phi^{el}_n$ depende slo de las variables electrnicas $\mb r$ y depende de las coordenadas nucleares $\mb R$ como parmetros.\\

Imponiendo que las soluciones de la ecuacin \ref{schnp} formen un conjunto completo ortonormal de funciones, entonces toda solucin $\Psi(\mb{r,R})$ de la ecuacin completa \ref{sch} se puede expandir en una serie de tales funciones. Para resolver la ecuacin \ref{sch} escogemos como ansatz :

\beq \Psi(\mb{r,R})=\sum_m \chi_m(\mb R)\phi^{el}_m(\mb {r,R}).\eeq

Ahora sustituimos esta funcin en la ecuacin \ref{sch}, multiplicando por $\phi^{el*}_n$ e integrando sobre las coordenadas electrnicas $\mb r$ se obtiene

\beq \int\left[ \phi^{el*}_n(\mb {r,R})(\hat H- E)\sum_m\chi_m(\mb R)\phi^{el}_m(\mb {r,R})\right]d\mb r=0.\eeq

Sustituyendo $\hat H=\hat H_0+\hat T_{nuc}$ en la ecuacin anterior y usando \ref{schnp} se obtiene:

\beq \left( E_n^{(0)}(\mb R) -E\right)\chi_n(\mb R)+\int\left[\phi^{el*}_n(\mb {r,R})\hat H'\sum_m\chi_m(\mb R)\phi^{el}_m(\mb {r,R})\right]d\mb r=0\eeq
y el ltimo trmino es explcitamente: 

\ber\int\phi^{el*}_n\left(\hat H'\sum_m\chi_m\phi^{el}_m\right)d\mb r &=&\int\left[\phi_n^*\sum_m (\hat H'\chi_m)\phi_m\right]d\mb r+\int \left[\phi^*_n\sum_m(\hat H'\phi_m)\chi_m\right]d\mb r\nnum
\\&&-\hbar^2\int\phi_n^*\left[\sum_k\frac{1}{M_k}\sum_m \frac{\partial}{\partial \mb R_k}\phi_m\frac{\partial}{\partial \mb R_k}\chi_m\right]d\mb r.\eer
Reduciendo obtenemos:
 
 \beq \left( E_n^{(0)}(\mb R)+\hat H'\right)\chi_n(\mb R)+\sum_mc_{nm}\chi_m(\mb R)=E\chi_n(\mb R),\label{schnuc}\eeq

donde los coeficientes $c_{nm}$ son de la forma:

\beq c_{nm}=\int\phi_n^*\hat H'\phi_md\mb r-\frac{\hbar^2}{2}\left [\int\phi_n^*\sum_k\frac{1}{M_k}\frac{\partial}{\partial \mb R_k}\phi_md\mb r\right]\frac{\partial}{\partial \mb R_k}.\eeq

Las ecuaciones \ref{schnp} y \ref{schnuc} forman un conjunto acoplado de ecuaciones para la funciones de onda electrnica y nuclear acopladas mediante los coeficientes $c_{mn}(\phi)$ que dependen de $\phi$:

\bse
\begin{align}
\hat H_0(\mb R)\phi^{el}_n(\mb {r,R})&=E_n^{(0)}(\mb R)\phi^{el}_n(\mb {r,R}),\label{equiva}\\
\hat H'\chi_n(\mb R)+\sum_mc_{nm}\chi_m(\mb R)&=\left(E- E_n^{(0)}(\mb R)\right)\chi_n(\mb R).\label{equivb}
\end{align}
\ese

Las ecuaciones  \ref{equiva} y \ref{equivb} son equivalentes a la ecuacin de Schrdinger \ref{ham}. La ecuacin \ref{equivb} describe el movimiento del ncleo con energa cintica $\hat H'$ en el potencial $E_n^{(0)}(\mb R)$ que es la solucin de la ecuacin \ref{equiva}. Los coeficientes $c_{nm}$ son elementos de matriz de acoplamiento y describen cmo se acoplan los estados $\phi_n$ y $\phi_m$ a travs de movimiento nuclear. \\

La aproximacin de \textit{Born-Oppenheimer} (BO) desprecia el acoplamiento entre el movimiento de los electrones y el movimiento nuclear, los coeficientes $c_{nm}$ son cero, por lo tanto

 \beq \left( E_n^{(0)}(\mb R)+\hat H'\right)\chi_n(\mb R)=E\chi_n(\mb R),\label{bonu}\eeq
donde

\beq \hat H_{\text{nuc}}=\hat H'+E_n^{(0)}(\mb R)=\hat T_{\text{nuc}}+U_n(\mb R).\eeq
Notamos justamente que la energa total de la molcula rgida  $E^{(o)}_n(\mb R)$ es la energa potencial    $U_n(\mb R)$ de los ncleos. Entonces la aproximacin BO separa  la ecuacin de Schrdinger \ref{sch} en dos ecuaciones desacopladas:

\beq \hat H_0\phi^{el}_n(\mb {r})=E_n^{(0)}\phi^{el}_n(\mb {r}),\label{boel}\eeq
que corresponde a la parte electrnica y
 \beq \left( \hat T_{\text{nuc}}+E^{(0)}_n\right)\chi_{n,i}(\mb R)=E_{n,i}\chi_{n,i}(\mb R)\label{bonu},\eeq
 
 que corresponde al ncleo acoplado.

\section{Mtodo de Hartree-Fock (HF)}

El mtodo de Hartree-Fock data de los aos 20, justo despus del planteamiento de la ecuacin de Schrdinger. En 1927  Douglas Hartree introdujo un procedimiento para aproximar funciones de onda de tomos e iones, conocido como mtodo de campo autoconsistente (SCF por sus siglas en ingls, \textit{Self-Consistent-Field}).

Utilizando la aproximacin de Born-Oppenheimer, el mtodo de Hartree consiste en aproximar el hamiltoniano electrnico de un sistema atmico de muchos electrones por un hamiltoniano compuesto que sea la suma de muchos hamiltonianos de electrn independiente\cite{messiah}. Para ello, cada electrn estar bajo la accin de un potencial efectivo para un electrn $i(1<i<N)$ que contiene el potencial coulombiano del ncleo ms el potencial promedio de la distribucin de carga de los otros $(N-1)$ electrones. \\

 De acuerdo con la ecuacin \ref{schnp}, $\hat H_0$ adems de la energa cintica del electrn incluye la interaccin ncleo-electrn, electrn-electrn y la interaccin ncleo-ncleo. Como $V_{\text{nuc,nuc}}$ es una constante para un conjunto de coordenadas nucleares $\{\mb R\}$ fijas, lo ignoraremos teniendo en cuenta que ste no cambia las autofunciones y slo corre los autovalores de la energa. Usando como notacin $\mb r_i$ las posiciones de los electrones y $r_{ij}=|\mb r_i-\mb r_j|$, los operadores:
 
 \beq \hat H_0=\sum _{i=1}^N  \hat  h_i+\sum_{i<j}^N \frac{1}{r_{ij}},\eeq
 
 con
 \beq \hat  h_i=\half \nabla_i^2-\upsilon(\mb r_i).\eeq
 
 y $\upsilon(\mb r_i)$ dado por la ecuacin \ref{v_nuc}. Ahora, el mtodo de Hartree- Fock \cite{fock} introduce consideraciones mecnico cunticas en el anlisis del problema de muchos electrones, tales como la condicin de antisimetra aplicada a la funcin de onda. Debido a que los electrones son fermiones, la funcin de onda total debe ser antisimtrica con respecto al intercambio de cualquier par de electrones. Esto es equivalente a exigir que los electrones cumplan el principio de exclusin de Pauli. Una funcin de onda con estas caractersticas se puede representar por medio del determinante de Slater \cite{messiah}. Como notacin se usan kets en la representacin de coordenadas:

\begin{align} \ket{\Phi_{HF}}&= \frac{1}{\sqrt{N!}}\left|\barrf \phi_1(\mb r_1)& \phi_2(\mb r_1)&\ldots& \phi_N(\mb r_1)\\
\phi_1(\mb r_2)& \phi_2(\mb r_2)&\ldots& \phi_N(\mb r_2)\\
\vdots&\vdots&\ddots&\vdots\\
\phi_1(\mb r_N)& \phi_2(\mb r_N)&\ldots& \phi_N(\mb r_N)\earrf\right|\\
&=\frac{1}{\sqrt{N!}}det[\phi_1\phi_2\ldots\phi_N].
\end{align}

El valor esperado de la energa es \cite{parr}

\beq E_{HF}= \bra{\phi_{HF}}\hat H_0\ket{\phi_{HF}}=\sum_{i=1}^{N}H_i+\frac{1}{2}\sum_{i,j=1}^{N}(J_{ij}-K_{ij})\label{hfe},\eeq

donde

\beq H_i=\int \phi_i^*(\mb r)[-\half\nabla^2+\upsilon(\mb r)]\phi_i(\mb r)d\mb r,\eeq

\beq J_{ij}=\int\int \phi_i(\mb r_1)\phi_i^*(\mb r_1)\frac{1}{r_{12}}\phi_j(\mb r_2)\phi_j^*(\mb r_2)d\mb r_1d\mb r_2,\eeq

\beq K_{ij}=\int\int \phi_i(\mb r_1)\phi_j^*(\mb r_1)\frac{1}{r_{12}}\phi_i(\mb r_2)\phi_j^*(\mb r_2)d\mb r_1d\mb r_2.\eeq

Estas integrales son  reales y $J_{ij}\geq K_{ij}\geq 0$. Las $J_{ij}$ son llamadas \textit{integrales Coulombianas} y las $K_{ij}$ son \textit{integrales de intercambio}.\\

Sujeta a la condicin de ortonormalizacin 

\beq \int \phi_i^*(\mb r) \phi_j(\mb r)d\mb r=\delta_{ij},\label{norm}\eeq

se minimiza la energa \ref{hfe} y se obtienen las ecuaciones diferenciales de Hartree-Fock:

\beq \hat F\phi_i(\mb r)=\sum _{j=1}^N\epsilon_{ij}\phi_j(\mb r),\label{hfeq}\eeq

donde el operador de Fock, $\hat F$, es

\beq\hat F=-\half\nabla^2+\upsilon+\hat g\label{fock}\eeq

y el operador Coulombiano-Intercambio, $\hat g$, es dado por 

\beq \hat g=\hat j-\hat k \eeq

tal que la forma en la que actan sobre una funcin arbitraria $m(\mb r)$ es

\beq \hat j(\mb r_1)m(\mb r_1)= \sum_{k=1}^N\int \phi_k^*(\mb r_2)\phi_k(\mb r_2)\frac{1}{r_{12}}m(\mb r_1)d\mb r_2\eeq

y

\beq  \hat k(\mb r_1)m(\mb r_1)= \sum_{k=1}^N\int \phi_k^*(\mb r_2)m(\mb r_2)\frac{1}{r_{12}}\phi_k(\mb r_1)d\mb r_2.\eeq
 
La matriz $\mb \epsilon$ est compuesta de los multiplicadores de Lagrange asociados con la condicin \ref{norm} y es una matriz hermitiana. Ahora multiplicando la ecuacin \ref{hfeq} por $\phi_i^*$ e integrando se obtiene la ecuacin para las energas orbitales:

\beq \epsilon_i=\epsilon_{ii}=\bra{\phi_i}\hat F\ket{\phi_i}=H_i+\sum_{j=1}^N(J_{ij}-K_{ij}).\eeq

Sumando sobre $i$ y comparando con la ecuacin \ref{hfe}, se obtiene:

\beq E_{HF}=\sum_{i=1}^N \epsilon_i-V_{\text{el,el}},\label{hfenergy}\eeq

donde $V_{\text{el,el}}$ es la energa total de repulsin electrn-electrn

\begin{align}
V_{\text{el,el}}&=\int \Phi_{HF}^*(\mb r^N)(\sum_{i<j}\frac{1}{r_{ij}})\Phi_{HF}(\mb r_N)d\mb r^N\\
&= \half \sum_{i,j=1}^N(J_{ij}-K_{ij}).
\end{align}

Sea $ W_{HF}$ la energa total molecular incluyendo la repulsin nuclear-nuclear:

\ber W_{HF}&=\sum_{i=1}^N \epsilon_i-V_{\text{el,el}}+V_{\text{nuc,nuc}}\\
&=\sum_{i=1}^N H_i+V_{\text{el,el}}+V_{\text{nuc,nuc}}.\eer

Es de resaltar que ni $E_{HF}$ ni $W_{HF}$ es igual a la suma de las energas orbitales.

Las ecuaciones de Hartree-Fock no son lineales por lo tanto no cumplen con el principio de superposicin, entonces  el procedimiento necesario para solucionar las ecuaciones de Hartree-Fock \ref{hfeq} numricamente es usando funciones de prueba en el principio variacional. A partir de las funciones de onda de un slo electrn $\phi_i^{(0)}$ ($1\leq i\leq N$) que se dan como funciones de prueba  se computa el potencial efectivo. Con esto se construye la matriz de Fock y se diagonaliza para obtener la energas $\epsilon_i$. Luego usando el principio variacional \cite{sakurai} se minimizan las energas para conseguir funciones de onda de un electrn mejoradas $\phi_i^{(1)}$. Este proceso se repite el nmero $k$ de iteraciones hasta que las $\phi_i^{(k)}$ no difieran mucho de las $\phi_i^{(k-1)}$. Estas funciones de onda optimizadas se denominan funciones de onda de \textit{campo autoconsistente}. Dicho procedimiento se conoce como mtodo de campo autoconsistente . \\



\section{Teora del Funcional de la Densidad (DFT)}

Despus de usar la aproximacin de Bohr-Oppenheimer el problema de muchos cuerpos es ms simple de tratar pero an es difcil de resolver. La teora del funcional de la densidad consiste en trabajar, sin prdida de rigor, con la densidad de electrones $\rho (\pmb r)$ como una variable bsica, en vez de la funcin de onda $\Psi(\pmb r_1 ,s_1,\pmb r_2 , s_2, \ldots ,\pmb r_n,s_n )$. La densidad $\rho$ es simplemente la densidad tridimensional de una sola partcula que se evidencia en experimentos de difraccin. Por otro lado la teora cuntica del estado base se puede poner en trminos de $\rho$. En la teora del funcional  se trabaja con las ecuaciones de Kohn-Sham \cite{ks} que son similares a las ecuaciones de Hartree-Fock, incluyen efectos de intercambio y efectos de correlacin. Este mtodo es empleado en muchos clculos realizados para slidos  \cite{solids} y ha ido creciendo su aplicacin en tomos y molculas \cite{parr}. El mtodo del funcional de la densidad se vuelve teora  con las publicaciones de Kohn y Hohenberg \cite{dft} (1964), en donde establecen con dos teoremas: la existencia de un funcional $E[\rho]$ y el un principio variacional para dichos funcionales. \\

 \textit{ Primer teorema (1964)}: \textit{El potencial externo $v(\mb r)$ es determinado, hasta una constante, por la densidad electrnica $\rho (\mb r)$}. $\rho$ determina tambin el nmero de electrones $N$ y con ste la funcin de onda del estado base $\Psi$ y por lo tanto todas las propiedades electrnicas del sistema. Es de resaltar que $\upsilon$ no se restringe al potencial coulombiano. \\
  
 La prueba de este teorema es bastante intuitiva. Si hubiesen dos potenciales $\upsilon$ y $\upsilon '$ que difieren por ms que una constante, a partir de los cuales se obtiene la misma $\rho$ para el estado base, entonces se tendran dos hamiltonianos $H$ y $H'$ cuyas densidades del estado base son la misma obtenida a partir de diferentes funciones de onda $\Psi$ y $\Psi '$. Si se toma  $\Psi '$ como la funcin de prueba para el problema de $\hat H$, se tendra entonces:
 
\ber E_0<\bra{\Psi'} \hat H\ket{\Psi '}&=\bra{\Psi'} \hat H'\ket{\Psi '}+\bra{\Psi'} \hat H-\hat H'\ket{\Psi '} \\
&=E_0'+\int \rho(\mb r)[\upsilon(\mb r)-\upsilon'(\mb r)]d\mb r\label{prueba1}\eer

donde $E_0$ y $E_0'$ son las energas del estado base para $\hat H$ y $\hat H'$ respectivamente. De la misma manera usamos $\Psi$ como funcin de prueba para el problema de $\hat H'$:

\ber E_0'<\bra{\Psi} \hat H'\ket{\Psi}&=\bra{\Psi} \hat H\ket{\Psi}+\bra{\Psi} \hat H'-\hat H\ket{\Psi} \\
&=E_0-\int \rho(\mb r)[\upsilon(\mb r)-\upsilon'(\mb r)]d\mb r\label{prueba2}\eer

Sumando las ecuaciones \ref{prueba1} y \ref{prueba2} se obtiene una contradiccin:

\beq E_0+E_0'<E_0'+E_0\eeq

Por lo tanto no pueden haber dos diferentes potenciales que den el mismo $\rho$ para sus estados base.\\

Como  $\rho$ determina $N$ y $v$ a partir de stos se pueden determinar las propiedades del estado base as como la energa cintica $T[\rho]$, la energa potencial $V[\rho]$ y la energa total $E[\rho]$  que ahora escribiremos como $E_v[\rho]$ para hacer explcita su dependencia con $\upsilon$\\

 \beq E_v[\rho]=T[\rho]+V_{ne}[\rho]+V_{ee}[\rho]=\int\rho(\mb r)v(\mb r)d\mb r+ F_{HK}[\rho]\label{eqhk}\eeq
 
 donde $F_{HK}$ es 
 
 \beq F_{HK}=T[\rho]+V_{ee}[\rho]\label{fhk}\eeq
 escribimos
  \beq V_{ee}[\rho]=J[\rho]+\text{trminos no clsicos}\label{prinvar}\eeq
 
 donde $J[\rho]$ es la repulsin clsica.\\
 
 \textit{Segundo teorema HK}:  \textit{Para una densidad de prueba} $\tilde\rho(\mb r)$, \textit{tal que} $\tilde\rho(\mb r)\geqslant0$ y $\int\tilde\rho(\mb r)d\mb r=N$

\beq E_0\leqslant E_v[\tilde\rho]\eeq
 \textit{donde} $E_v[\tilde\rho]$ \textit{es el funcional energa de la ecuacin} \ref{eqhk}. Este es anlogo al principio variacional para las funciones de onda.\\
 
 Para probar este teorema, teniendo en cuenta  que $\tilde\rho$ determina su propio $\tilde\upsilon$ y la funcin de onda $\tilde\Psi$  puede ser tomada como funcin de prueba para el problema de inters que tenga el potencial externo $\upsilon$:

\beq \bra{\tilde\Psi}\hat H\ket{\tilde\Psi}=\int\tilde\rho(\mb r)\upsilon(\mb r)d\mb r+F_{HK}[\tilde\rho]=E_v[\tilde\rho]\geq E_v[\rho].\eeq

Asumiendo la diferenciabilidad de $E_v[\rho]$ y recordando que el principio variacional requiere que la densidad del estado base satisfaga:

\beq \delta\left\{E_v[\rho]-\mu\left[\int\rho(\mb r)d\mb r-N\right]\right\}=0\label{delta},\eeq

de las que obtenemos las ecuaciones de Euler Langrange \cite{finn}

\beq \mu=\frac{\delta E_v[\rho]}{\delta\rho(\mb r)}=\upsilon(\mb r)+\frac{\delta F_{HK}[\rho]}{\delta\rho(\mb r)}.\eeq

Es de notar que si conocemos $F_{HK}[\rho]$, la ecuacin \ref{delta} sera una ecuacin exacta para la densidad del estado electrnico. Como $F_{HK}[\rho]$ en \ref{fhk}es definida independiente del potencial externo $\upsilon(\mb r)$ por lo que a $F_{HK}[\rho]$ se conoce como \textit{potencial universal de } $\rho(\mb r)$, una vez se tenga la forma explcita de $F_{HK}[\rho]$, bien sea aproximada  o exacta, se puede aplicar este mtodo a cualquier sistema.


 \subsection{Mtodo de Kohn y Sham}

Los teoremas de Hohenberg y Kohn nos garantizan que si conocemos la densidad electrnica para el estado base de un sistema $\rho_0(\mb r)$, es posible calcular todas las propiedades del sistema a partir de sta sin tener que usar la funcin de onda. Sin embargo no nos dicen cmo calcular $E_0$ a partir de $\rho_0$, (pues \ref{fhk} es desconocido) y tampoco sabemos como obtener $\rho_0$ sin primero encontrar la funcin de onda.\\

Kohn y Sham en 1965, especulando si $\rho(\mb r)$ puede ser descompuesta de manera nica en trminos de orbitales monoelectrnicos tal que d un nico valor $T[\rho]$,  introdujeron un sistema auxiliar o ficticio  llamado \textit{sistema Kohn-Sham}, compuesto por $N$ partculas no interactuantes, tal que experimenta el mismo potencial externo que el sistema de inters $\upsilon(\mb{ r})$.\\

En DFT la energa cintica del sistema se expresa en trminos del funcional de la densidad electrnica. Como notacin se usa, en la representacin de coordenadas,  $\psi_i(\mb r)=\braket{\mb r}{\psi_i}$:

\beq T_s[\rho]=\sum_i^N\bra{\psi_i}\hat{T}\ket{\psi_i}=\sum_i^N\bra{\psi_i}-\frac{1}{2}\nabla^2\ket{\psi_i}\label{funcT}\eeq

y

\beq \rho (\mb r)=\sum_i^N |\psi_i(\mb r)|^2\label{ksden}\eeq

Kohn y Sham reescribieron la ecuacin de Hohenberg-Kohn \ref{eqhk} introduciendo las siguientes definiciones:

\beq\Delta\bar{T}[\rho]=\bar{T}[\rho]-\bar{T}_s[\rho]\eeq

siendo $\Delta\bar{T}$ la diferencia entre el promedio de la energa cintica del estado base del sistema de inters y el sistema de referencia ficticio. Sea:

\beq \Delta\bar{V}_{ee}[\rho]=\bar{V}_{ee}[\rho]-\frac{1}{2}\int\int\frac{\rho(\mb r_1)\rho(\mb r_2)}{r_{12}}d\mb r_1d\mb r_2\eeq

siendo la ltima parte de la ecuacin, la expresin de la repulsin electrnica en unidades atmicas considerando una distribucin de carga uniforme, con $r_{12}=|\mb r_1-\mb r_2|$. Ahora reescribimos \ref{eqhk} usando las definiciones anteriores:

\beq E_{\upsilon}[\rho]=\int\rho_0(\mb r)\upsilon(\mb r)d\mb r+\bar{T}_s[\rho]+\frac{1}{2}\int\int\frac{\rho(\mb r_1)\rho(\mb r_2)}{r_{12}}d\mb r_1 d\mb r_2+\Delta\bar{T}[\rho]+\Delta\bar{V}_{ee}[\rho]\eeq

Los funcionales $\Delta \bar{T}$ y $\Delta \bar{V}_{ee}$ son desconocidos. A partir de stos se puede definir el \textit{funcional de la energa de correlacin e intercambio} $E_{xc}[\rho]$ como:

\beq E_{xc}[\rho]=\Delta \bar{T}[\rho]+\Delta \bar{V_{ee}}[\rho]\eeq 

por lo tanto:

\beq E_{\upsilon}[\rho]=\int\rho(\mb r)\upsilon(\mb r)d\mb r+\bar{T}_s[\rho]+\frac{1}{2}\int\int\frac{\rho(\mb r_1)\rho(\mb r_2)}{r_{12}}d\mb r_1 d\mb r_2+E_{xc}[\rho]\label{kse}\eeq

La clave para obtener una buena precisin en el mtodo de Kohn-Sham est tambin en el aproximar bien a $E_{xc}$. Finalmente, el teorema variacional de Hohenberg-Kohn dice que se puede encontrar la energa del estado base variando $\rho$  sujeto a la condicin $\int\rho(\mb r) d\mb r = N$ tal que minimize el funcional $E_{\upsilon}[\rho]$. 
Esto es equivalente a, en vez de variar $\rho$, variar los orbitales $\psi_i$ que determinan $\rho$ en donde adems se adiciona la siguiente condicin para que el funcional de la energa cintica \ref{funcT} siga siendo vlido:

\beq  \int \psi_i^*(\mb r)\psi_j(\mb r) d\mb r= \delta_{ij}\eeq

Por lo tanto, as como se puede mostrar que los orbitales ortonormales que minimizan la energa de Hartree-Fock \ref{hfe} satisfacen la ecuacin de Fock \ref{hfeq}, se puede mostrar que los orbitales que minimizan la ecuacin \ref{kse} satisfacen (ver prueba en seccin 7.2 \cite{parr}):

\beq \left[T_s+\upsilon(\mb r)+\int\frac{\rho(\mb r_2)}{r_{i2}}d \mb r_2+\upsilon_{xc}\right]\psi_i=\epsilon_i\psi_i\label{kseq},\eeq

que es la ecuacin de Kohn y Sham y $\psi_i$ los orbitales de Kohn-Sham, $\upsilon_{xc}$ es el \textit{potencial de intercambio y correlacin} obtenido de:

\beq \upsilon_{xc}(\mb r)= \frac{\delta E_{xc}[\rho(\mb r)]}{\delta \rho(\mb r)}\eeq

Como el funcional $E_{xc}$ es desconocido se usan diversas aproximaciones, cuya precisin ha sido estudiada usndolos para el clculo de propiedades que puedan ser comparadas con datos experimentales en diferentes molculas. Los orbitales de Kohn-Sham $\psi_i$, son orbitales para el sistema ficticio de electrones no interactuantes, por lo tanto estrictamente hablando, no tienen significado fsico ms all de permitir obtener la densidad \ref{ksden}.



\chapter{Rompimiento de Simetra en Molculas}

\section{Transicin de Peierls }

La transicin de Peierls es cualitativamente una distorsin de la red peridica de un cristal unidimensional. El presente tratamiento se sigue de \cite{fowler}.

\subsection{Teorema de Peierls}

Rudolf Peierls descubri en 1930 que una cadena 1-dimensional de iones igualmente espaciados es \textit{inestable}\cite{peierls}.

Para entender la distorsin de Peierls se considera el modelo simple de una cadena de $N$ iones igualmente distanciados por $a$. La longitud total de la cadena es $L=Na$:

\hspace{-0.7cm}\begin{center}
\includegraphics[height=0.2\textheight]{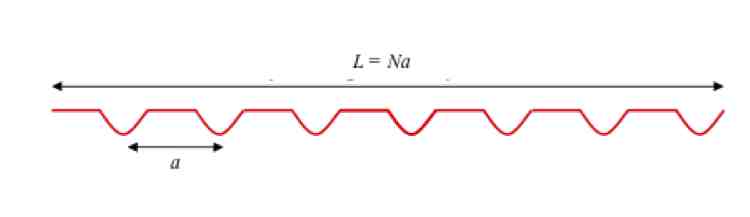}
\label{chain}
\end{center}

Para dicho sistema consideramos el siguiente hamiltoniano:

\beq H=H_0+V=\frac{p^2}{2m}+V(x)\eeq

donde $H_0$ es el hamiltoniano de un gas de electrones no interactuantes en una lnea de longitud $L$ y $V$ es un potencial peridico debido a los iones, es decir, $V(x+a)=V(x)$.

Las autofunciones de $H_0$ son ondas planas, imponiendo condiciones de frontera peridicas:

\beq \ket{k}=\frac{1}{\sqrt L}e^{ikx}\quad \textit{con}\quad k=\frac{2\pi n}{L},\label{val_k}\eeq

siendo $n$ un entero. Donde la energa :

\beq H_0\ket{k}=E_0\ket{k}, \quad \textit{con}\quad E_0=\frac{\hbar^2 k^2}{2m}.\eeq

Suponiendo que hay dos electrones por cada in, ser un total de $2N$ electrones y de acuerdo  a los posibles valores de $k$ la densidad en el espacio de las $k $ debe ser de $L/2\pi$. Para este sistema sabemos que a la temperatura de Fermi \cite{kittel} existe un mximo valor de la energa que corresponde a los valores $+ k_F$ y $-k_F$.  De acuerdo a la ecuacin \ref{val_k}, los posibles valores de $k$ estn espaciados $2\pi/L$, por lo tanto,  el nmero de $k$ permitidas entre $- k_F$ y $+ k_F$ es $Lk_F/\pi$. Como dos electrones pueden tener el mismo valor de $k$ con diferente espn por lo tanto $N=Lk_F/\pi=L/a$, as $k_F=\pm\pi/a$.\\


Por otro lado usando el teorema de Fourier, el potencial, considerandolo como una funcin peridica contnua, puede ser expresado como:

\beq V(x)= \sum_{q=0}^\infty u_q e^{i\frac{2\pi}{a}qx}\eeq

Ahora usaremos la teora de perturbaciones:

\beq H=H_0+\lambda V\eeq

Los elementos de matriz del potencial en la base de los kets propios del hamiltoniano no perturbado son:

\beq V_{kk'}=\bra{k}V\ket{k'} \label{elements}\eeq
\beq V_{kk'}=\frac{1}{L}\int_0^L e^{i(k-k')x}V(x)dx\eeq
\beq V_{kk'}=\frac{1}{L}\sum_{q=0}^\infty u_q\int_0^L e^{i(k-k'-\frac{2\pi}{a}q)x}dx\eeq

En donde los nicos trminos que sobreviven son aquellos que cumplen 
\beq q=a(k-k')/2\pi=0,\pm 1, \pm 2, ...\label{qvalues}\eeq

De esta manera podemos calcular la correcciones a la energa:\\

i) A primer orden $k=k'$ tenemos el potencial promediado en el espacio cuya constante $u_o$ se puede hacer cero y no obtenemos correccin a la energa a primer orden.\\

ii) A segundo orden:\\

De la ecuacin \ref{elements} debemos considerar los elementos de matriz con $k\neq k'$. La diferencia entre los correspondientes estados es:

\beq \frac{\hbar^2}{2m}\left\{ \left( k-\frac{2\pi q}{a}\right)^2-k^2\right\}=\frac{2\pi\hbar^2q}{ma}\left(\frac{\pi q}{ma}-k\right)\label{difener}\eeq

ste trmino se anula cuando $k=\pi q/a$ y en tal caso $k'=-\pi q/a $, por lo tanto en la vecindad de esos estados la teora de perturbaciones no converge. 
Por otro lado para estados lejanos a estos casos excepcionales la cantidad \ref{difener} es finita, por lo tanto en esa zona la correccin a la energa a segundo orden en $V$, asumiendo que la diferencia de energas de los estados $k$ y $k'$ es mayor que $V_{kk'}$,   es:

\beq E_k^2= \sum_{q} \frac{\mid V_{k(k-(2\pi q/a))}\mid^2}{E_k^0-E_{k-(2\pi q/a)}^0}\eeq

En la vecindad de los estados degenerados usaremos la teora de perturbaciones para el caso degenerado, para ello diagonalizamos el hamiltoniano completo o perturbado en el subespacio expandido por $\ket{k}^0$ y $\ket{k-K}^0 $, con $K=2\pi /a$

\beq\bmat 
E_k^0 & V_K^*\\
V_K  & E_{k-K}^0
\emat\eeq

 Diagonalizando obtenemos los valores propios:
 
 \beq E_\pm=\frac{1}{2}(E_k^0+E_{k-K}^0)\pm\sqrt{\left(\frac{E_k^0-E_{k-K}^0}{2}\right)^2+|V_K|^2} \eeq
 
 Si $|E_k^0-E_{k-K}^0|\ggg |V_K|$ las correcciones resultantes son cercanas a $E_k^0$ y $E_{k-K}^0$ respectivamente.  Por otro lado si $k$  se aproxima a $\pi/a$ entonces  $|E_k^0-E_{k-K}^0|$ es del orden de $|V_K|$ y las energas se alejan de los valores no perturbados, es decir, si $k\to\pi/a$ por debajo, $E_k^0<E_{k-K}$ y la energa ms pequea baja por la perturbacin $E_k=E_-<E_k^0$ por lo tanto se dice que la perturbacin causa que los niveles de energa se repelen.\\

 \begin{figure}[h!]\centering
\includegraphics[height=0.27\textheight]{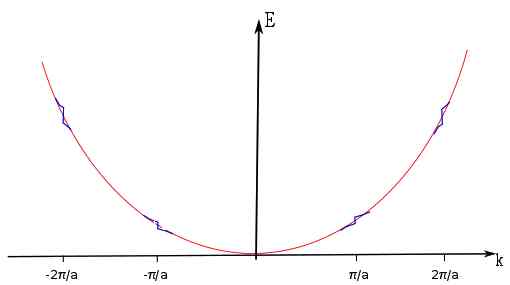}
\caption{Ilustra el cambio en la energa de los electrones libre no interactuantes debido a la perturbacin peridica. Se muestran los casos $q=\pm 1,\pm 2$}
\label{poten}
\end{figure}

  Para $k=\pi/a$, $E_{k-K}^0=E_k^0$, se levanta la degeneracin para dar $E_\pm =E_{\pi/a}^0\pm|V_K|$ como lo muestra la figura \ref{poten} con una brecha o gap de $2 |V_K|$. La banda ms baja, es decir,  el conjunto de energas permitidas ms bajas, est en el rango de $k=-\pi/a$ a $k=\pi/a$.  Recordando que los valores permitidos de $k$ son $k=2\pi n/L$, el espaciamiento entre los diferentes valores de $k$ es de $2\pi/L$ por lo tanto en la banda de energa ms baja el nmero total de valores permitidos de $k$ es $L/a$ que es igual al nmero de iones. Para el caso en consideracin de un cristal uni-dimensional divalente (con dos electrones por in) sta banda se llenara completamente teniendo en cuenta los dos estados de spin posibles por cada electrn.  Si se considera  un cristal monovalente (con un electrn por in), la banda de energa ms baja slo se llenar a la mitad, por lo tanto se esperara que a temperatura T=0K un material monovalente unidimensional sera un conductor. Contrario a esto, en 1976 se sintetizaron varios materiales orgnicos \cite{onedim-conduct} que dispuestos de cierta manera, reproducan un conductor unidimensional y al enfriarlo sorprendente el material se volvi un aislante. La explicacin a este fenmeno lo da Peierls \cite{peierls} con la dimerizacin en donde los tomos pasan de una red igualmente espaciada a una red donde los tomos se acomodan en pares, duplicando as el periodo de la misma. \\

 \subsection{Disminucin en la energa electrnica debido a dimerizacin de la red.}

 \begin{center}
\includegraphics[height=0.07\textheight]{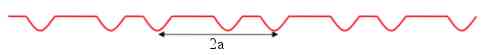}
\end{center}
 
 La mayor contribucin en el cambio de la energa viene de valores cercanos a $k=\pi/2a$ y simtricamente a $k=-\pi/2a$ para el caso de $q=1$ en la ecuacin \ref{qvalues} teniendo en cuenta que el periodo ahora es . Se hace el mismo tratamiento para los dems valores de $q$. 
 
 \beq \Delta E=2\int_0^{\pi/2a}(E_k^0-E_-)g(k) dk\eeq
 
 donde el 2 se debe a que se incluye la contribucin simtrica de la brecha, $g(k)$ es la densidad de estados en el espacio $k$ que es $L/\pi$ teniendo en cuenta el espn  y siendo $K=\pi/a$ para este caso:
 
 \beq  \Delta E =2\int_0^{\pi/2a}\left(\frac{1}{2}(E_k^0-E_{k-K})+\sqrt{\left(\frac{E_k^0-E_{k-K}}{2}\right)^2+|V_K|^2}\right) \frac{L}{\pi} dk\eeq

 \beq E_k^0-E_{k-K}= \frac{\hbar^2k^2}{2m}-\frac{\hbar^2(k-K)^2}{2m}\eeq
 
  \beq E_k^0-E_{k-K}= bk-c\eeq
  
  donde $b=\hbar^2\pi/ma$ y $c=\hbar^2\pi^2/2ma^2$. Ahora definiendo la variable $s=bk-c$ que es una variable negativa dentro de rango considerado para $k$, se tiene:
  
\beq\Delta E= \frac{2L}{\pi b}\int_{-c}^0 \left( s+\sqrt{s^2+|V_K|^2}\right)ds  \eeq
 
 Cabe resaltar que en ausencia de potencial $\Delta E$ es cero. Resolviendo la integral se obtiene:
 
 \beq \Delta E = \frac{c^3}{3}+\frac{c}{2}\sqrt{c^2 + V_K^2} + \frac{V_K^2}{2} \log{\left[\frac{V_K}{-c+\sqrt{c^2+V^2_K}}\right]}\label{dimeri} \eeq

   Para valores de $V_K$ pequeos notamos que el trmino logartmico es grande (negativo) y por lo tanto el ltimo trmino de la ecuacin \ref{dimeri} tambin lo es. Por otro lado el costo energtico para llevar a cabo la dimerizacin se modela como energa elstica que debe ser proporcional a $V_K^2$. Comparando  la disminucin de energa debido al doblamiento del perodo con la energa elstica para poder efectuar dicho doblamiento, se puede concluir que es mucho ms lo que disminuye la energa con la dimerizacin que lo que tiene que invertir la molcula para realizarla, por lo tanto la dimerizacin tiene mayor probabilidad de ocurrir y es as como una cadena de iones igualmente espaciada con un electrn por in es inestable.
 



\section{Efecto Jahn-Teller}

El efecto Jahn-Teller describe la distorsin de una molcula no lineal si sus estados electrnicos son degenerados. Dicho efecto nombrado en honor  a Herman Arthur Jahn y Edward Jahn Teller quienes lo probaron usando teora de grupos. En 1934 E. Teller y Lev Landau estudiaban la estabilidad de la molcula $CO_2$ \cite{teller-landau}. El estudiante de Teller, Rudolph Renner demostr que una molcula lineal que tiene estados electrnicos degenerados, debe ser estable, a lo cual Landau objet  argumentando que en los estados degenerados la simetra sobre la cual la degeneracin est basada se destruye. Jahn y Teller empezaron a trabajar en ello demostrando que el nico caso en el que  no se cumple es en el caso de la molcula lineal.

El efecto Jahn Teller puede ser visto como un ejemplo de la interaccin electrn-fonn, que es la interaccin entre electrones y iones los cuales intercambian energa destruyendo o creando fonones. Esto quiere decir que mientras el conjunto de iones pasa de un estado vibracional a otro, un electrn pasa de un estado electrnico a otro (de una banda a otra) agregando o restando energa y momento de tal manera que la energa total del sistema permanezca constante.  En el efecto Jahn-Teller debe haber una multiplicidad entre los estados electrnicos con uno o ms modos normales de vibracin y el teorema de Jahn-Teller dice que casi cualquier conjunto de estados electrnicos degenerados asociados con una configuracin electrnica habr alguna interaccin de rompimiento de simetra en la cual la distorsin se asocia con la eliminacin de la degeneracin en los estados electrnicos. La prueba de ste teorema est basada en teora de grupos que afirma que tomando en cuenta todos los posibles puntos de grupo de simetras bajo los cuales un sistema tal como  una molcula puede ser invariante y tambin tomando en cuenta todos los modos normales de cada sistema, clasificados por su simetra. Como todos los estados electrnicos de tal sistema pueden ser clasificados por su simetra y cada uno identificado  por una representacin irreducible del punto de grupo para una configuracin inica simtrica. Se puede mostrar que la interaccin electrn-fonn es permitida por consideraciones de simetra en casi todos los casos. Las excepciones son molculas lineales y molculas con estados electrnicos degenerados de Kramers.

La interaccin es permitida pero se hace la siguiente suposicin: que cualquiera que sea permitido puede realmente ocurrir. La distorsin capaz de levantar la degeneracin puede ser clasificada por simetra usando representaciones irreducibles, la teora de grupos nos dice cuales son. Para sistemas con nmero par de electrones la representacin irreducible de simetra se identifica por la distorsin que debe ocurrir en el cuadro simtrico formado usando los estados electrnicos de la representacin irreducible; para sistemas con nmero impar de electrones se usa el cuadro antisimtrico.


\chapter{Aspectos Computacionales}

Hemos visto que los mtodos a usar son bsicamente mtodos variacionales que usan estados o funciones iniciales de prueba. La exactitud de dichos mtodos depende en gran parte de la buena eleccin de dichas funciones, pero es difcil saber cuando se ha hecho una buena eleccin. 

\section{Aproximacin de Combinacin Lineal de Orbitales Atmicos}

En la aproximacin de Combinacin Lineal de Orbitales Atmicos (LCAO por sus siglas en ingls \textit{Linear Combination of Atomic Orbitals)} se aproxima la funcin de onda molecular $\phi$, que representa el estado de una molcula, por medio de una combinacin lineal de las funciones de onda cuyo mdulo al cuadrado representa la densidad electrnica de cada tomo, dichas funciones de onda se llaman orbitales atmicos. Cuando se realizan clculos moleculares, es comn el uso de una base compuesta por un nmero finito de orbitales atmicos, centrado en cada ncleo atmico dentro de la molcula (combinacin lineal de orbitales atmicos de prueba o ansatz). Inicialmente, estos orbitales atmicos eran tpicamente orbitales tipo Slater, introducidos por John Slater en 1930 \cite{sto},  que corresponde a un conjunto de funciones que decae exponencialmente con la distancia de los ncleos. Ms tarde Samuel Francis Boys en 1950 \cite{fboys} demostr que es conveniente que estos orbitales de tipo Slater sean aproximados como combinaciones lineales de orbitales gaussianos, debido a que es ms fcil calcular integrales con funciones de base gaussiana u orbitales de tipo gaussiano (GTO por sus siglas en ingls, \textit{Gaussian Type Orbitals}), lo condujo a un enorme ahorro de esfuerzo computacional \cite{pople}. Existen varios conjuntos de bases tipo GTO, el ms pequeo de stos se llama conjunto de bases mnimas, y son normalmente compuestos por el nmero mnimo de funciones de base necesarios para representar todos los electrones de cada tomo. El mayor de ellos puede contener docenas a cientos de funciones de base en cada tomo. \\

Matemticamente, para molculas poliatmicas con $n $ tomos, los orbitales moleculares de Kohn-Sham  $\psi$, de la ecuacin \ref{kseq}, se pueden aproximar formalmente como:

\beq \psi=\sum_i^nc_i\varphi_i\label{lcao}\eeq

en donde las funciones $\varphi$ son en general funciones de un slo electrn. Usando el mtodo variacional, se optimiza $\phi$ usando las condiciones para obtener la mnima energa:

\beq \frac{\partial}{\partial c_i}\left (\int\psi^*H\psi d\tau\right)=0; \quad   i=1,2,\ldots,n \eeq

De dnde se obtiene un sistema de $n$ ecuaciones con $n$ parmetros:

\beq c_1(H_{11}-ES_{11})+c_2(H_{12}-ES_{12})+\ldots+c_n(H_{1n}-ES_{1n})=0,\nonumber\eeq
\beq c_1(H_{21}-ES_{21})+c_2(H_{22}-ES_{22})+\ldots+c_n(H_{1n}-ES_{1n})=0,\nonumber\eeq
\beq\vdots\label{sistemaec}\eeq
\beq c_1(H_{n1}-ES_{n1})+c_2(H_{n2}-ES_{n2})+\ldots+c_n(H_{nn}-ES_{nn})=0\nonumber\eeq

en donde:

\beq H_{ik}=\int \varphi_i^*H\varphi_k d\tau\quad\text{y}\quad S_{ik}=\int\varphi_i^*\varphi_kd\tau\label{integrales}\eeq

Buscando las races del determinante de la matriz formada por los coeficientes del sistema de ecuaciones \ref{sistemaec} se obtienen los $n$ valores de la energa, para lo cual se deben calcular las integrales \ref{integrales} previamente y finalmente se pueden calcular los coeficientes $c_i$.

\section{Bases}

Siguiendo \cite{lectstand} har un resumen de las bases estndar que han sido desarrolladas, no sin antes mencionar que ste en un campo en desarrollo constante, pues diferentes grupos de investigacin se dedican a optimizar y desarrollar nuevas bases optimizando diferentes parmetros de dichas bases asociados con la contraccin de la distribucin electrnica, polarizacin, entre otros \cite{demtro}.

\subsection{Conjuntos de bases de orbitales tipo Slater} 

STO por sus siglas en ingls \textit{Slater type orbital}. Son conjuntos de bases mnimas denotados por STO-$n$0$g$ donde $n$ primitivos orbitales Gauss se montan en un slo orbital tipo Slater (STO), $n$ inicialmente toma los valores 2-6. \\

Armnicos esfricos Slater, caracterizados por dos enteros $L, M$ ($-L\le M \le L$):

\beq\psi(\mb r)=Ar^Le^{-\zeta r}Y_L^M(\theta,\phi),\eeq

donde $A$ es una constante de normalizacin. En coordenadas cartesianas:

\beq\psi(\mb r)=x^ly^mz^ne^{-\zeta r},\eeq

caracterizado po 3 enteros $l, m, n$; $L=l+m+n$

\subsection{Conjuntos de bases de orbitales tipo Gaussiano}

El uso de orbitales Gaussianos en teora de la estructura electrnica fue propuesto por Boys \cite{fboys} en 1950. La razn principal por la cual se usan bases con funciones Gaussianas se debe al \textit{Teorema del producto graussiano} \cite{teogauss}, que garantiza que el producto de dos GTO centrados en dos diferentes tomos es una suma finita de Gaussianas centradas a lo largo del eje que une los dos centros. De esta manera, integrales con cuatro se pueden reducir a la suma de dos integrales con dos centros y luego a la suma finita de integrales de un centro.\\

Armnicos esfricos Gaussianos, caracterizados por dos enteros $L, M$ ($-L\le M \le L$):

\beq \psi(\mb r)= Br^Le^{-\zeta r^2}Y_L^M(\theta, \phi),\eeq

donde $B$ es una constante de normalizacin. GTO en coordenadas cartesianas, centrado en un tomo de coordenadas $\mb C$:

\beq  \psi(\mb r) = D (x-C_x)^l(y-C_y)^m(z-C_z)^ne^{-\zeta(\mb r-\mb C)^2}\eeq

donde $D$ es una constante de normalizacin. Cuando $l=m=n=0$ se tiene un orbital $s$-GTO, cuando $l=m=n=1$ se tiene un orbital $p$-GTO y as sucesivamente.

\subsection{Conjuntos de bases Pople} 

En este conjunto de bases se aproximan los orbitales tipo Slater por funciones Gaussianas primitivas que tienen la forma:

\beq g_l(x,y,z)=N_lx^ay^bz^c e^{\alpha_l r^2}\label{primgauss}\eeq

Donde $N_l$ y $\alpha_l$ son constantes y $a,b$ y $c$ son enteros. Cada funcin base en $\varphi_i$ de la ecuacin \ref{lcao} es representada como una combinacin lineal de un nmero pequeo de estas gaussianas primitivas. 

\beq \varphi_i=\sum_{l=1}^{M_i} g_l(x,y,z)d_{li} \eeq

donde $d_{li}$ es un coeficiente fijo y $M_i$ es el numero de gaussianas primitivas usadas para representar las funciones base.

La nomenclatura para las bases Pople es $K-L1G$ o $K-L11G$ donde $K$ y $L$ son enteros.  El primer conjunto representa el conjunto de bases con divisin de valencia doble zeta y el segundo el conjunto de bases con divisin de valencia triple zeta. $G$ indica simplemente que se estn usando gaussianas primitivas. $K$ indica el nmero de gaussianas primitivas que se van a usar para construir los orbitales internos, es decir, para los orbitales internos $M_i=K$. $L$ indica el nmero de gaussianas primitivas que se van a usar para construir los orbitales de valencia, es decir, para los orbitales de valencia, $M_i=L$.  Para el caso  doble zeta, el primer nmero 1 indica que con una funcin se construirn los orbitales adicionales de valencia con diferentes parmetros. Finalmente para el caso triple zeta, el segundo nmero 1, indica que con una funcin gaussiana primitiva se construirn los segundos orbitales adicionales.  Un ejemplo para ilustrar, es el tomo de carbono usando la base 6-311G: Se usarn 6 gaussianas primitivas para construir el orbital $1s$, 3 gaussianas primitivas para construir, en cada caso,  los orbitales $2s$ y $2p(3)$; una funcin gaussiana primitiva para construir, en cada caso, los orbitales $2s'$ y $2p'(3)$; finalmente, una funcin gaussiana primitiva para construir, en cada caso, los orbitales $2s''$ y $2p''(3)$

\subsubsection{Conjunto mnimo de bases}

Este conjunto se construye usando una funcin base por cada orbital de cada tomo que compone la molcula. As un orbital no est completamente ocupado se incluyen todas la funciones de dicho orbital, por ejemplo, si se ocupa parcialmente el orbital tipo $p$ se deben incluir las 3 funciones tipo $p$.

Para ilustrar mejor, usar el ejemplo de la molcula de metano $\text{CH}_4$:

\begin{table}[htb!]
\small
    \begin{tabular}{|c|c|c|c|}
    \hline
    çtomo $\longrightarrow$ Funciones & No. de Funciones & No. de çtomos  & Total No. de  \\ 
        $\quad\quad\quad\quad\quad$  Base &  Base      &          &Funciones Base\\ \hline
        C  $\longrightarrow$ $1s 2s 2p(3)$ & 5 & 1 & 5 \\\hline
        H  $\longrightarrow$ $1s $& 1 & 4 & 4 \\\hline 
    \end{tabular}
    \caption{En el conjunto de bases mnima, para la molcula de metano, se usan: una funcin base para cada orbital del tomo de carbono y una para el tomo de hidrgeno, en total 9 funciones base para la molcula. }
    \end{table}

\subsection{Conjunto de bases Doble Zeta}

Entre ms funciones base se usen en la LCAO mejor resultado obtendremos. Las bases doble zeta (DZ) incrementan el nmero de funciones usando dos funciones base por cada tipo de orbital encontrado en los tomos por separado.

Este tipo de bases usa el mejoramiento en LCAO que tiene que ver con modificar la densidad electrnica alrededor de los tomos, pues sta no es la misma en caso de la molcula que el caso de los tomos aislados debido a la interaccin interatmica.

Continuando con nuestro ejemplo, para la molcula de metano, usando el conjunto de bases DZ se tienen las funciones mostradas en la tabla \ref{meth_dz}.

\begin{table}[htb!]
\fontsize{9.5}{16}\selectfont
    \begin{tabular}{|c|c|c|c|}
    \hline
    çtomo $\longrightarrow$ Funciones & No. de Funciones & No. de çtomos  & Total No. de  \\ 
        $\quad\quad\quad\quad\quad$  Base &  Base      &          &Funciones Base\\ \hline
        C  $\longrightarrow$ $1s 1s' 2s 2s' 2p(3) 2p'(3)$ & 10 & 1 & 10 \\\hline
        H  $\longrightarrow$ $1s 1s''$& 2 & 4 & 8 \\\hline 
    \end{tabular}
    \caption{En el conjunto de bases doble zeta para la molcula de metano se usan: dos funciones base para cada orbital del tomo de carbono y  dos para el tomo de hidrgeno, en total 18 funciones base para la molcula. }
    \label{meth_dz}
    \end{table}

\subsection{Conjunto de bases Triple Zeta}

El conjunto de bases triple zeta (TZ) incrementa el nmero de funciones, al igual que en el caso anterior, pero  en este caso usando tres por cada orbital ocupado en los tomos separados. Por ejemplo en la molcula de metano se usan 15 funciones para el tomo de carbono y 3 para el tomo de hidrgeno, en total 27 funciones base.

Cabe mencionar que existen bases, de ste tipo, ms grandes como cuadruple-zeta.

\subsection{Conjunto de bases con divisin de valencia}

Con el objetivo de disminuir el tiempo de clculo e intentando no perder precisin se usan este tipo de bases. Asumiendo que en un determinado tomo los electrones  que estn ms cerca del ncleo se ven menos afectados por la presencia de otros tomos que los electrones de valencia, stos son tratados con un conjunto mnimo de bases mientras que los electrones de valencia son tratados con unas bases ms grandes. En particular si los electrones internos son tratados con el conjunto de bases mnimas y los de valencia con el conjunto de bases DZ, este conjunto de bases se denominan bases con divisin de valencia doble zeta. Con el mismo razonamiento se tiene el conjunto de bases con divisin de valencia triple zeta etc.

Para la molcula de metano el conjunto de bases con divisin de valencia doble zeta usan las funciones mostradas en la tabla \ref{meth_dv}

\begin{table}[htb!]
\fontsize{9}{16}\selectfont
    \begin{tabular}{|c|c|c|c|}
    \hline
    çtomo $\longrightarrow$ Funciones & No. de Funciones & No. de çtomos  & Total No. de  \\ 
        $\quad\quad\quad\quad\quad$  Base &  Base      &          &Funciones Base\\ \hline
        C  $\longrightarrow$ $1s  2s 2s' 2p(3) 2p'(3) $& 9 & 1 & 9 \\\hline
        H  $\longrightarrow$ $1s 1s''$ & 2 & 4 & 8 \\\hline 
    \end{tabular}
    \caption{En el conjunto de bases con divisin de valencia doble zeta para la molcula de metano se usan:  una funcin para $1s$, dos funciones base para cada orbital de valencia del tomo de carbono y  dos funciones para el tomo de hidrgeno, en total 17 funciones base para la molcula. }
    \label{meth_dv}
    \end{table}

\subsection{Funciones base de Polarizacin}

En la bsqueda de obtener mejor precisin aumentando la base se incluyen funciones de polarizacin en los conjuntos de base. Una funcin de polarizacin es cualquier  orbital diferente a los orbitales que componen la molcula por separado. Por ejemplo para el tomo de hidrgeno el nico orbital ocupado es el tipo $s$, por lo tanto funciones de polarizacin seran tipo $p$ o $d$. El uso de estas funciones permite representar mejor la densidad electrnica de la molcula.

\subsection{Funciones base de Difusin}

Estas son funciones extra (usualmente de tipo $p$ o $s$) que se adicionan para representar distribuciones electrnicas anchas. Estas son especialmente importantes en la representacin de la densidad de electrones en aniones (donde puede haber enlaces particularmente largos con densidad de electrones, repartida en una amplia regin).

\subsection{Conjunto de funciones de ondas planas}

ste conjunto de bases es particularmente usado en clculos peridicos, donde existe una celda unitaria que replica la red con periodicidad  $\mb R$. Las ondas planas contienen la periodicidad de la red:

\beq \psi (\mb r)= \frac{1}{\Omega} \sum_{\mb G}\phi_n(\mb G)e^{i\mb G.\mb r}\eeq
donde $\Omega $ es el volumen de la celda unitaria y $\mb G$ es el vector de la red recproca.
\beq e^{i\mb G.(\mb r+\mb R)}=e^{\mb G.\mb r}.\eeq

Para clculos con ondas planas hay un factor de truncamiento, en el que slo aquellos vectores de la red recproca, cuya energa sea ms baja que la predefinida como energa de corte (Cutoff) $E_{cut}$, se mantienen en la expansin, mientras que el resto de los coeficientes se hacen cero. Adems de reducir el costo computacional, este truncamiento limita los efectos de la orientacin de la celda unitaria en el resultado del clculo. Por otro la densidad $\rho (\mb r)$ debe tener la simetra translacional que tambin se expande en ondas planas:

\beq\rho (\mb r)=\sum_i\psi_i^*(\mb r)\psi_i(\mb r) = \sum_{\mb G}\tilde{\rho} (\mb G)e^{i\mb G.\mb r}\eeq

\subsection{Pseudopotenciales}

Normalmente los electrones que estn en el centro no tienen gran influencia en las propiedades electrnicas que son principalmente dadas por los electrones de valencia (no siempre es vlido). Tomando ventaja de ste hecho se puede simplificar las descripcin de un tomo. La idea de los pseudopotenciales es poder describir explcitamente slo los electrones de valencia y reemplazar el efecto de los electrones del centro con un potencial modificado. Los electrones de valencia deben tener los mismos autovalores que para el caso en los que se tienen todos los electrones:

\beq \int_{|\mb r|< L}|\psi_{pseudo}(\mb r)|^2d\mb r= \int_{|\mb r|< L}|\psi_{all_el}(\mb r)|^2d\mb r\eeq

para todos los $L$  fuera de la regin del centro ($L>r_{centro}$).


\section{Funcionales}

Recordemos que para poder resolver el problema de muchos cuerpos en el marco de DFT es necesario contar con una expresin apropiada para $E_{xc}[\rho]$ y en vista de que no se cuenta con una expresin exacta se requieren aproximaciones. En esta seccin se  mencionarn los funcionales usados para los clculos que se realizaron en el presente trabajo. Como vimos anteriormente $E_{xc}$ tiene en cuenta la diferencia entre la repulsin electrnica clsica y cuntica, tambin incluye la diferencia de la energa electrnica entre el sistema ficticio no interactuante y el sistema real. Sin embargo en la prctica no se intenta obtener estas porciones explcitamente. La bsqueda de una expresin de $E_{xc}$ ha representado un desafo en los ltimos aos. En muchas aproximaciones aparecen parmetros empricos que introducen correcciones a la energa cintica.\\

Una de las aproximaciones ms simples es la aproximacin local propuesta por Kohn y Sham en 1965 basada en un gas uniforme de electrones, \textit{aproximacin local de la densidad}  (LDA por su siglas en ingls \textit{Local Density Approximation}):

\beq E_{xc}^{LDA}[\rho]=\int \rho(\mb r)\epsilon_{xc}(\rho)d\mb r\eeq

donde $\epsilon_{xc}(\rho)$ es la energa de intercambio y correlacin por partcula de un gas uniforme de electrones de densidad $\rho$. Para este caso el correspondiente potencial de correlacin de intercambio es:

\beq \upsilon_{xc}^{LDA}(\mb r)=\frac{\delta E_	{xc}^{LDA}}{\delta \rho(\mb r)}=\epsilon_{xc}(\rho(\mb r))+\rho(\mb r)\frac{\delta \epsilon_{xc}(\rho)}{\delta \rho}\eeq

y las ecuaciones de Kohn y Sham \ref{kseq} quedan:

\beq \left[  -\frac{1}{2}\nabla^2+\upsilon(\mb r)+\int\frac{\rho(\mb r')}{|\mb r-\mb r'|}d \mb r' +\upsilon_{xc}^{LDA}(\mb r)  \right]\psi_i=\epsilon_i\psi_i\label{kslda}\eeq

La solucin autoconsistente \ref{kslda} define  lo que es conocido como el mtodo LDA.


La funcin $\epsilon_{xc}(\rho)$ se puede dividir en dos contribuciones, la parte de correlacin y la parte de intercambio,

\beq \epsilon_{xc}(\rho)=\epsilon_c(\rho)+\epsilon_x(\rho)\eeq

La parte de intercambio est dada por el funcional de la energa de intercambio de Thomas-Fermi \cite{parr}

\beq \epsilon_x(\rho)=-C_x\rho(\mb r)^{1/3}, \quad\quad C_x=\frac{1}{4}\left(\frac{3}{\pi}\right)^{1/3}\eeq

Normalmente los funcionales se denotan con algn acrnimo en el que conste la inicial de los autores y el ao de publicacin. 

Fock en 1930\cite{fock} postula, teniendo en cuenta la antisimetrizacin de la funcin de onda, lo que se conoce tambin como el funcional ''exacto'':

\beq  E_x^{Exacto}=E_x^{F30}=-\frac{1}{2}\sum_i\sum_j\int\int\frac{\phi_i*(\mb r)\phi_j*(\mb r')\phi_j(\mb r)\phi_i(\mb r')}{|\mb r-\mb r'|}d\mb r d\mb r',\eeq

Siguiendo el trabajo de Thomas-Fermi, Dirac mostr que el funcional de intercambio correspondiente al gas uniforme de electrones tiene la forma sencilla:

\beq E_x^{D30}[\rho(\mb r)]= -\frac{3}{4}(\frac{3}{\pi})^{1/3}\int\rho^{4/3}(\mb r)d\mb r\eeq

LDA es aplicable a sistemas cuya densidad vare lentamente con $\mb r$. Por otro lado, antes de Kohn-Sham, Slater en 1951 \cite{hfslater} propuso un mtodo simplificado del mtodo de Hartree-Fock, aproximando el operador no local de Fock \ref{fock} por un operador local simple usando el modelo de gas uniforme de electrones dando como resultado la ecuacin \textit{Hartree-Fock-Slater}:

\beq \left[  -\frac{1}{2}\nabla^2+\upsilon(\mb r)+\int\frac{\rho(\mb r')}{|\mb r-\mb r'|}d \mb r' +\upsilon_{x\alpha}(\mb r)  \right]\psi_i=\epsilon_i\psi_i\label{kslda_l}\eeq

con el potencial local $\upsilon_{x\alpha}$

\beq \upsilon_{x\alpha}=-\frac{3}{2}\left\{\frac{3}{\pi}\rho(\mb r)\right\}^{1/3}\eeq

Durante la dcada de los 80 hubo numerosos intentos de construir un funcional capaz de reproducir resultados precisos destacndose los trabajos de Becke en 1986 (B86) \cite{beck1,beck2} y De Pristo-Kress en 1987 (DK87) \cite{DK}, en este ltimo caso, su aplicacin a tomos y molculas ha sido escasa. \\

En LDA cuando se considera la polarizacin del spin (LSDA por sus siglas en ingls) el funcional tiene la forma:

\beq E_x^{LSDA}=\frac{3}{4}\left(\frac{6}{\pi}\right)^{1/3}\int[(\rho^\alpha)^{4/3}+(\rho^\beta)^{4/3}]\mb r\eeq

Intentos posteriores en la bsqueda del funcional de intercambio se centraron en amortiguar el incremento del integrando. As Becke \cite{beck3}, en 1988, mostr que el incremento del integrando se poda atenuar condicionando que el funcional lleve a la densidad de la energa de intercambio exacta en las regiones asintticas:


\ber E_x^{B88}[\rho(\mb r)]&=&{E_x^{LSDA}}[\rho(\mb r)]\nonumber\\ 
&&-b\sum_{i=\alpha,\beta}\int \frac{(\rho^i)^{4/3}\chi_i^2}{1+6b\chi_i\sinh^{-1}\chi_i} d\mb r\eer

donde $\chi_i=|\nabla\rho^i|/(\rho^i)^{4/3}$. Se fija el parmetro  emprico $b=0,0042$ de tal forma que reproducen datos experimentales de los gases He y Rn. Este funcional es ampliamente usado en la actualidad por su efectividad y es la base de casi cualquier funcional de intercambio posteriormente desarrollado.

Otro grupo de contribuciones al desarrollo de funcionales de intercambio es debido a Perdew et al. \cite{perdew}  cuya expresin es:

\beq E_x^{PW86}[\rho(\mb r)]= \int \epsilon_x^{D30}(1+1,2975s^2+14s^4+0,2s^6)^{1/15}d\mb r\eeq

con 
\beq s= \frac{|\nabla\rho(\mb r)|}{(3\pi^2)^{1/3}\rho(\mb r)^{4/3}}\eeq

Los funcionales de intercambio mencionados anteriormente y otros que no se han incluido se pueden enmarcar dentro de la aproximacin conocida como \textit{aproximacin del gradiente generalizado} (GGA por sus siglas en ingls), expresado como:

\beq E_x^{GGA}[\rho(\mb r)]=\int\epsilon_x[\rho(\mb r), |\nabla\rho(\mb r)|]d\mb r\eeq

El siguiente paso en el diseo de funcionales es el de extender las dependencias del integrando a la funcin laplaciana de la densidad \cite{fila}, aunque no han sido muy exitosos debido a las fluctuaciones de la segunda derivada de la densidad. 
\\

Por otra parte, para el funcional de correlacin existen clculos Monte-Carlo realizados por Ceperley y Alder \cite{ceperley} para dicha expresin. Tambin se han obtenido expresiones analticas para valores lmites de la densidad\cite{vosko} usando el modelo de gas uniforme de electrones, para ello sea:

\beq r_s=\left[\frac{3}{4\pi\rho(\mb r)}\right]^{1/3},\quad\quad \xi=\frac{\rho_\alpha(\mb r)-\rho_\beta(\mb r)}{\rho_\alpha(\mb r)+\rho_\beta(\mb r)}\eeq

donde $r_s$ es el radio de una esfera cuyo volumen es el volumen efectivo de un electrn\cite{parr}, conocido como radio de Wigner y $\xi$ es la polarizacin de espn. Vosko et al. \cite{vosko} han usado resultados numricos para obtener una forma especfica del integrando $\epsilon_c(r_s,\xi)$ con los lmites:

\ber &&\epsilon_c(r_s,\xi)=c_0(\xi)\ln r_s-c_1(\xi)+c_2(\xi9r_s\ln r_s-c_3(\xi)+\cdots;\\
&&\forall r_s< 1, \quad \textit{lmite de alta densidad}\nonumber\eer

y

\ber \epsilon_c(r_s,\xi)=-\frac{d_0(\xi)}{r_s}+\frac{d_1(\xi)}{r_s^{3/2}}+\cdots;\\
\forall r_s\geqslant 1, \quad \textit{lmite de baja densidad}\nonumber\eer

Los coeficientes $(c_i,d_i)$ han sido calculados de una forma ms precisa\cite{coef1,coef2}. Definiendo la funcin $f(\xi)$:

\beq f(\xi)=\frac{1}{2}(2^{1/3}-1)^{-1}\{(1+\xi)^{4/3}+(1-\xi)^{4/3}-2\}\eeq

S. H. Vosko y colaboradores establecieron la expresin del integrando \cite{vosko}

\beq \epsilon_c^{VNW}(r_s,\xi)=\epsilon_c(r_s)+\alpha(r_s)\left[\frac{f(\xi)}{f''(0)}\right][1+\beta(r_s)\xi^4]\eeq

en el artculo original \cite{vosko} se encuentran tablas de los valores de $\alpha(r_s)$ y $\beta(r_s)$.

Tambin existen funcionales de correlacin tipo GGA. Respecto a funcionales $E_c^{GGA}$ una de las propuestas ms usada es la derivada por Perdew \cite{perdewc86}. En 1988 Lee, Yang y Parr \cite{lyp88} abandonaron el modelo de gas uniforme de electrones para usar el Helio, el sistema ms simple con efectos de correlacin, proponiendo as el funcional:

\ber E_c^{LYP88}[\rho(\mb r)]&=&-a\int \frac{1}{1+d\rho^{-1/3}(\mb r)}\{\rho(\mb r)+b\rho^{-2/3}(\mb r)[C_{TF}\rho^{5/3}(\mb r)\nonumber\\
&&-2t_w+\frac{1}{9}\left(t_w+\frac{1}{2}\nabla^2\rho/\mb r)\right)]e^{c\rho^{-1/3}}\}d\mb r\eer

donde $t_w=\frac{1}{8}\frac{|\nabla\rho(\mb r)|^2}{\rho(\mb r)}-\frac{1}{8}\nabla^2\rho(\mb r)$ y $C_{TF}$ es la constante del funcional de Thomas-Fermi.

\subsection{Funcionales Hbridos}

Son aquellos consistentes de la combinacin de los diferentes funcionales mencionados u otros que no se incluyen en este documento por no ser de nuestro inters. Las combinaciones ms comunes  son las que usan el funcional de intercambio B88 con los funcionales de correlacin.

Cabe mencionar que la contribucin debida al intercambio es mayor que la debida a la correlacin, por lo tanto es importante contar con una buena aproximacin del funcional de intercambio.\\

El funcional hbrido que se usa principalmente en este trabajo es el B3LYP definido como:

\beq E_{xc}^{B3LYP}=(1-a)E_x^{LDA}+aE_x^{Exacto}+b \Delta E_x^{B88}+(1-c)E_c^{VWN}+cE_c^{LYP88}\eeq

donde los coeficientes $a,b$ y $c$ se han optimizado usando diferentes sistemas como $G_2$ entre otros\cite{coef3}. $\Delta E_x^{B88}$ es el funcional $E_x^{B88}$ sin la contribucin LSDA.

En esta lnea de los funcionales hbridos se han visto posteriores desarrollos y mejoras\cite{docto}.

\section{Procedimiento SCF}

Para acelerar los clculos de DFT y HF las iteraciones se hacen en dos fases:

\begin{itemize}

\item La densidad se acerca a la convergencia de $10^{-5}$ utilizando integrales con exactitud de seis dgitos y una integracin en la red modesta FineGrid en los clculos de DFT. Esto se refiere al mtodo de computacin y uso de integrales de dos electrones, por default es FineGrid consta de 72 cascarones radiales y 302 puntos angulares por cascarn. Este paso se termina despus de 21 iteraciones incluso si no est totalmente convergente. Este paso se omite de forma predeterminada si los tomos de metales de transicin estn presentes.

\item La densidad converge a $10^{-8}$ usando integrales de precisin de hasta 10 dgitos permitiendo un total de 128 ciclos para el segundo paso.

\end{itemize}

La mayor contribucin a la energa total viene de los electrones internos, no de los de enlace. Si la base de un tomo es variacionalmente deficiente en zonas internas(p.ej. 1$s$), un mtodo variacional molecular recupera mucha energa corrigiendo esa zona deficiente con las bases de los tomos vecinos.   El resultado es un acortamiento de las distancias de enlace y un aumento de las energas de enlace o disociacin, ambos irregulares y anmalos. La solucin es utilizar bases que no slo sean correctas en las zonas de enlace sino tambin en las internas. Una correccin ms sencilla y menos fiable es el mtodo del contrapeso, que consiste en calcular la energa de enlace en cada punto de seales de error como la diferencia entre la energa total y las energas atmicas calculadas con la base molecular.

\section{Cdigos usados en este trabajo}

\begin{itemize}

\item \textbf{Gaussian 09:} El cdigo es inicialmente publicado por Pople y sus colaboradores  \cite{pople70}. desde entonces ha contado con varias actualizaciones, de las cuales la ltima es la 09.  El cdigo es capaz de calcular las energas, geometras, frecuencias de vibracin, estados de transicin, caminos de reaccin, estados excitados y una variedad de propiedades basadas en diversas funciones de onda no correlacionadas y correlacionadas usando diversos mtodos como HF, DFT, teora de perturbaciones a segundo orden (MP2 \cite{moller}), entre otros \cite{gaussian}. Incorpora todos los conjuntos de bases mencionados anteriormente, entre otros.\\

El algoritmo de optimizacin incluido en \textit{Gaussian} es el ``algoritmo de Berny'', elaborado por Bernhard Schlegel \cite{gaussian}. Este algoritmo utiliza las fuerzas que actan sobre los tomos de una estructura dada, junto con la matriz de la segunda derivada (llamada matriz Hessiana) para predecir las estructuras energticamente ms favorables y as optimizar la estructura molecular hacia el prximo mnimo local en la superficie de energa potencial. Como el clculo explcito de la matriz de las segundas derivadas es bastante costoso, el algoritmo de Berny construye un aproximado de la matriz Hessiana al comienzo del procedimiento de optimizacin mediante la aplicacin de un campo de fuerza y a continuacin se utilizan las energas y primeras derivadas calculadas a lo largo del camino de optimizacin para actualizar esta matriz Hessiana aproximada. El xito del proceso de optimizacin por lo tanto, depende en cierta medida de qu tan bien la aproximacin de la matriz  hessiana representa la verdadera situacin en un momento dado. Para muchos sistemas, la hessiana aproximada funciona bastante bien, pero en algunos casos una mejor hessiana tiene que ser utilizada. Para ms informacin visitar  la pgina http://www.gaussian.com/.

\item \textbf{NWchem:} Este programa proporciona muchos mtodos para calcular las propiedades de los sistemas moleculares incluyendo clculos peridicos utilizando las descripciones de la mecnica cuntica estndar de la funcin de onda electrnica y la densidad \cite{nwchem}. Usa conjuntos de bases mencionados anteriormente, entre otros, e incorpora mtodos como Hartree-Fock (HF), DFT, Excited-State Calculations (CIS, TDHF, TDDFT), Real-time TDDFT, entre otros  \cite{nwchem}.  Para ms informacin visitar  la pgina http://www.nwchem-sw.org/.

\item \textbf{CP2K:} CP2K es un conjunto de mdulos, que comprende una variedad de mtodos de simulacin moleculares en diferentes niveles de precisin, que van desde ab initio DFT, a semi-emprica aproximacin NDDO, a hamiltonianos clsicos. Se utiliza comunmente para predecir energas, estructuras moleculares, frecuencias de vibracin de sistemas moleculares, mecanismos de reaccin, y es adecuado idealmente para llevar a cabo simulaciones de dinmica molecular \cite{cp2k}. 

Utiliza un enfoque mixto de ondas Gaussianas y ondas planas (GPW por sus siglas en ingls, \textit{Gaussian Plane Waves}). Para ms informacin visitar  la pgina http://www.cp2k.org/.

\end{itemize}


\chapter{Resultados}\label{results}

\section{Geometra}

Usando el mtodo del funcional de la densidad DFT se hicieron clculos de optimizacin para diferentes estructuras. En algunos casos se us tambin el mtodo de Hartree-Fock (HF) para verificar que el comportamiento cualitativo coincide.

\subsection{Hojuelas Aromticas}

   Son molculas aromticas que se caracterizan por tener uno o ms anillos hexagonales planos de tomos  de carbono o bencenos con terminaciones con tomos de hidrgeno\cite{aromatic, aromaticity} .\\

El punto de partida fue un trabajo previo \cite{js1} en el que se estudia la capacidad de adsorcin de Litio en hojuelas o laminillas pequeas de grafeno tratadas como molculas aromticas  con hidrgenos en cada enlace externo. En este estudio se muestra cmo los electrones pueden ser transferidos a la superficie en regiones localizadas y adems el rompimiento de simetra cuando los Litios son adsorbidos de lados opuestos a la superficie de la laminilla.  Consideramos molculas aromticas con simetra de reflexin. Los resultados de los clculos revelan una distorsin fuera del plano de la molcula que implica un rompimiento de la simetra de reflexin.

\begin{figure}[htbp!]
$\begin{array}{ccc}
\hspace*{-0.2in} \includegraphics[height=0.15\textheight]{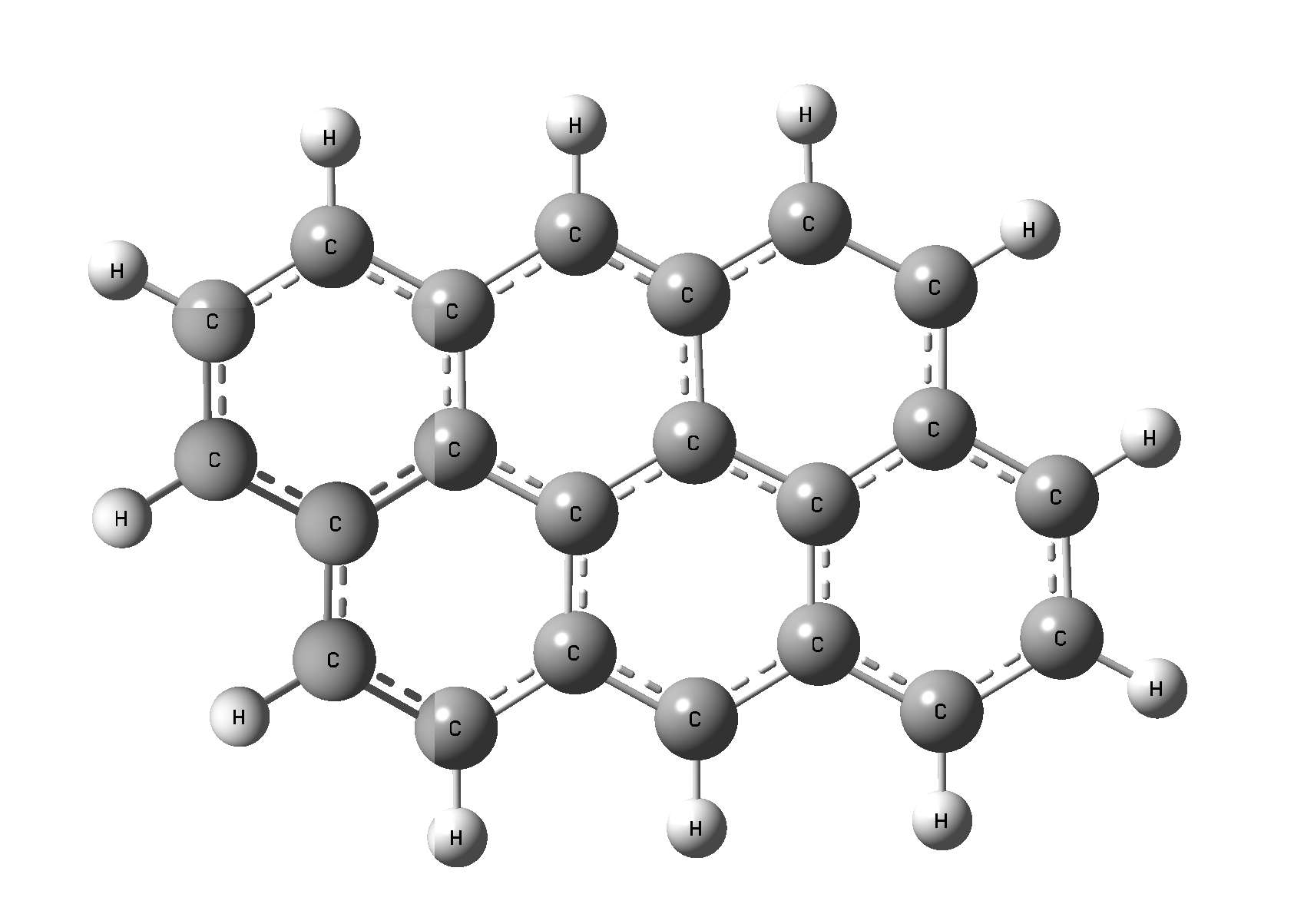}&
\hspace*{-0.3in} \includegraphics[height=0.18\textheight]{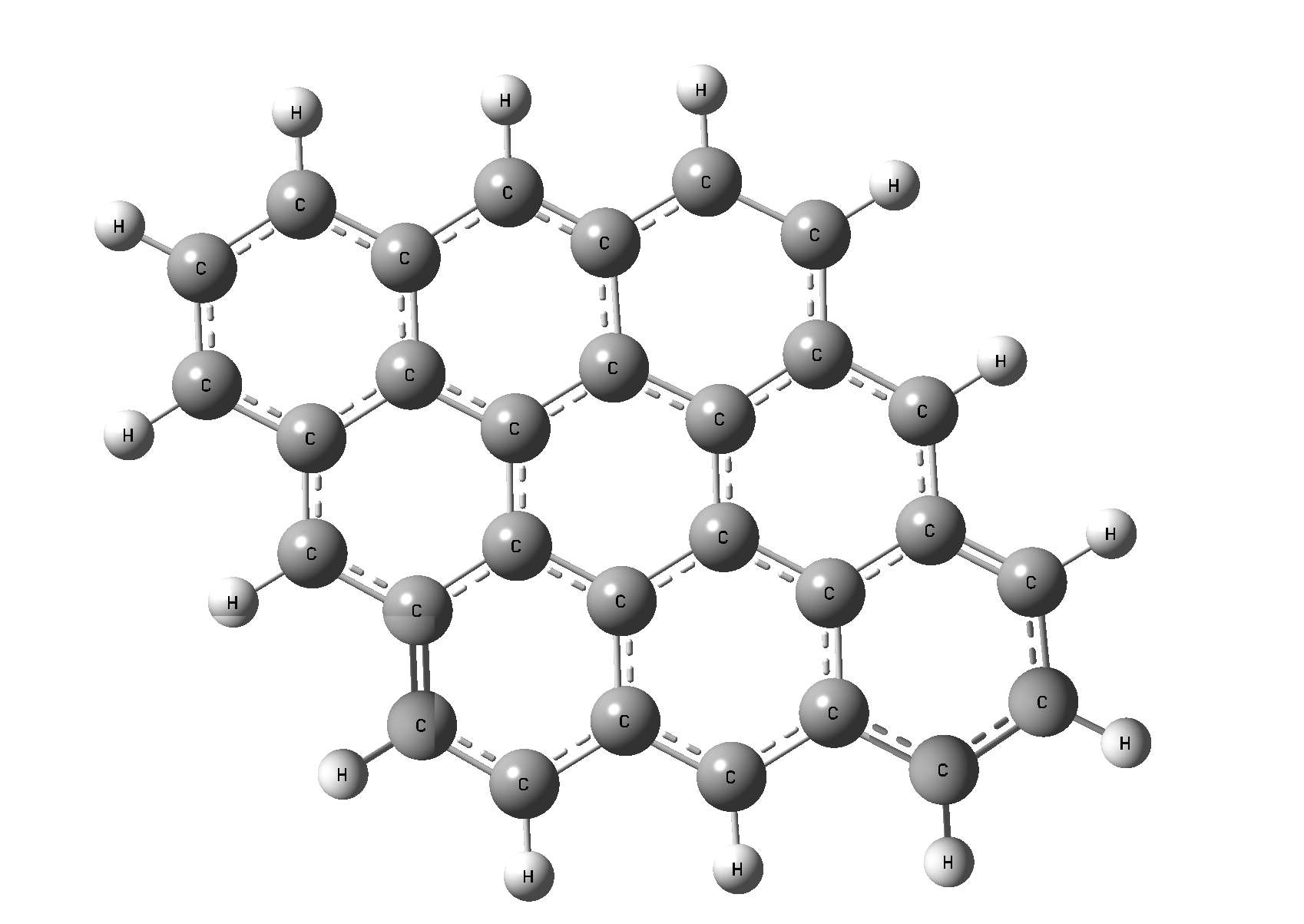}&
\hspace*{-0.3in} \includegraphics[height=0.2\textheight]{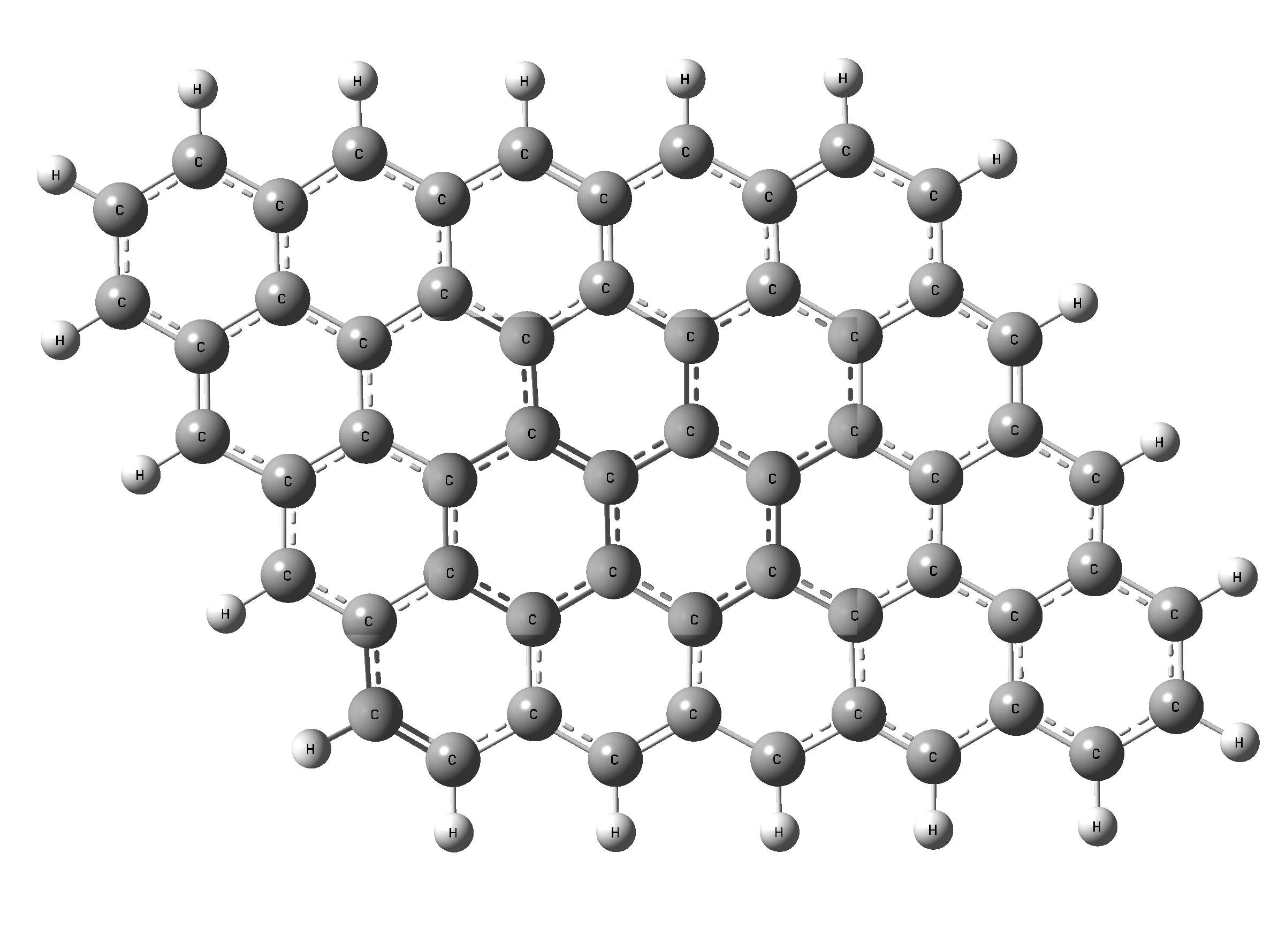}
\end{array}$
\caption{Hojuelas aromticas de 6, 9 y 20 anillos de benceno.}
\label{flakes}
 \end{figure}

Para un tomo adsorbido  cualquier hamiltoniano Born-Oppenheimer  va a romper la simetra especular en el plano de los anillos de carbono, pero si dos tomos se adsorben en los lados opuestos de la hoja podra existir una configuracin simtrica  \cite{article_flakes}. Esta configuracin simtrica nuclear es inestable en cualquiera de las aproximaciones que hemos utilizado, es decir, en los clculos de Hartree Fock  con y sin algunas correlaciones de electrones, as como en los clculos del funcional de la densidad. Para grafeno se muestra este efecto aumentando el tamao de las hojuelas de 6 a 20 anillos con el fin de asegurarnos de que no vemos ms efectos de tamao finito, en la figura  \ref{flakes} se muestran tres de ellas. El rompimiento de la simetra se manifiesta a lo largo de este rango.

\begin{figure}[htb!]
        \begin{subfigure}[b]{0.5\textwidth}
       \hspace*{0.2in}  \includegraphics[height=0.1\textheight]{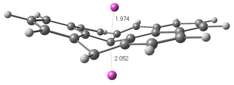}  
       \caption{}
	\end{subfigure}
	\begin{subfigure}[b]{0.5\textwidth}
        \hspace*{0.2in}  \includegraphics[height=0.1\textheight]{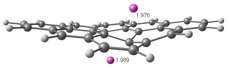} 
        \caption{}
         \end{subfigure}
          \caption{Estructura optimizada de adsorcin de dos tomos Litios sobre ($a$) una hojuela de seis anillos y ($b$) hojuela de 9 anillos de benceno. }
         \label{fl_double}
 \end{figure}

Se realizaron los clculos de optimizacin de la geometra  usando el mtodo de HF con las bases mnimas Pople orbitales tipo Slater, STO-3g* (STO por su siglas en ingls) simulados, por una superposicin de tres Gaussianas  y tambin se usaron las bases 3-21g*. Se realizaron clculos de nivel superior con DFT para geometras especficas con el fin de confirmar la consistencia de los resultados de campo medio y mtodo de DFT con el funcional hbrido B3LYP y  el conjunto de bases 3-21g*. Todos los clculos se llevaron a cabo con el cdigo \textit{Gaussian} \cite{gaussian} . Encontramos una fuerte deformacin de la hojuela de grafeno al adsorber un par de Litios de lados opuestos como se muestran en las figuras \ref{fl_double}($a$),  \ref{fl_double}($b$)  \cite{article_flakes} para los casos de hojuelas de 6 y 9 bencenos o anillos respectivamente.\\

\begin{figure}[htb!]
        \begin{subfigure}[b]{1\textwidth}
       \hspace*{-0.1in}  \includegraphics[height=0.14\textheight]{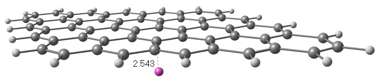}  
       \caption{}
	\end{subfigure}\\
	\begin{subfigure}[b]{1\textwidth}
        \hspace*{-0.1in}  \includegraphics[height=0.14\textheight]{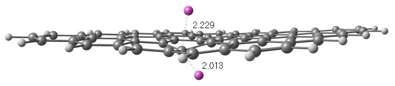} 
        \caption{}
         \end{subfigure}
          \caption{Resultado de la doble adsorcin de tomos de ($a$)uno Litio y ($b$) dos Litios  a la hojuela de 20 anillos. }
         \label{double_hoj}
 \end{figure}

Para explorar si es una caracterstica de las hojas de grafeno y no slo de pequeas molculas aromticas se presenta la adsorcin de tomos de Li la hojuela de 20 anillos en la figura \ref{double_hoj} en donde se muestran los resultados de una y dos adsorciones de Litio y se observa que los efectos de una sola adsorcin son mucho ms dbiles que el doble de adsorcin. Teniendo  en cuenta que las distancias de la hoja son ms grandes las deformaciones son mucho ms pequeas para una y doble adsorcin. 

 Se calcularon las energas de adsorcin de uno y dos Litios en las hojuelas aromticas como se muestra en la tabla \ref{tablefl}, donde se observa que, mientras que la doble adsorcin causa la distorsin ms fuerte, la parte principal de la energa de la doble adsorcin ya es producida por la adsorcin de un slo tomo de Litio. \\
 
Un anlisis de los resultados nos lleva a concluir que mientras que para una pequea hojuela la deformacin debido a la adsorcin de un par de Litios es grande para una hojuela mucho ms grande o infinita la deformacin ser mucho ms pequea.\\
\phantom{aaa}\\
\phantom{aaa}\\
\phantom{aaa}\\

\begin{table}[htb!]
\centering
\begin{tabular}{ccc}
\hline
\textbf{No. Anillos} &\textbf{Primera Adsorcin}& \textbf{Segunda Adsorcin}\\
\hline
 6&	59.07	&11.13\\
7&	99.30	&21.23\\
8&	103.72	&16.32\\
9&	109.28	&22.23\\
17&	287.71	&56.88\\
18	&275.84	&62.23\\
19&	279.43	&81.77\\
20	&290.56	&90.07\\
\hline
\end{tabular}
\caption{Energas de adsorcin de uno y dos tomos de Litio en diferentes hojuelas aromticas. La energa est en unidades de kcal/mol.}
\label{tablefl}
\end{table}

 \begin{figure}[htb!]
  \centering
\includegraphics[height=0.15\textheight]{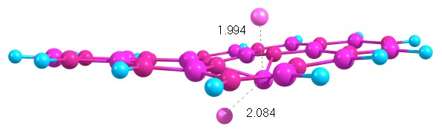}
\caption{Resultado de la doble adsorcin de tomos de Litio a la hojuela de Boro-Nitrgeno de 20 anillos} 
\label{flake9dBN}
 \end{figure}

El ltimo resultado de esta seccin tiene que ver con se ha obtenido la misma deformacin con la doble adsorcin de Litio en una hojuela de 9 anillos de  Boro-Nitrgeno, que se muestra en la figura \ref{flake9dBN}, calculada a nivel DFT y funcional B3LYP. La energa de adsorcin de un slo  Li para el BNS es 15,92 kcal/mol, mientras que la energa de doble adsorcin es 37,3 kcal/mol.\\

 \clearpage


\subsection{Poliacenos}

Los poliacenos, son una clase particular de molculas aromticas compuestas de anillos de benceno dispuestos linealmente con una frmula general C$_{4n+2}$H$_{2n+4}$ donde $n$ es el nmero de anillos de benceno. Varias molculas de sta clase han sido de inters por sus propiedades electrnicas, termodinmicas y pticas para el desarrollo de nuevos materiales\cite{poly1}-\cite{poly5}. Adems varios de ellos han sido sintetizados experimentalmente\cite{polyexp1}-\cite{polyexp3}. \\

 Los poliacenos son una importante clase de compuestos orgnicos, los ms pequeos que son el naftaleno y el antraceno son estables y se obtienen del petrleo.  Los siguientes en tamao son el tetraceno y el pentaceno que son semiconductores orgnicos y han sido utilizados transistores orgnicos \cite{ofet} (OFET por sus siglas en ingls, \textit{Organic Field-Effect Transistor}), diodos orgnicos de emisin de luz \cite{oled}(OLED por sus siglas en ingls, \textit{Organic Light-Emitting Diode}) y clulas fotovoltaicas orgnicas \cite{opv} (OPV por sus siglas en ingls, \textit{Organic Photovoltaics}); por otro lado el pentaceno ha sido recientemente visualizado \cite{pentacene}, imagen que se muestra en la figura \ref{penta}.  Poliacenos ms grandes como el hexaceno, heptaceno, octaceno y nonaceno tienen movilidades ms grandes pero son propensos a la degradacin por oxidacin as que el inters en poliacenos ms grandes sigue vigente.\\

Los poliacenos son los sistemas que estudiamos buscando simplificar el sistema y as obtener un claro entendimiento de los efectos de la adsorcin de Litio y las causas del rompimiento espontneo de simetra. Con ste objetivo se realizaron clculos numricos usando el cdigo de Gaussian\cite{gaussian} para Hartree-Fock y DFT. Se hicieron clculos desde antraceno (compuesto de tres anillos) hasta nonaceno (compuesto de nueve anillos) en donde encontramos que los resultados de HF y DFT  son similares cualitativamente en cuanto a que se observa deformacin hasta el heptaceno debido a la adsorcin de un par de Litios de lados opuestos sobre el mismo anillo del poliaceno aunque el ngulo de deformacin no es el mismo en ambos casos.\\

\begin{figure}[htb!]
        \begin{subfigure}[b]{0.5\textwidth}
       \hspace*{0.2in}  \includegraphics[height=0.1\textheight]{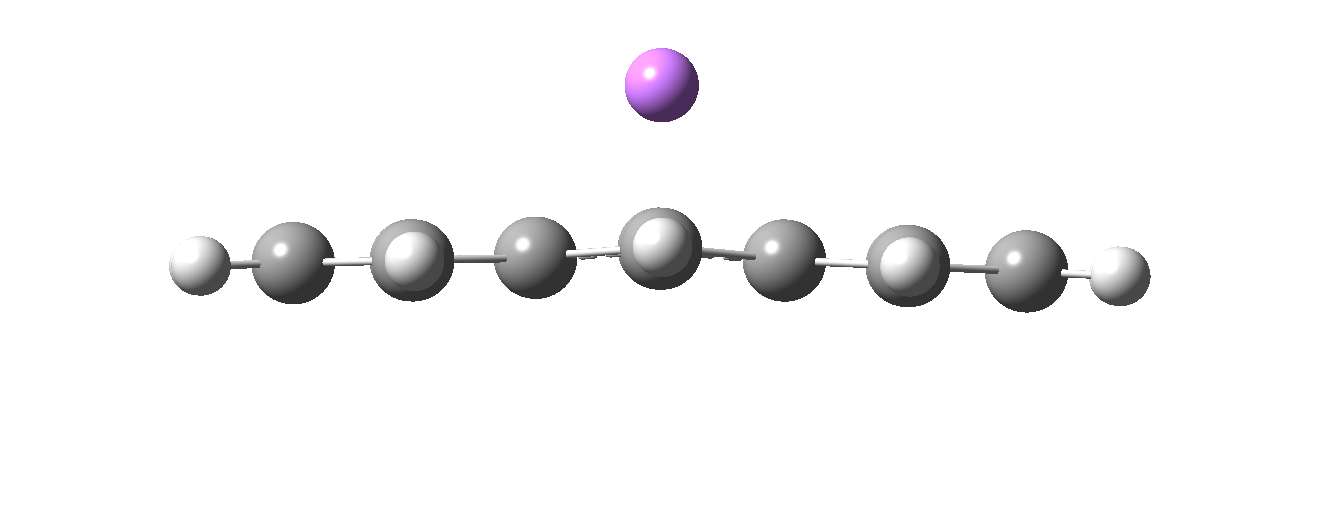}   
       \caption{}
	\end{subfigure}
	\begin{subfigure}[b]{0.5\textwidth}
        \hspace*{0.2in}  \includegraphics[height=0.1\textheight]{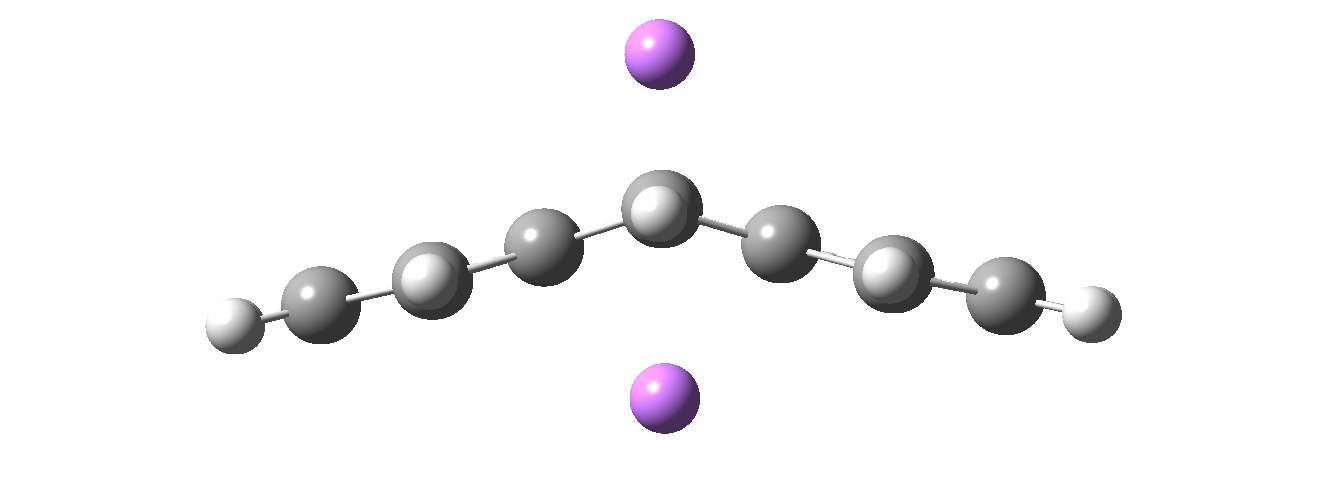} 
        \caption{}
         \end{subfigure}
          \caption{adsorcin de ($a$) uno Litio y ($b$) dos Litios sobre el anillo central del antraceno. La estructura ha sido optimizada usando el mtodo de DFT con el funcional B3LYP y el conjunto de bases 3-21g*.}
         \label{antraceno}
 \end{figure}

Luego de una amplia exploracin, encontramos que una manera eficiente de encontrar la configuracin que nos da la mnima energa era poner primero uno de los Litios, optimizarlo (en la figura \ref{antraceno}($a$) se muestra la estrucutura optimizada para el caso del antraceno) y luego adicionar el segundo Litio sobre el mismo anillo en el lado opuesto para nuevamente optimizarlo (en la figura \ref{antraceno}($b$) se muestra la estrucutura optimizada para el caso del antraceno). Para todos los casos que muestran la deformacin encontramos: Primero, que los Litios permanecen en el eje perpendicular al plano del anillo, justo en el centro del mismo, a diferencia por ejemplo del hidrgeno que siempre se ubica sobre un carbono \cite{hidro}. Segundo que en la doble adsorcin de Litio la direccin del doblamiento est condicionada por el lado en el que se pone el primer Litio, pues siempre se dobla el poliaceno en direccin opuesta a la posicin del primer Litio \cite{articulo}, como se muestra el caso del antraceno en la figura \ref{antraceno} .\\

\begin{figure}[htb!]
$\begin{array}{cc}

\includegraphics[height=0.23\textheight]{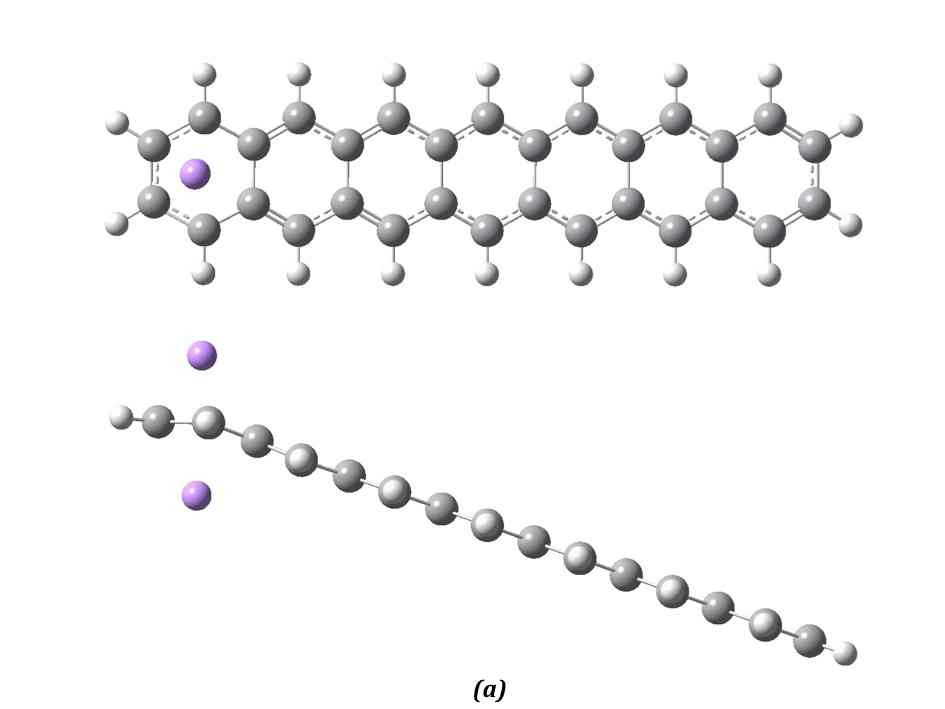}&

\includegraphics[height=0.23\textheight]{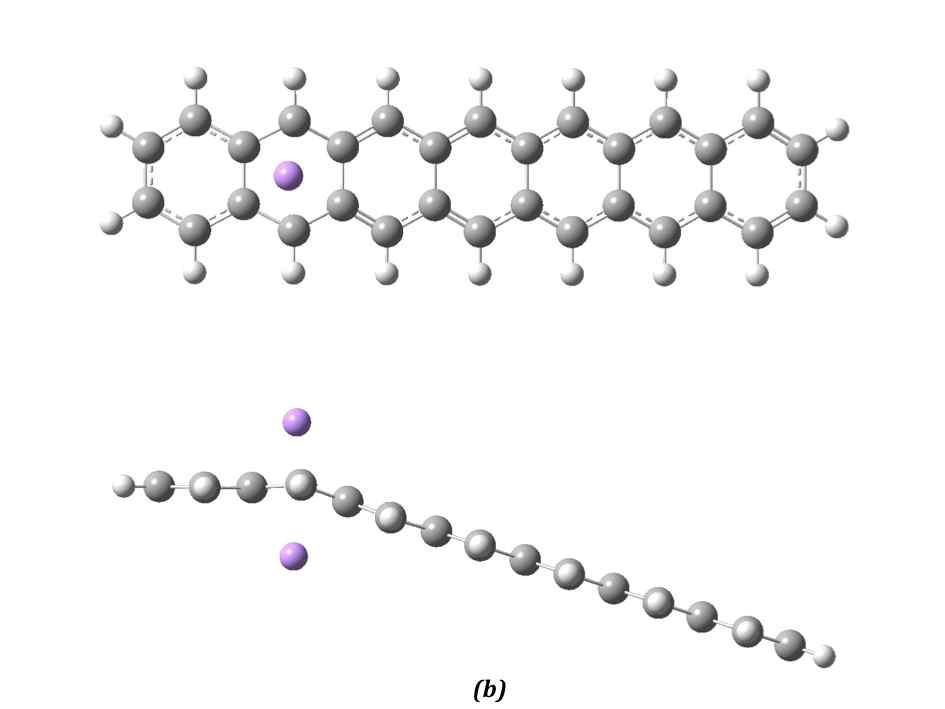}\\

  \includegraphics[height=0.23\textheight]{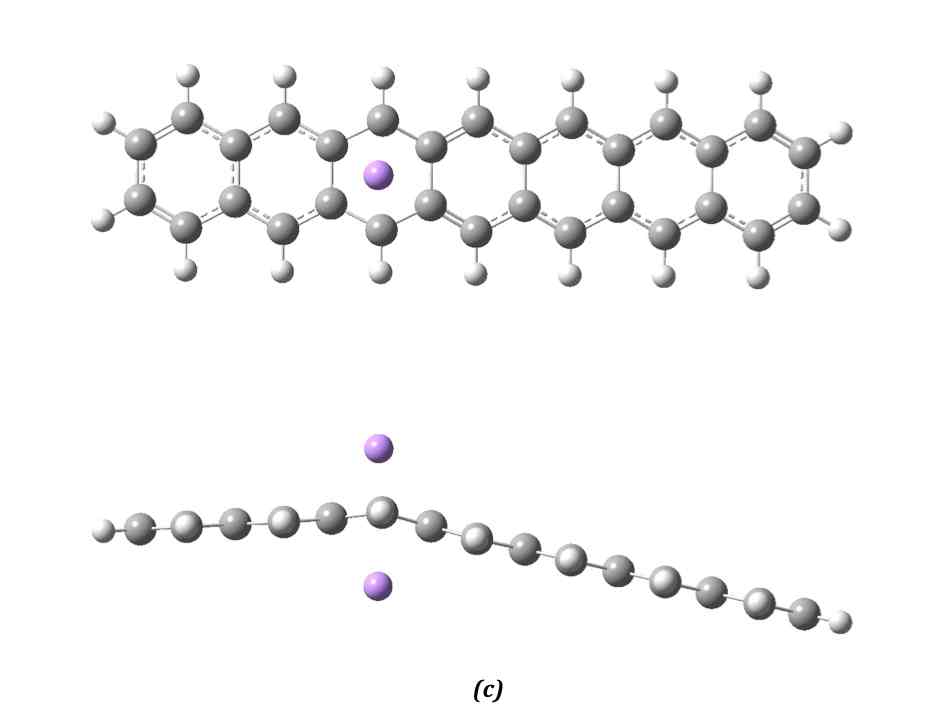}&
 
  \includegraphics[height=0.23\textheight]{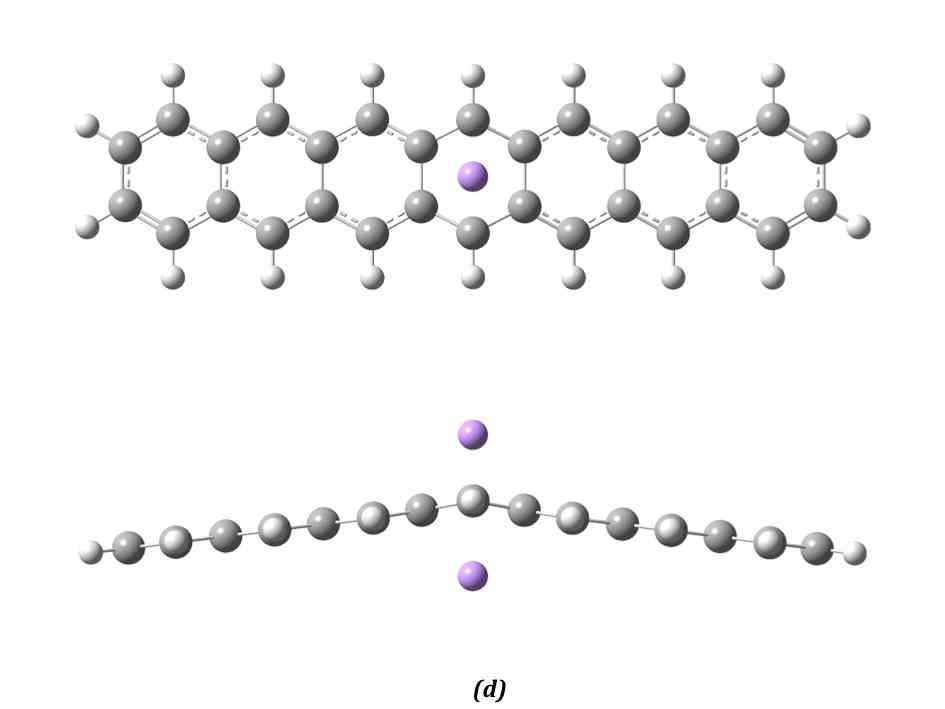}
\end{array}$
\caption{Heptaceno con: ($a$) 
 Par de tomos de Litio en el centro del primer anillo del heptaceno, ($b$) 
 Par de tomos de Litio en el centro del segundo anillo del heptaceno, ($c$) 
 Par de tomos de Litio en el centro del tercer anillo del heptaceno,  ($d$) 
 Par de tomos de Litio en el centro del cuarto anillo del heptaceno. Estructuras optimizadas usando el mtodo de DFT. }
\label{surp}
\end{figure}

Por otro lado, con el objetivo de poder comparar los resultados de los mtodos de HF y DFT, realizamos los clculos descritos anteriormente situando el par de Litios en los diferentes anillos del heptaceno como se ilustra en la figura \ref{surp}.\\ 

\begin{figure}[htb!]
$\begin{array}{cc}
 \hspace*{-0.2in} \includegraphics[height=0.27\textheight]{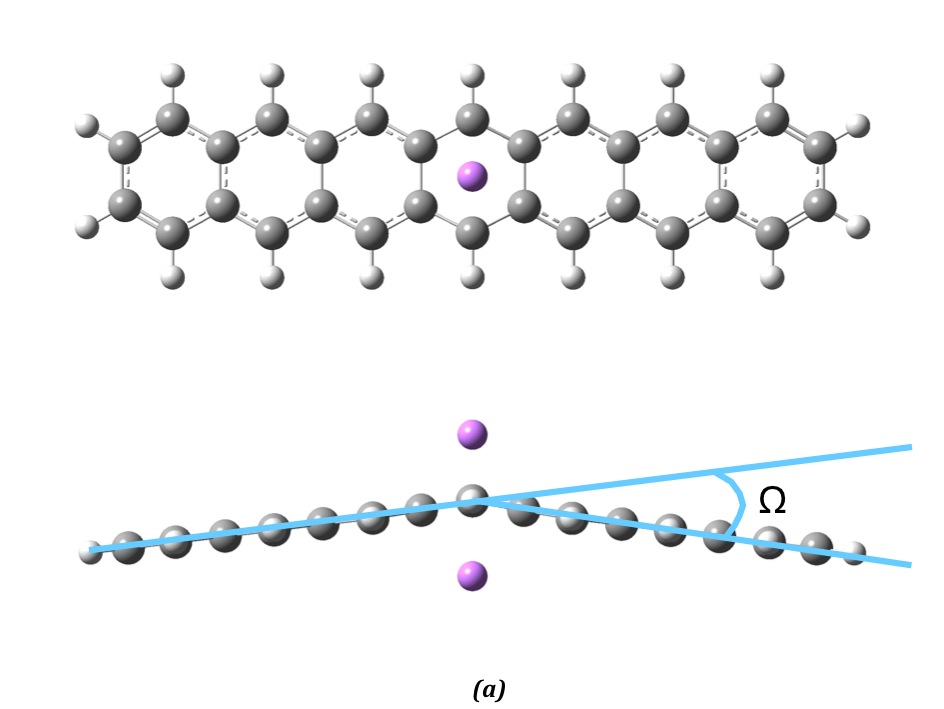}&
 \hspace*{-0.2in} \includegraphics[height=0.27\textheight]{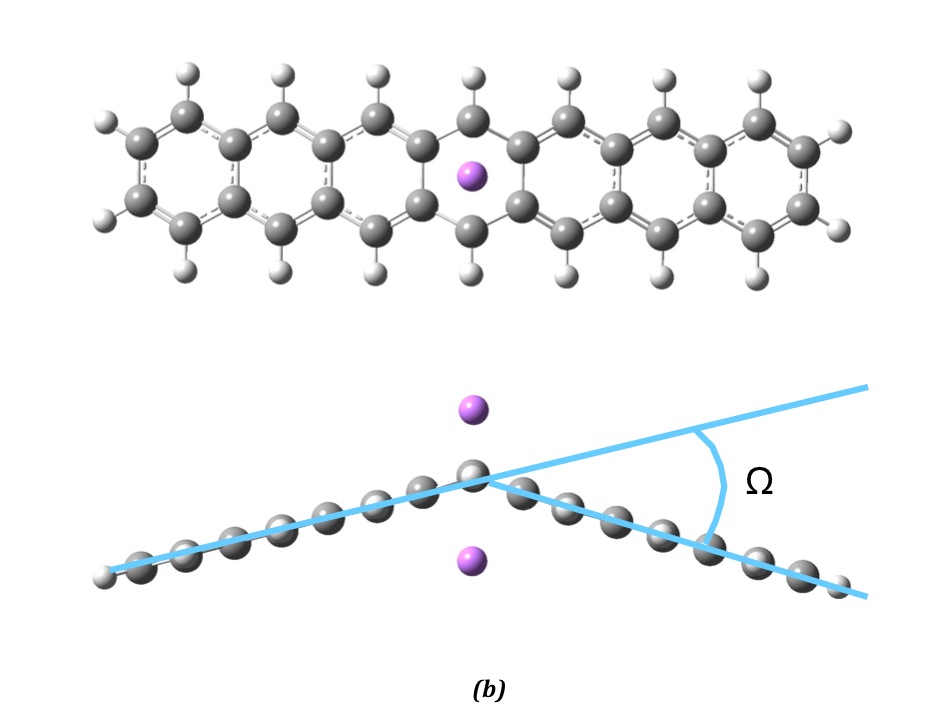}\\
\end{array}$
\caption{Adsorcin de dos tomos Litios en el anillo central del heptaceno .  ($a$)Se muestra el clculo de DFT y se ilustra la medida del ngulo de doblamiento. ($b$) Se muestra el clculo de HF y se  ilustra la medida del ngulo de doblamiento.}
\label{comparison}
\end{figure}

Para el mtodo de DFT se us el funcional B3LYP y el conjunto de bases 3-21g*, el mismo conjunto de bases que se us en el mtodo de HF. En ambos casos se obtuvieron resultados cualitativamente similares. En ambos casos medimos el ngulo de deformacin ilustrado en la figura \ref{comparison} para el caso del anillo central del heptaceno y cuyos valores para cada posicin del par de Litios en el heptaceno, denotada por el nmero del anillo contando de izquierda a derecha,  son tabulados en la tabla \ref{table}. \\

De la figura \ref{comparison} es evidente que el ngulo entre la configuracin plana y la calculada es ms grande en el caso de Hartree Fock; este hecho parece indicar que la ruptura de la simetra es un efecto de campo medio, que en realidad es un tanto atenuado por correlaciones. De la tabla \ref{table} se observa que para HF el ngulo disminuye a medida de que la configuracin es menos simtrica, es decir a medida que la adsorcin se aleja del centro, mientras el cambio en el ngulo, en los clculos hechos con DFT, el cambio es pequeo. Teniendo en cuenta que no hay ninguna razn fsica para que el ngulo cambie abruptamente como se observa en los clculos hechos con HF decidimos seguir con los clculos hechos con DFT.

\begin{table}\centering
\begin{tabular}{ccc}
\hline
&\textbf{Angulo de deformacin $\Omega$}& \\
\hline
 \textbf{POSICION} & \textbf{HF}
 & \textbf{DFT}\\
 \hline
 1&19.17&19.77\\
2 & 28.17 & 21.06 \\
3 & 34.77 & 20.80 \\
4 &36.80 & 19.76\\
\hline
\end{tabular}
\caption{çngulo de deformacin para diferentes posiciones del par de tomos de Litio en el heptaceno denotadas de acuerdo al anillo como 1, 2, 3 y 4 (centro).}
\label{table}
\end{table}

\begin{figure}[htb!]
 \hspace*{-0.3in} \includegraphics[height=0.17\textheight]{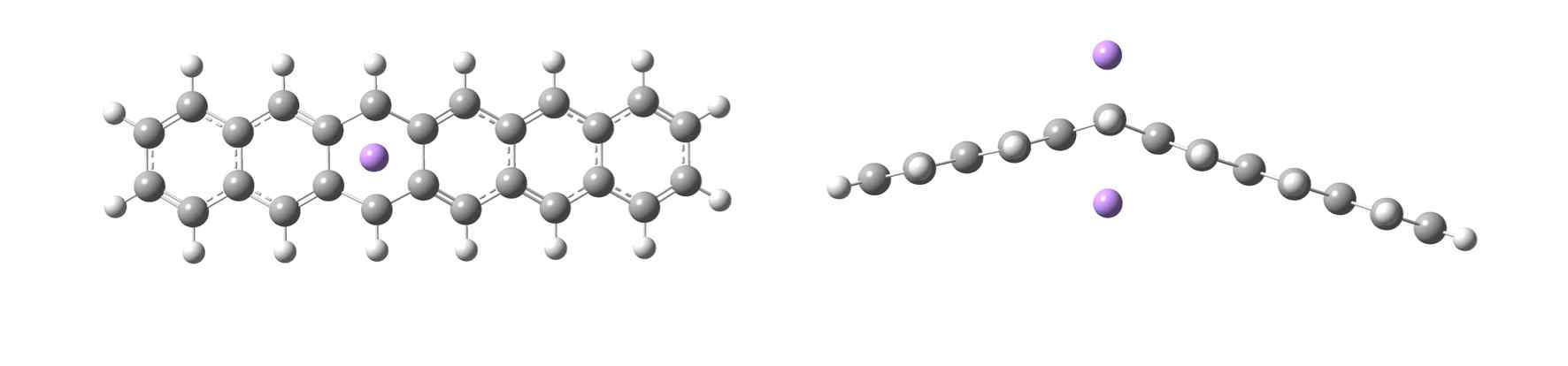}
 \caption{Hexaceno con dos tomos de Litio adsorbidos en el tercer anillo en lados opuestos. }
 \label{hexacene}
\end{figure}

Los poliacenos muestran simetra a lo largo de la cadena, es decir, translaciones finitas y reflexiones, as como rotaciones de 180 grados alrededor del centro de la cadena. El grupo total de simetra es un producto directo de las simetras 1-D en el eje de la cadena y aquellas perpendiculares a la cadena. Si tomamos el punto de vista, que la funcin principal de los tomos adsorbidos es de donar electrones localizados a la cadena, entonces, la densidad de electrones ya no es invariante bajo traslacin. Es importante mencionar que la deformacin se presenta a pesar de que en el poliaceno no exista la configuracin ms simtrica, es decir, para el caso de poliacenos con un nmero par de anillos (no hay anillo central), como ejemplo se muestra en la figura \ref{hexacene} el caso del hexaceno, clculo realizado con DFT (conjunto de bases 6-31g*).

\begin{figure}[htb!]
\includegraphics[height=0.4\textheight]{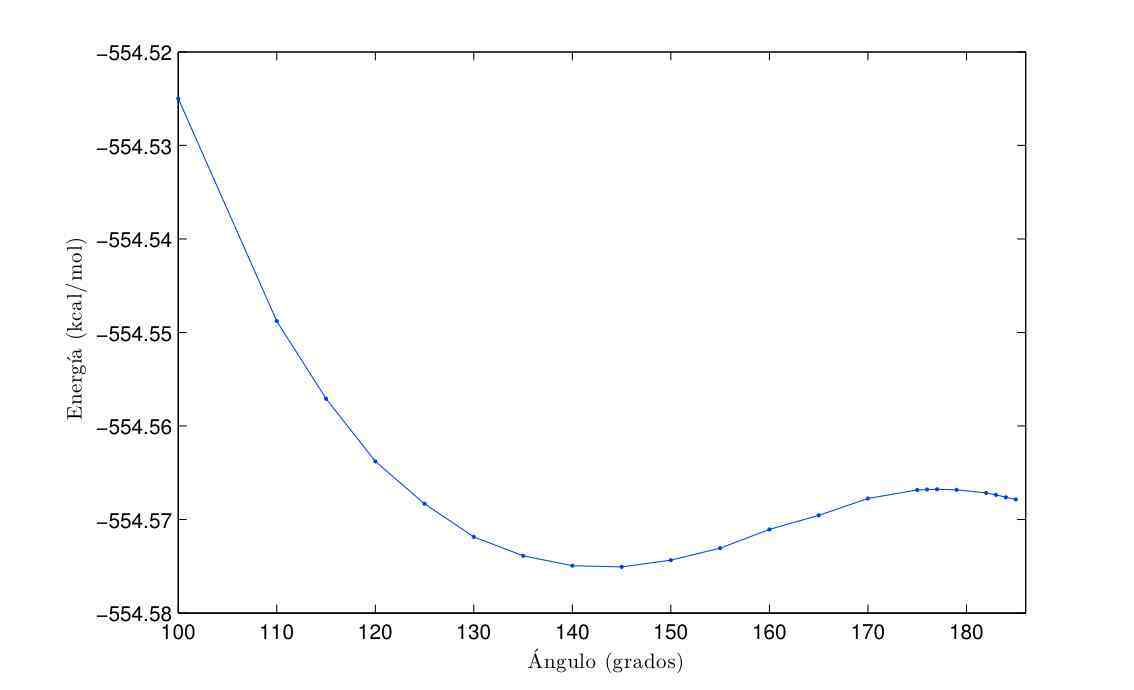}
 \caption{Potencial del antraceno en funcin del ngulo $\pi -\Omega$,}
\label{potencial}
\end{figure}

 Exploramos un poco ms la doble adsorcin de Litio en  el antraceno calculando el   en funcin del ngulo interno $\pi-\Omega$ en la figura \ref{comparison} usando el mtodo DFT con el funcional B3LYP y el conjunto de bases  6-31g*. El potencial se obtuvo optimizando la estructura en la que se vara el ngulo mencionado, fijando las coordenadas de los tomos de carbono e hidrgeno pero dejando las coordenadas de los Litios libres. Con el procedimiento anterior se obtuvo la  grfica \ref{potencial}, en donde se observa que el mnimo est al ngulo de 147¼  y un mximo local  alrededor de 180¼. Cabe mencionar que tambin se usaron bases ms grandes como 6-311g*, 6-311g** y 6-311+g**, en donde obtuvimos resultados cualitativamente similares  y hasta la base 6-311g* sin cambios cualitativos, con la diferencia que el costo computacional es ms grande con las dos ltimas bases. De la grfica \ref{potencial} se puede concluir que no estamos frente a un efecto Janh-Teller, pues para este caso se esperara que el potencial efectivo fuera lineal debido a la degeneracin del estado base para la configuracin plana.
 
 Dichas bases y mtodos usados en sta tesis han sido ampliamente usados, y los resultados corroborados en sistemas similares por distintos grupos de investigacin, slo por mencionar algunos\cite{respaldo1,respaldo2}, sin olvidar estudios estadsticos de los errores debido a la  superposicin de bases realizados para el cdigo de \textit{Gaussian}\cite{Ab_error} que dan respaldo al uso de las mismas.Tambin cabe mencionar que los resultados presentados aqu para el antraceno han sido calibrados con clculos de perturbaciones a segundo orden (MP2) \cite{
 moller}, usando las mismas bases, con lo cual se han obtenido resultados similares.\\

\begin{figure}[htb!]
        \begin{subfigure}[b]{0.5\textwidth}
       \includegraphics[height=0.12\textheight]{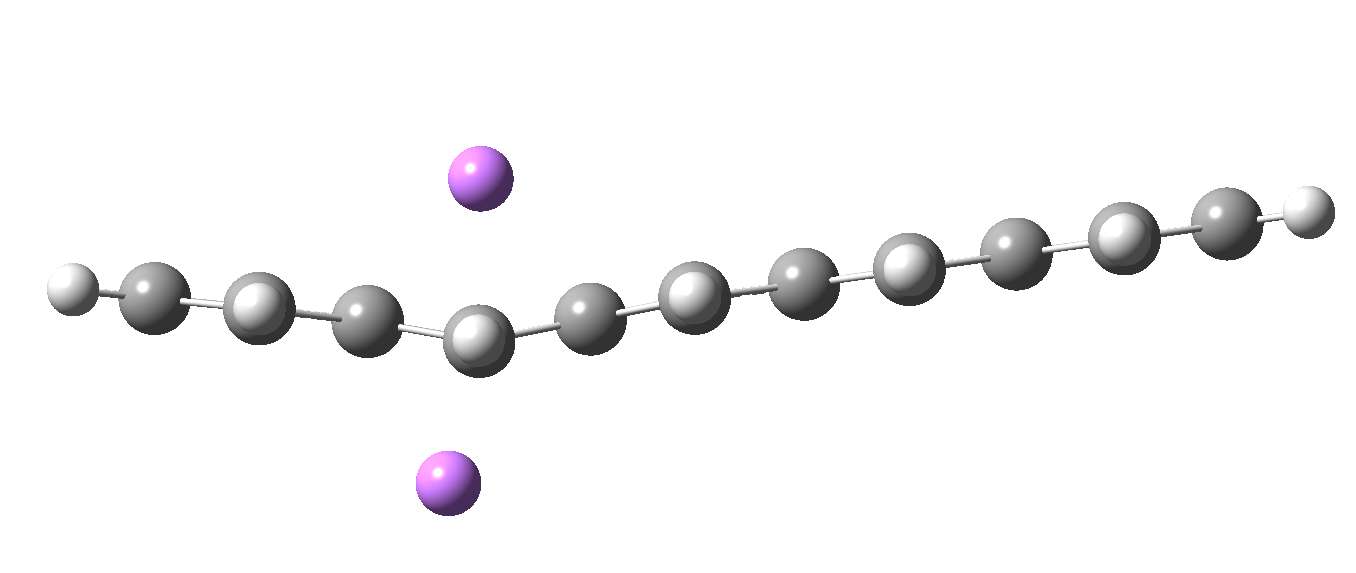}  \caption{}
	\end{subfigure}
       	\begin{subfigure}[b]{0.5\textwidth}
        \includegraphics[height=0.12\textheight]{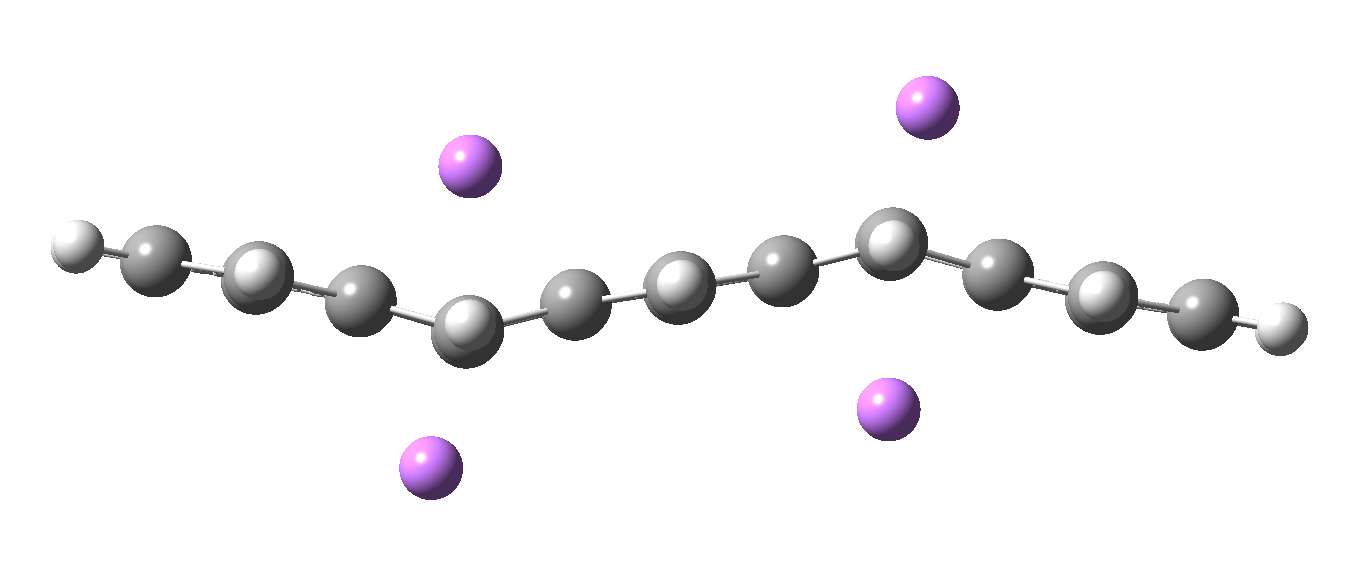}  \caption{}
         \end{subfigure}
          \caption{Se muestra la configuracin de mnima energa obtenida en dos pasos, ($a$) adicionando el primer par al pentaceno sobre el segundo anillo y luego ($b$) adicionando el segundo par  al cuarto anillo.}
          \label{pairV}
 \end{figure}

El siguiente paso fue aumentar el nmero de pares de Litios sobre los poliacenos, para estos clculos de DFT con el mismo funcional y el conjunto de bases  6-311g*.  Inicialmente se hicieron clculos poniendo los pares de Litios sobre anillos adyacentes en pentaceno y heptaceno. En estos casos los mnimos encontrados no muestran ninguna simetra. Los resultados sorprendentes los obtuvimos cuando pusimos los diferentes pares en anillos alternados. El procedimiento llevado a cabo en dichos clculos fue el siguiente: Se probaron diferentes condiciones iniciales variando la distancia de los Litios al pentaceno, los pasos y ordenacin para obtener la configuracin de mnima energa, con esto ltimo me refiero a los pasos y orden que se tomaron para optimizar la estructura. Para ilustrar mejor lo anterior se muestra la figura \ref{pairV} que corresponde a la estructura optimizada en dos pasos, primero se adiciona el primer par en el segundo anillo del pentaceno y se optimiza, la estructura optimizada se muestra en la figura \ref{pairV}($a$); luego a partir de sta configuracin se adiciona el segundo par en el cuarto anillo y nuevamente se optimiza, obteniendo la configuracin que se muestra en la figura \ref{pairV}($b$).\\

\begin{figure}[htb!]
        \begin{subfigure}[b]{0.5\textwidth}
       \includegraphics[height=0.12\textheight]{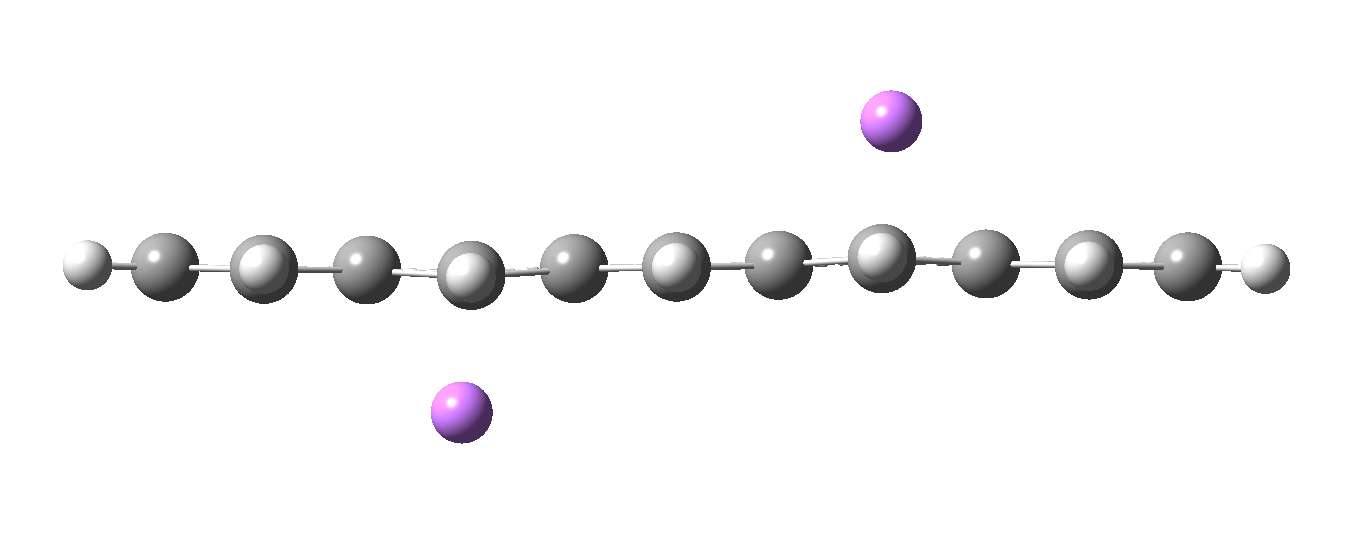}  \caption{}
	\end{subfigure}
       	\begin{subfigure}[b]{0.5\textwidth}
        \includegraphics[height=0.12\textheight]{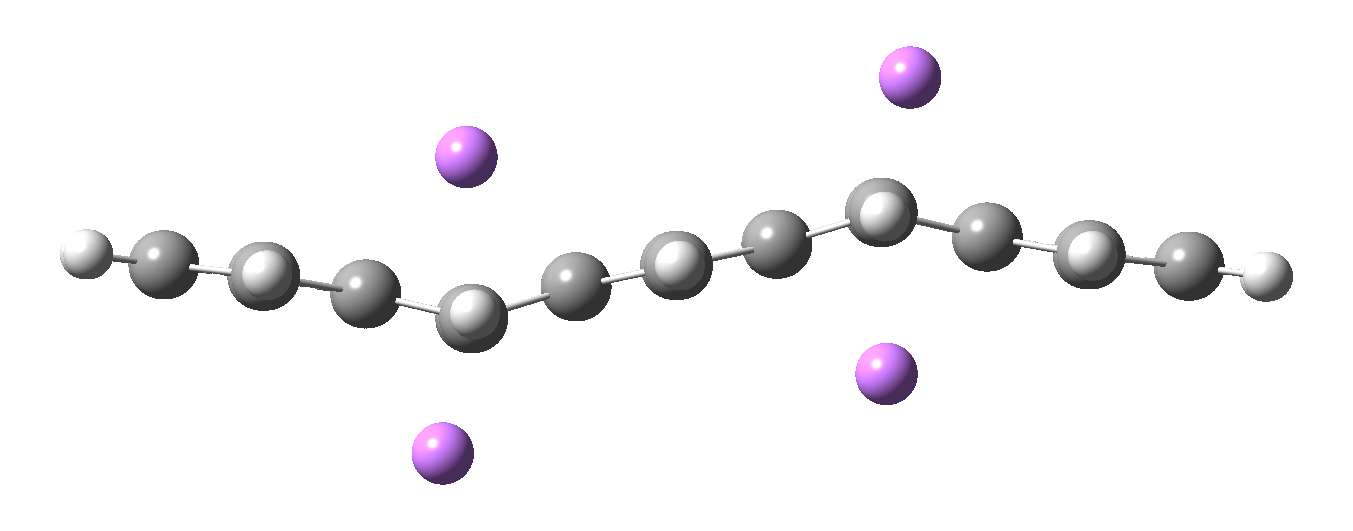}  \caption{}
         \end{subfigure}
          \caption{Se muestra la configuracin de mnima energa obtenida en dos pasos, ($a$) adicionando el primer par al pentaceno sobre el segundo y cuarto anillo del mismo en lados opuestos y luego ($b$) adicionando el segundo par  en los lugares restantes.}
          \label{diagV}
 \end{figure}

Otro camino en dos pasos es el se muestra en la figura \ref{diagV}, en donde el primer paso es poner un Litio de un lado en el segundo anillo del pentaceno y otro Litio en el lado opuesto del cuarto anillo y se optimiza la configuracin cuyo resultado se muestra en la figura \ref{diagV}($a$); luego se adiciona el segundo par en los anillos 2 y 4 y nuevamente se optimiza dando como resultado la configuracin que se muestra en la figura \ref{diagV}($b$). En este caso, como en el anteriormente descrito, la configuracin de mnima energa para la adsorcin de cuatro Litios en el pentaceno es el zigzag, con una energa de adsorcin de -101.2 kcal/mol.\\

\begin{figure}[htb!]
        \begin{subfigure}[b]{0.5\textwidth}
       \includegraphics[height=0.12\textheight]{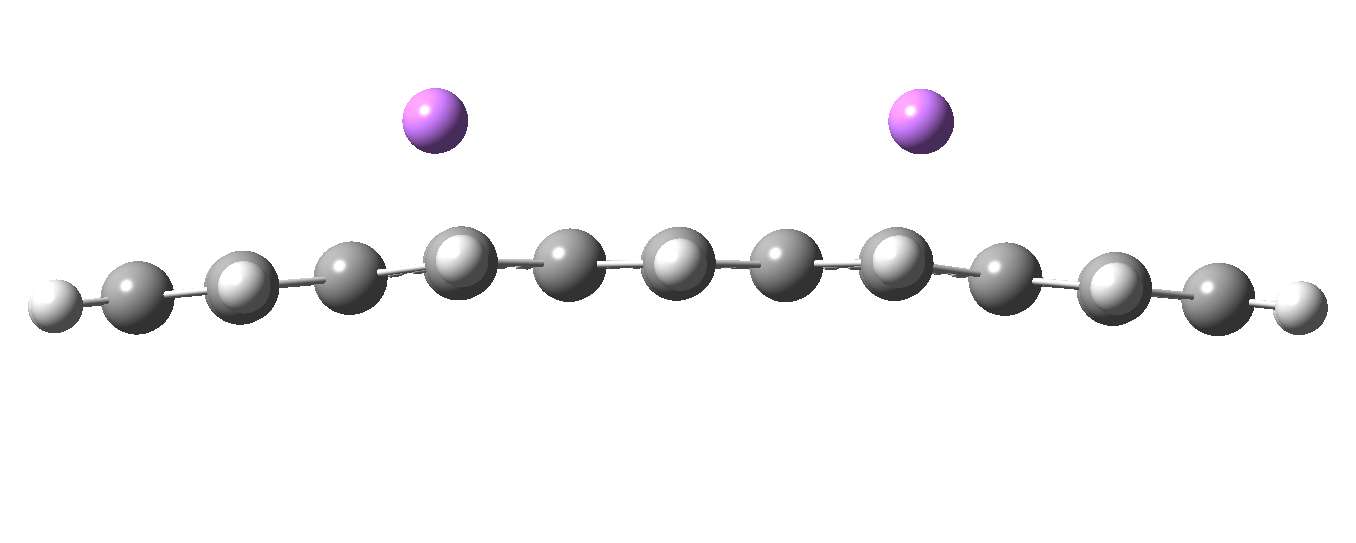}  \caption{}
	\end{subfigure}
       	\begin{subfigure}[b]{0.5\textwidth}
        \includegraphics[height=0.12\textheight]{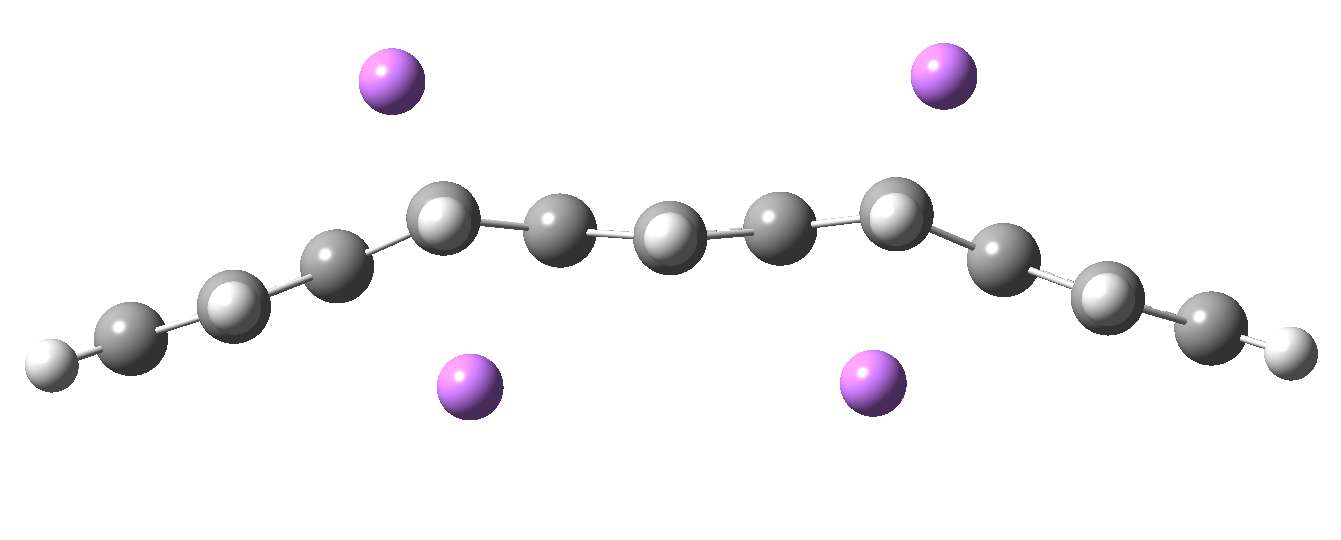}  \caption{}
         \end{subfigure}
          \caption{Se muestra la configuracin de mnima energa obtenida en dos pasos, ($a$) adicionando el primer par al pentaceno sobre el segundo y cuarto anillo del mismo lado y luego ($b$) adicionando el segundo par  en los lugares restantes al pentaceno.}
          \label{up_downV}
 \end{figure}

El ltimo camino posible en la optimizacin de dos pasos es poner el primer par del mismo lado en el segundo y cuarto anillo, optimizando se obtiene la configuracin mostrada en la figura \ref{up_downV}($a$); luego se adiciona el segundo par en lado opuesto al anterior y optimizando se obtiene la configuracin mostrada en \ref{up_downV}($b$). Esta configuracin final tiene una energa de adsorcin de -99.0 kcal/mol, que corresponde a una energa mayor a la obtenida con los caminos anteriormente mencionados, es decir es un mnimo local.  \\
 
\begin{figure}[htb!]
        \begin{subfigure}[b]{0.5\textwidth}
       \includegraphics[height=0.142\textheight]{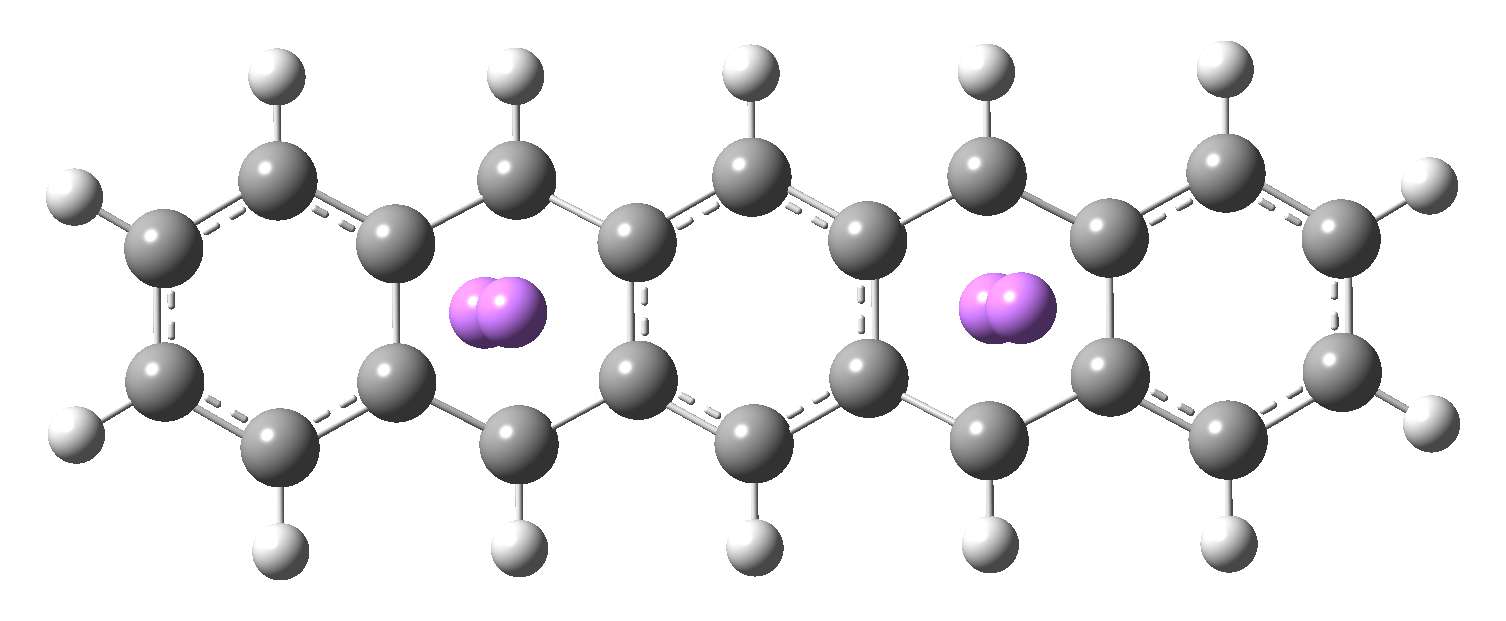}
	\end{subfigure}
       	\begin{subfigure}[b]{0.5\textwidth}
        \includegraphics[height=0.142\textheight]{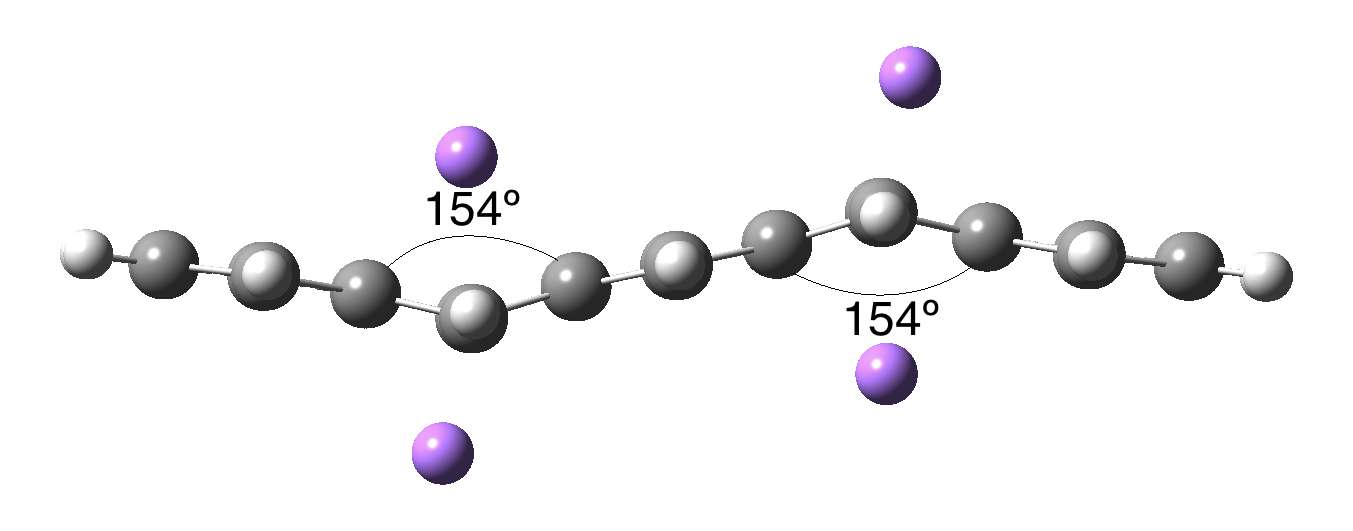} 
         \end{subfigure}
          \caption{Se muestra la configuracin de mnima energa para el pentaceno con adsorcin de cuatro Litios en el segundo y cuarto anillo.}
          \label{zigV}
 \end{figure}

  Esto es slo una muestra de los diferentes caminos y pasos que se probaron para encontrar la configuracin de mnima energa, cabe mencionar que para el caso del pentaceno tambin se realizaron optimizaciones de un tomo de Litio por paso con diferentes secuencias, as como la optimizacin de los cuatro litios al mismo tiempo y diferentes combinaciones entre adsorcin de un Litio y luego tres Litios.  Dando como resultado la configuracin de mnima energa igual a las mostradas en las figuras \ref{diagV}($b$) y  \ref{pairV}($b$) y  \ref{zigV}, en esta ltima podemos apreciar la vista frontal y lateral de la configuracin as como la medida de los ngulos del zigzag obtenido. ste mismo procedimento se llev a cabo con todos los clculos mostrados en este trabajo. \\
 \\

\begin{figure}[htb!]
        \begin{subfigure}[b]{0.5\textwidth}
        \hspace*{-0.1in}\includegraphics[height=0.14\textheight]{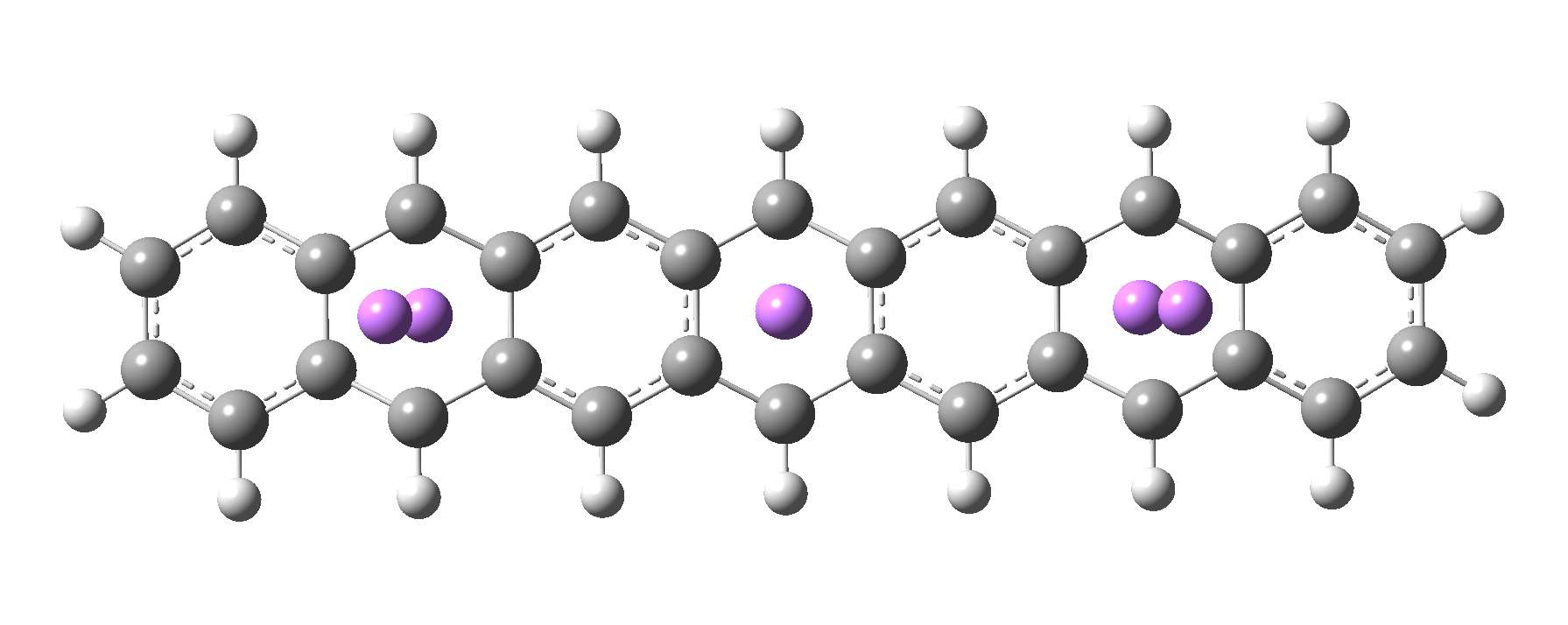} 
	\end{subfigure}
       	\begin{subfigure}[b]{0.5\textwidth}
       \hspace*{-0.1in}\includegraphics[height=0.14\textheight]{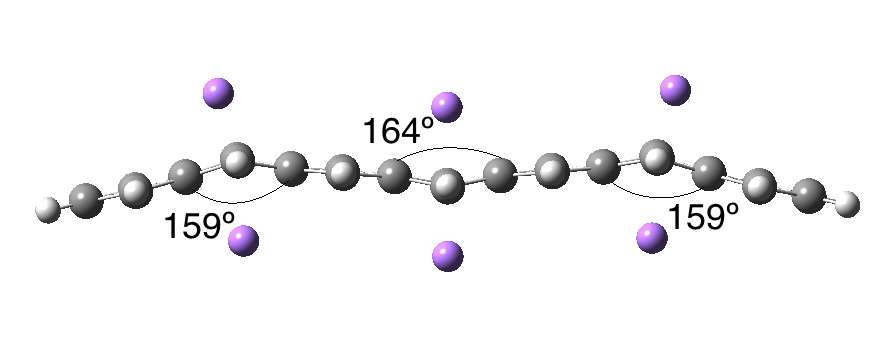}  
         \end{subfigure}
          \caption{Se muestra la configuracin de mnima energa para el heptaceno con adsorcin de seis Litios en el segundo, cuarto y sexto anillo.}
          \label{minVII}
 \end{figure}

Anlogamente se realizaron clculos para el heptaceno con tres pares de Litios, cada uno en el segundo, cuarto y sexto anillo respectivamente y cuya configuracin de mnima energa, mostrada en la figura \ref{minVII}, nuevamente obtuvimos el zigzag. Para el caso del nonaceno obtuvimos que el zigzag es un mnimo local no absoluto. Estos clculos fueron reproducidos con el programa NWchem \cite{nwchem} con el cual obtuvimos los mismos resultados cualitativos y cuantitativos.\\

\begin{figure}[htb!]
 \begin{subfigure}[b]{0.5\textwidth}
     \hspace*{0.7in}\includegraphics[height=0.12\textheight]{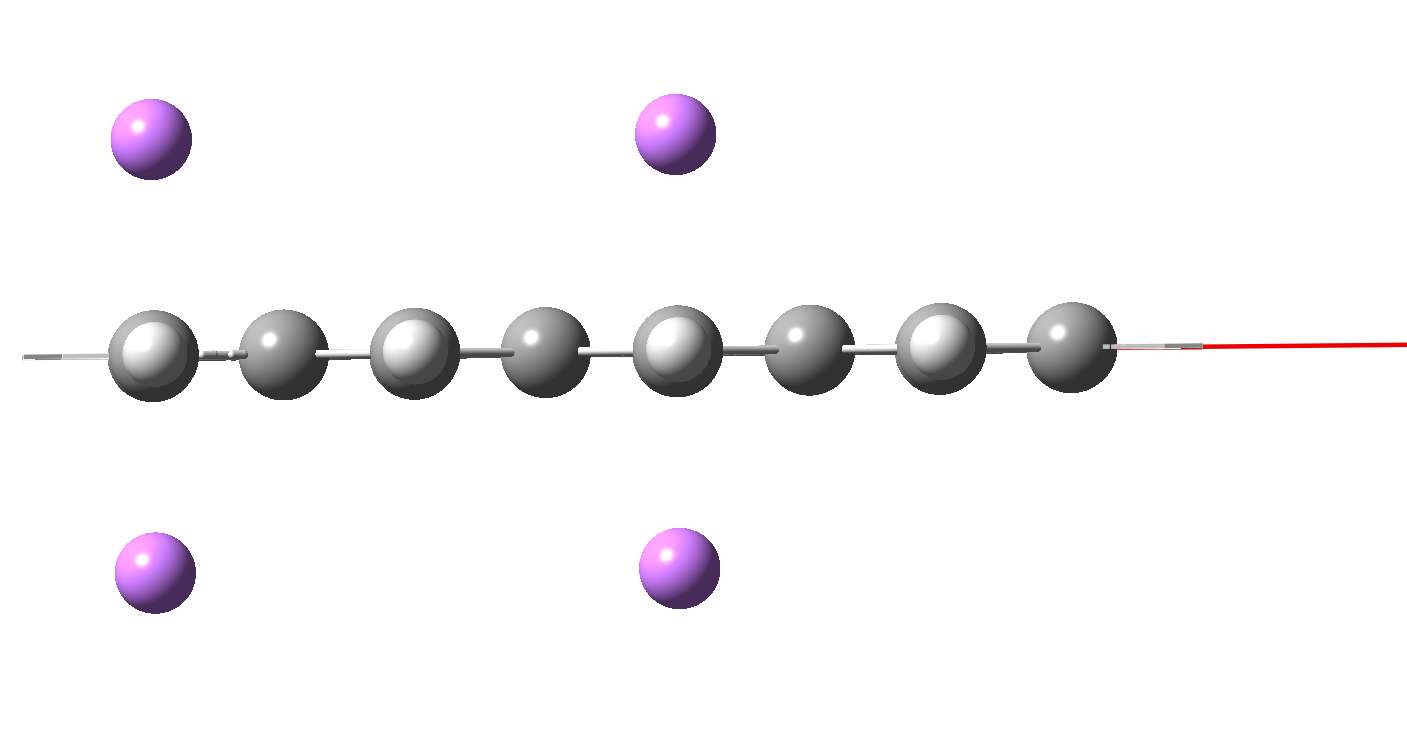}	\caption{}
         \end{subfigure}
	 \begin{subfigure}[b]{0.5\textwidth}
    \hspace*{0.5in} \includegraphics[height=0.13\textheight]{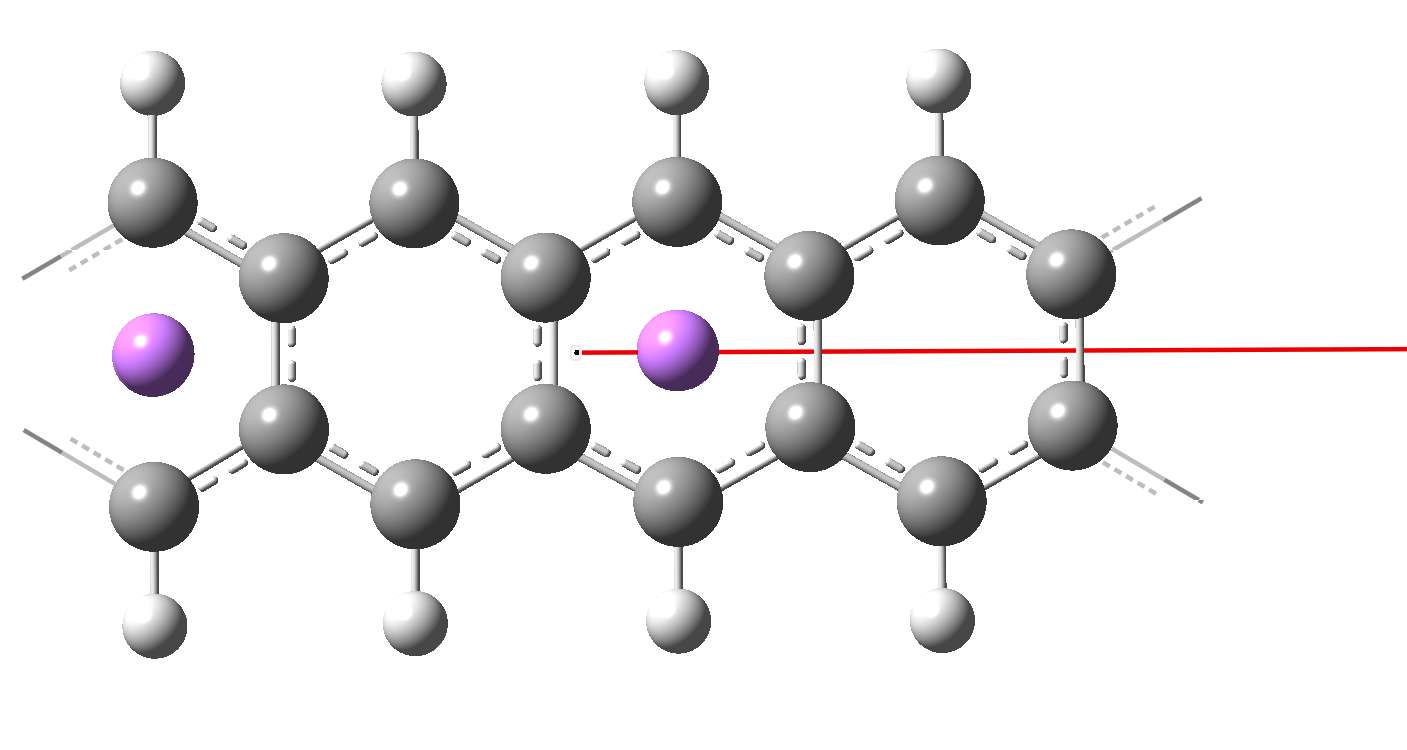}  	\caption{}
	\end{subfigure}\\

        \begin{subfigure}[b]{0.3\textwidth}
        \includegraphics[height=0.13\textheight]{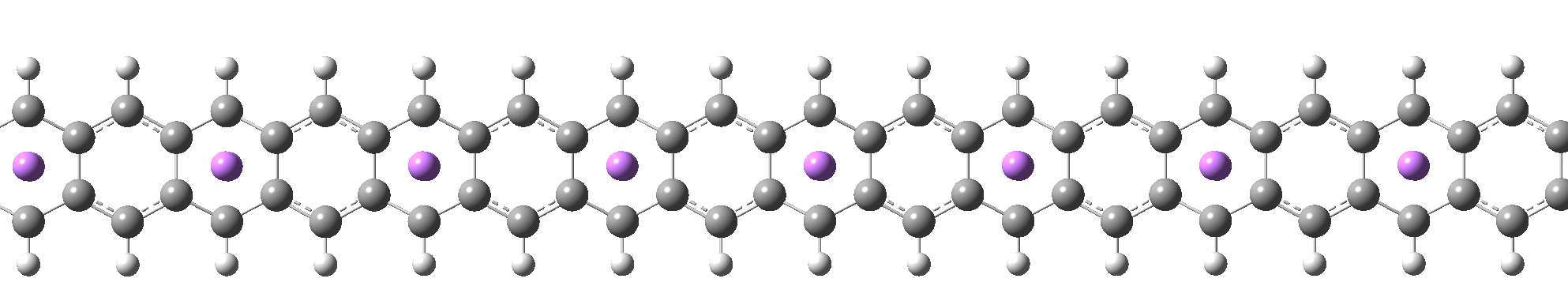}    
	\end{subfigure}\\
        	\begin{subfigure}[b]{1\textwidth}
         \includegraphics[height=0.13\textheight]{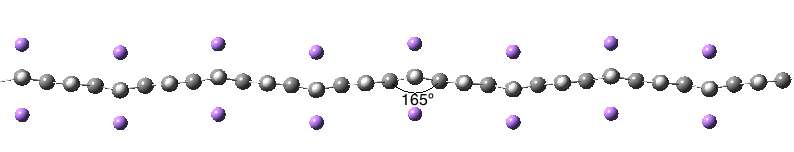}   \caption{}
         \end{subfigure}
         \caption{Clculo peridico usando una celda unitaria con cuatro Litios mostrada desde una vista ($a$) lateral y ($b$) frontal. ($c$) Celda unitaria optimizada repetida cuatro veces. }
         \label{period}
\end{figure}

Pensando en extender cada vez ms el poliaceno se realizaron clculos peridicos usando el programa Gaussian con el mtodo de DFT, con el potencial B3LYP y el conjunto de bases 6-311g. En la figura \ref{period}($a$) y \ref{period}($b$) se ilustra la celda unitaria que contiene cuatro Litios. La configuracin de la mnima energa es la ilustrada en la figura \ref{period}($c$).\\

Se obtuvieron los mismos resultados con el programa cp2k \cite{cp2k}, que se realizaron en la Universidad ETH, en Zurich, Suiza. Para ste caso se us el potencial BLYP\cite{gth} y las bases MOLOPT\cite{molopt}.
 
El zigzag obtenido en los diferentes poliacenos y en particular en este clculo peridico nos brinda la evidencia de que estamos frente a una transicin de Peierls en donde hay un rompimiento de simetra de traslacin y un aumento en el periodo, de un periodo de dos anillos a uno de cuatro anillos de la cadena. \\

\textbf{Adsorcin de otros alcalinos en poliacenos}\\

Tambin se hicieron clculos para la adsorcin de los dems alcalinos (Na, K,  Rb, Cs y Fr ) en el anillo central del antraceno y pentaceno y, a excepcin del Sodio, el poliaceno siempre permaneci plano y los alcalinos sobre el eje que va  del centro del anillo y perpendicular al plano del poliaceno. \\

\begin{figure}[htb!]
        \begin{subfigure}[b]{0.5\textwidth}
     \hspace*{0.5in}   \includegraphics[height=0.13\textheight]{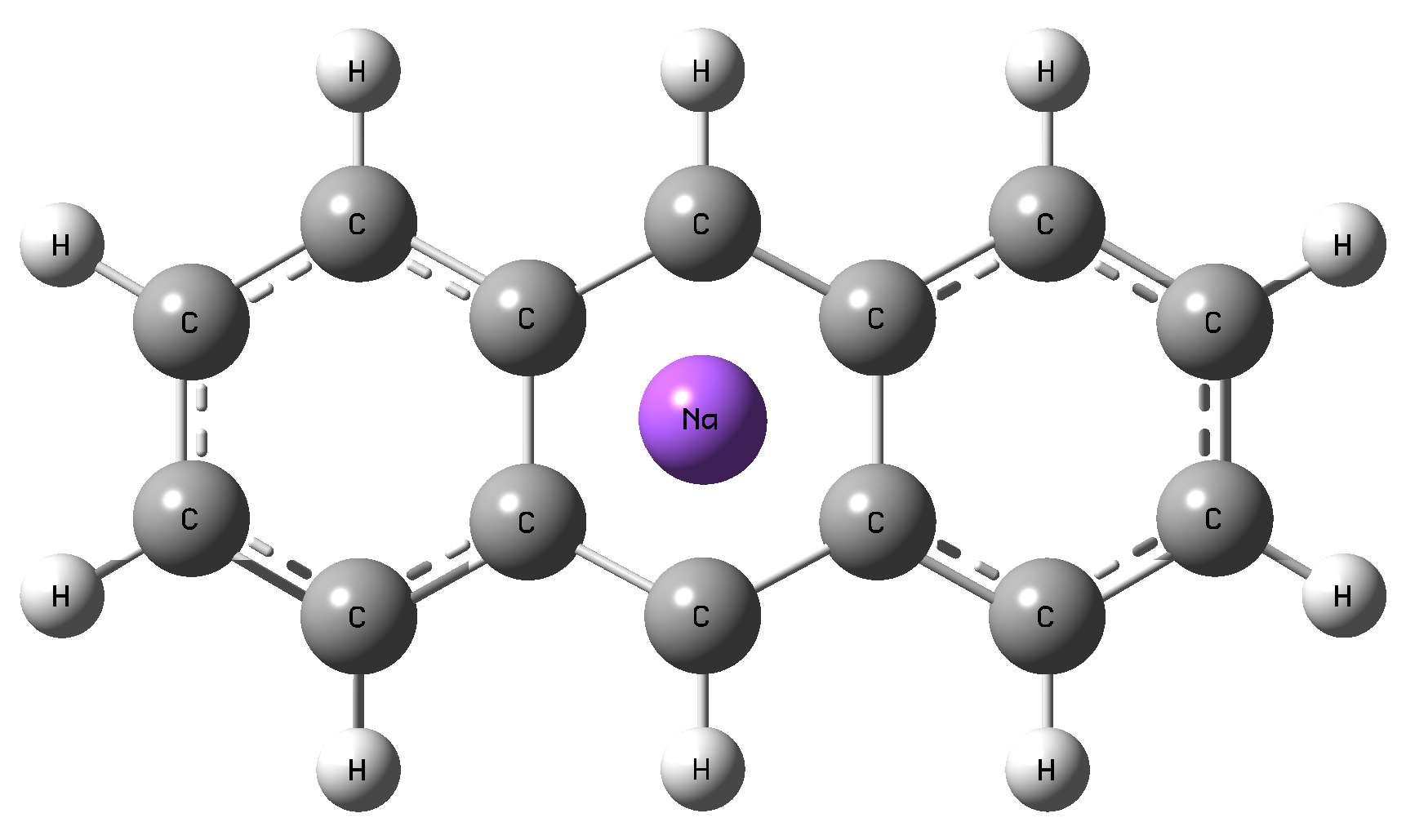} 
	\end{subfigure}
       	\begin{subfigure}[b]{0.5\textwidth}
        \hspace*{0.5in}    \includegraphics[height=0.13\textheight]{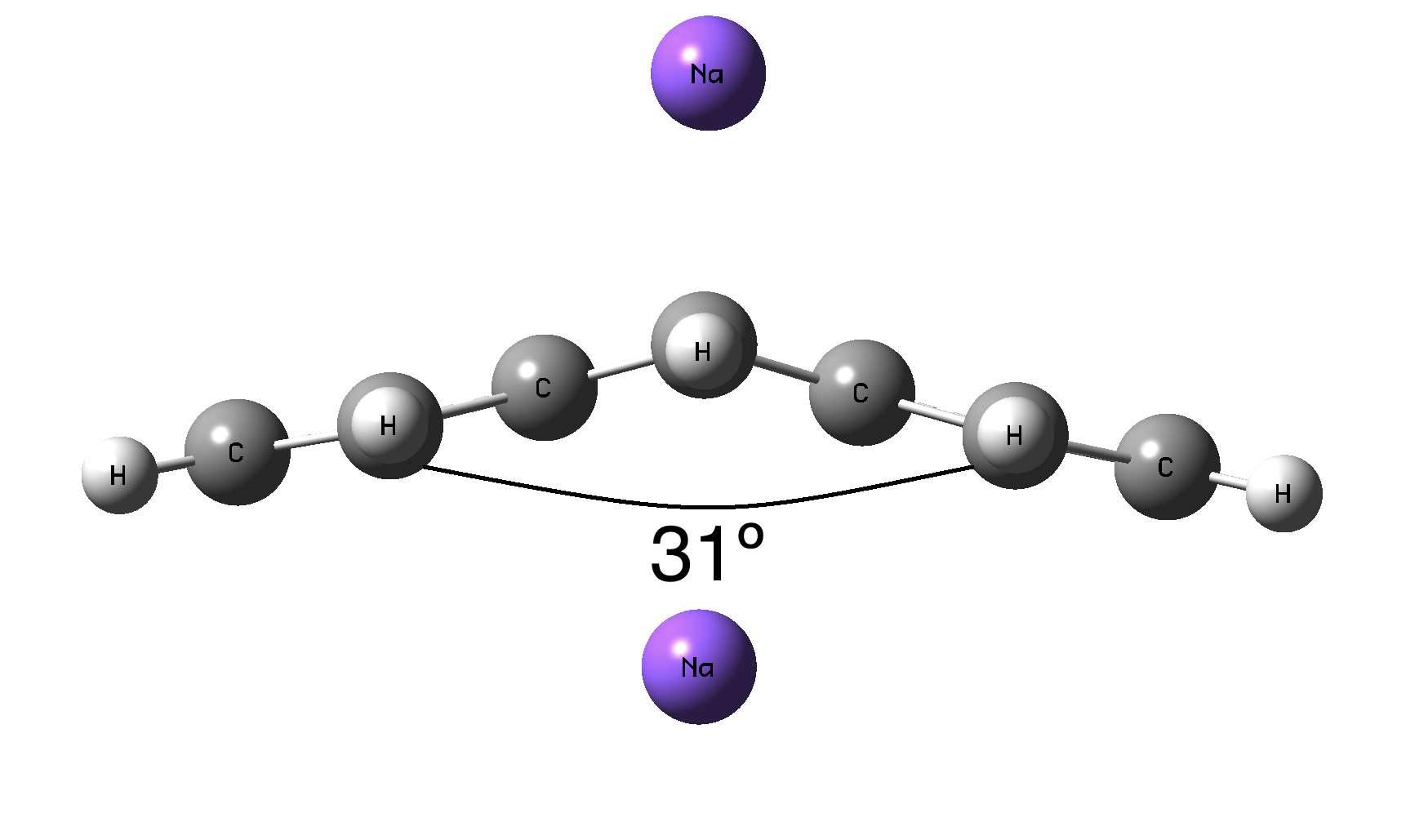}  
         \end{subfigure}
          \caption{Se muestra la configuracin de mnima energa para de la adsorcin de un par de Sodios en el anillo central del antraceno, obtenida del clculo realizado con el mtodo de DFT, con el potencial B3LYP y el conjunto de bases 6-311g(d).}
          \label{naantra}
 \end{figure}

\begin{figure}[htb!]
        \begin{subfigure}[b]{0.5\textwidth}
     \includegraphics[height=0.13\textheight]{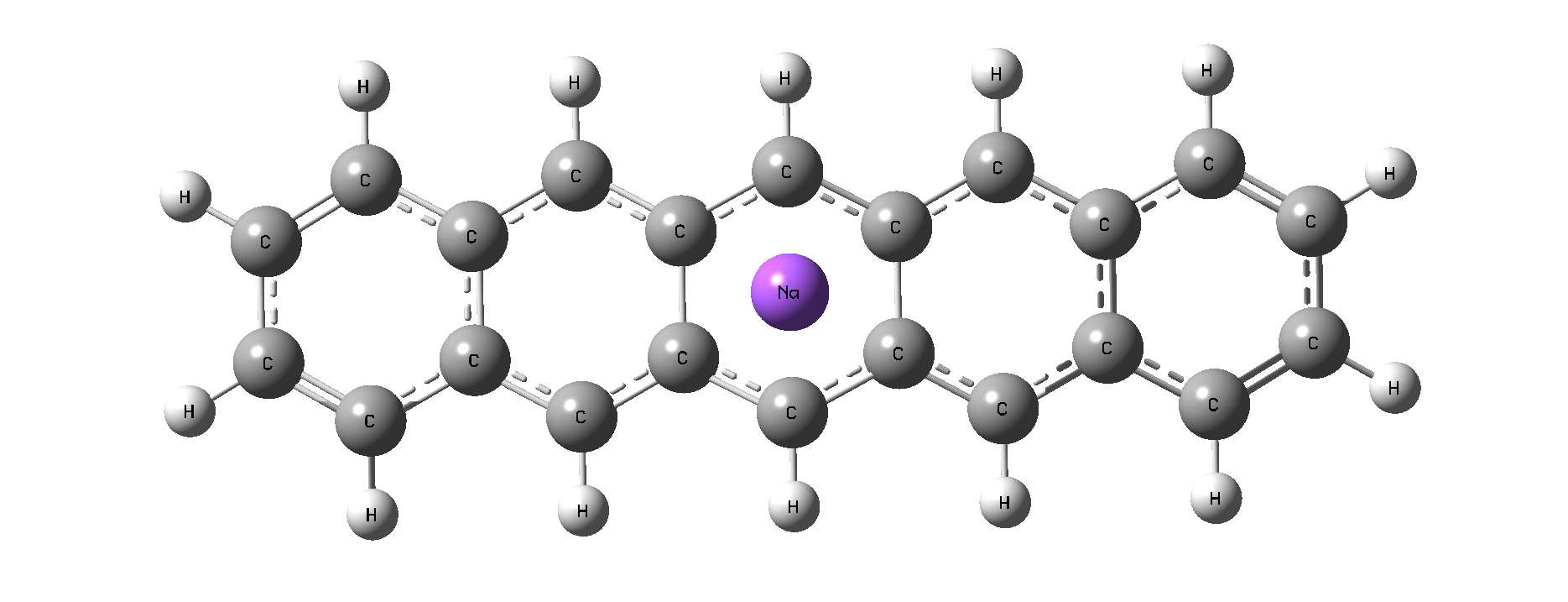} 
	\end{subfigure}
       	\begin{subfigure}[b]{0.5\textwidth}
         \includegraphics[height=0.13\textheight]{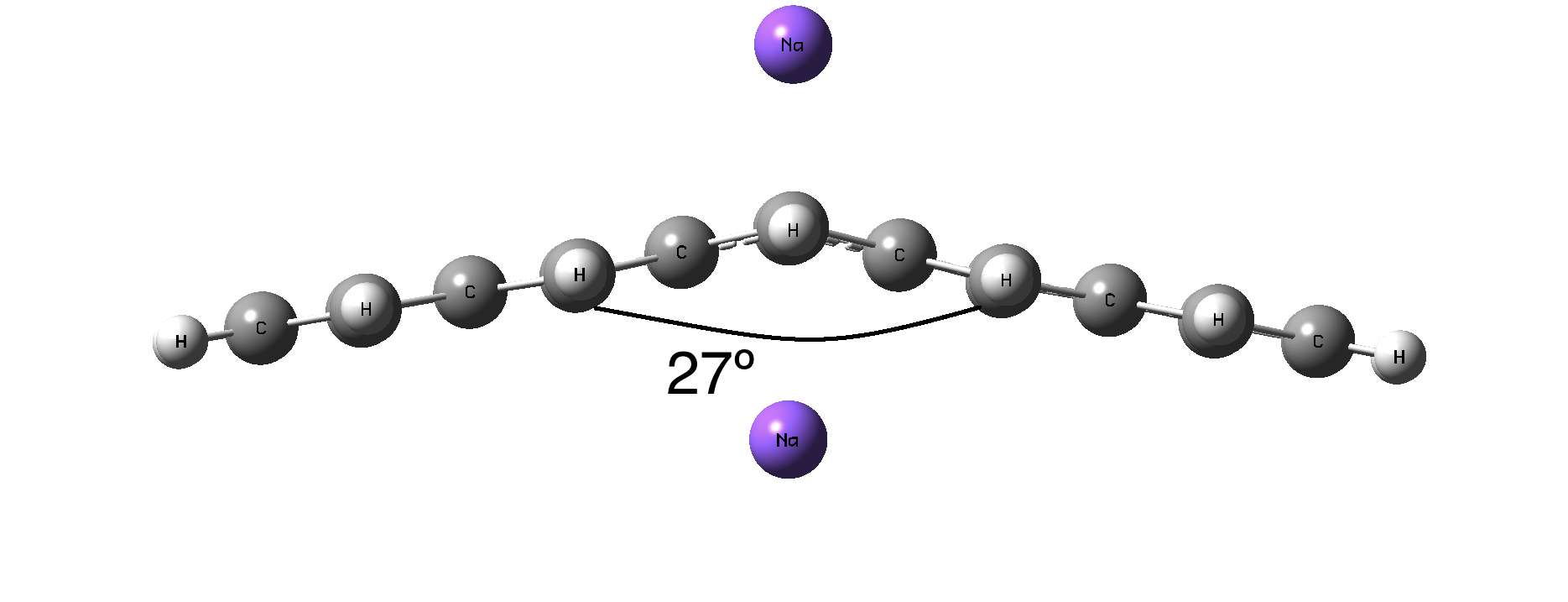}  
         \end{subfigure}
          \caption{Se muestra la configuracin de mnima energa para de la adsorcin de un par de Sodios en el anillo central del pentaceno, obtenida del clculo realizado con el mtodo de DFT, con el potencial B3LYP y el conjunto de bases 6-311g(d).}
          \label{napenta}
 \end{figure}

En cuanto al Sodio, se calcul la doble adsorcin de ste en lados opuestos al anillo central del antraceno, y del pentaceno, realizado con el mtodo de DFT, con el potencial B3LYP y el conjunto de bases 6-311g(d). Las configuraciones de mnima energa en el caso del antraceno y pentaceno se muestran en la figuras \ref{naantra} y \ref{napenta} respectivamente.\\

\begin{figure}[htb!]
        \begin{subfigure}[b]{0.5\textwidth}
     \hspace*{0.5in}   \includegraphics[height=0.14\textheight]{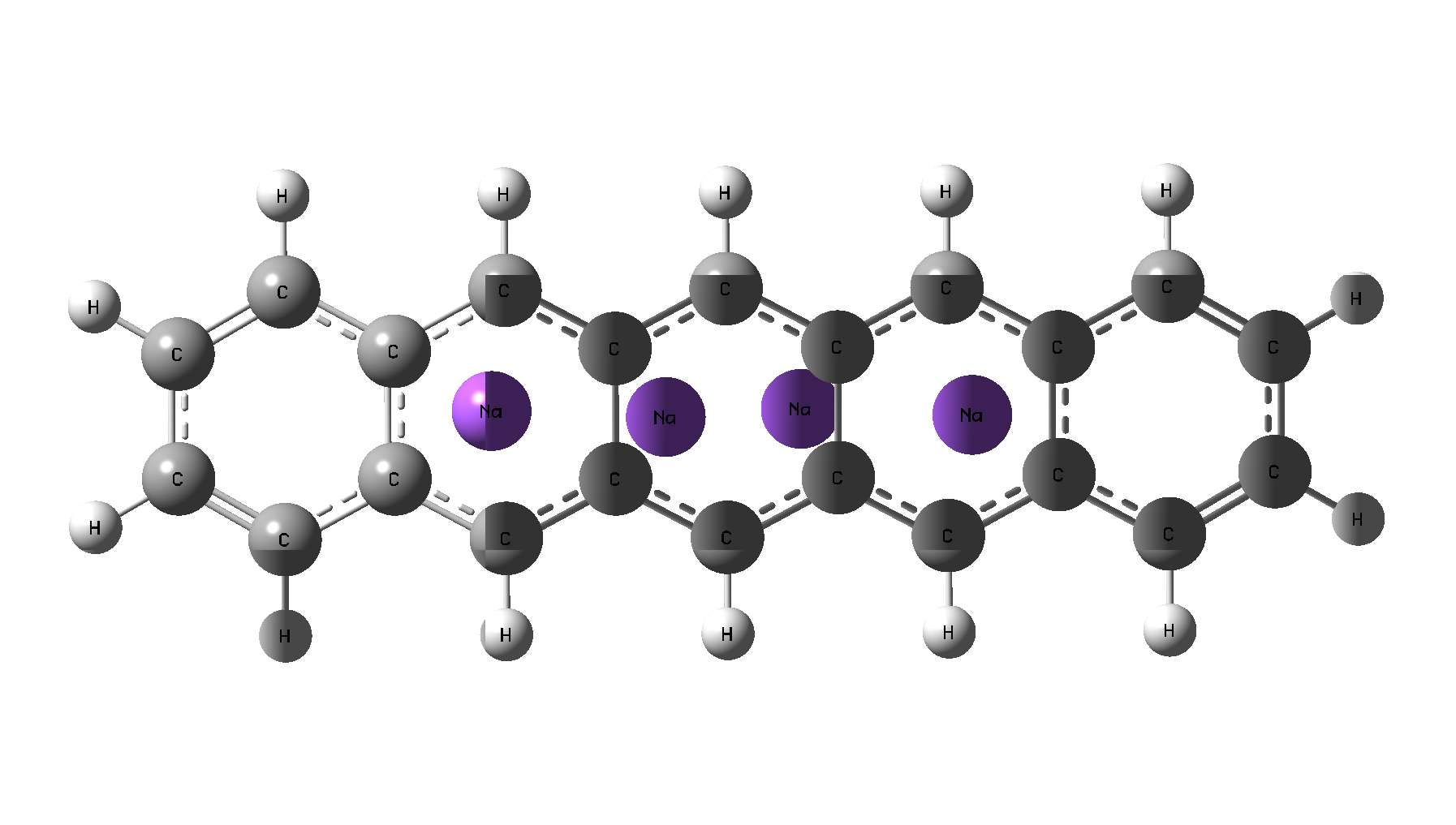} 
	\end{subfigure}
       	\begin{subfigure}[b]{0.5\textwidth}
        \hspace*{0.5in}    \includegraphics[height=0.14\textheight]{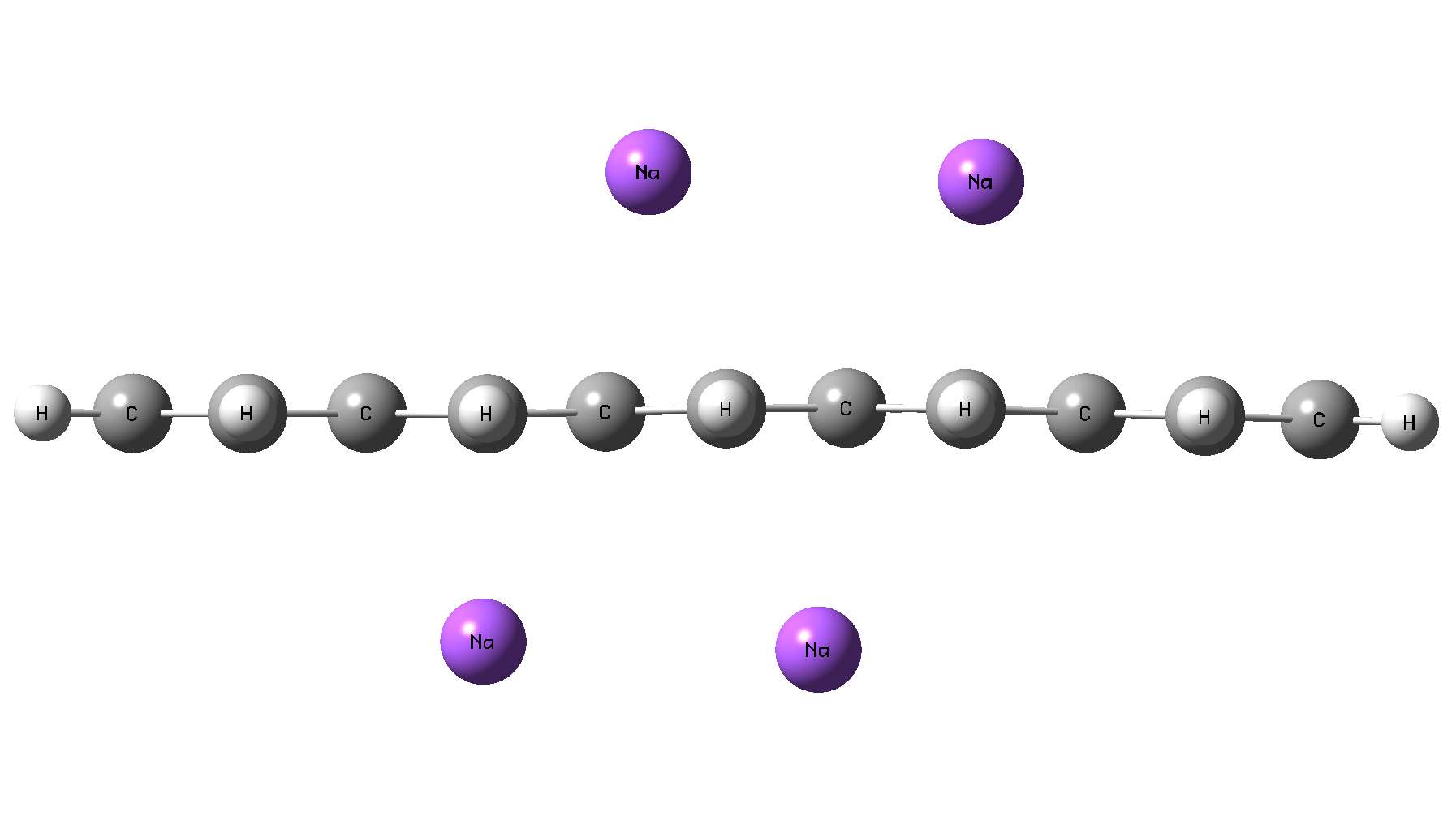}  
         \end{subfigure}
          \caption{Se muestra la configuracin de mnima energa para de la adsorcin de dos pares de Sodios en el segundo y cuarto anillo  del pentceno, obtenida del clculo realizado con el mtodo de DFT, con el potencial B3LYP y el conjunto de bases 6-311g(d).}
          \label{fourna_penta}
 \end{figure}

\begin{figure}[htb!]
        \begin{subfigure}[b]{0.5\textwidth}
     \includegraphics[height=0.13\textheight]{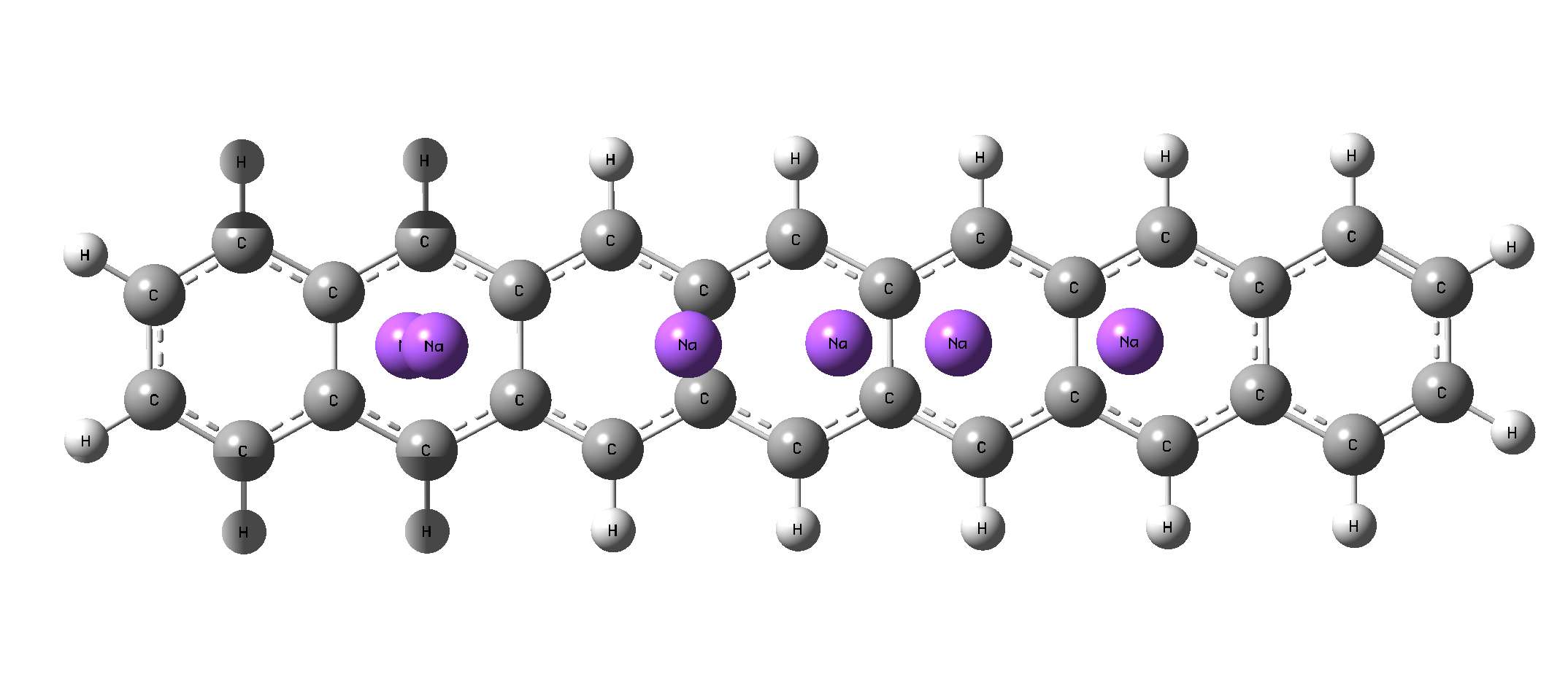} 
	\end{subfigure}
       	\begin{subfigure}[b]{0.5\textwidth}
         \includegraphics[height=0.13\textheight]{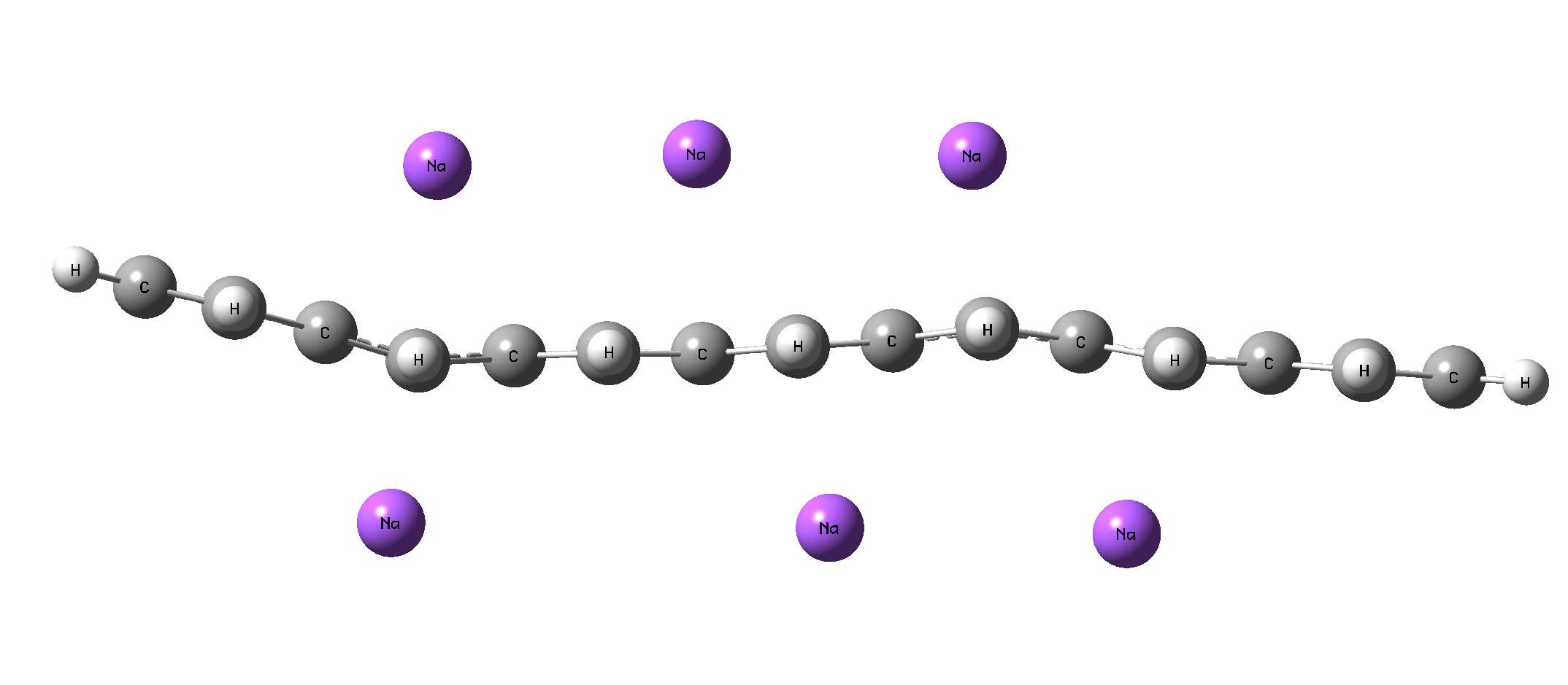}  
         \end{subfigure}
          \caption{Se muestra la configuracin de mnima energa para de la adsorcin de tres pares de Sodios en el el segundo, cuarto y sexto anillo  del heptaceno, obtenida del clculo realizado con el mtodo de DFT, con el potencial B3LYP y el conjunto de bases 6-311g(d).}
          \label{sixna_hepta}
 \end{figure}

Debido a que un par de Sodios deforman al antraceno y pentaceno de una forma parecida a como lo hacen un par de Litios en dichos poliacenos, se explor la mltiple adsorcin de forma similar que como se hizo con el Litio.  Con el mismo mtodo, funcional y conjunto de bases que en el caso anterior se obtuvo la configuracin de mnima energa para la adsorcin de dos pares de Sodios en el segundo y cuarto anillo del pentaceno, que se muestra en la figura \ref{fourna_penta}. Tambin se obtuvo la configuracin de mnima energa para la adsorcin de tres pares de Sodios en el el segundo, cuarto y sexto anillo  del heptaceno, que se muestra en la figura \ref{sixna_hepta}.  Los resultados obtenidos indican que no se produce el mismo efecto que la adsorcin de pares de litios en poliacenos, para el caso de adsorcin del Sodio y por lo tanto no se puede pensar en transicin de Peierls para la adsorcin de sodio en poliacenos.

\clearpage

\subsection{Poli-para-fenilenos}

\begin{figure}[htb!]
       \hspace*{0.7in}   \includegraphics[height=0.5\textheight]{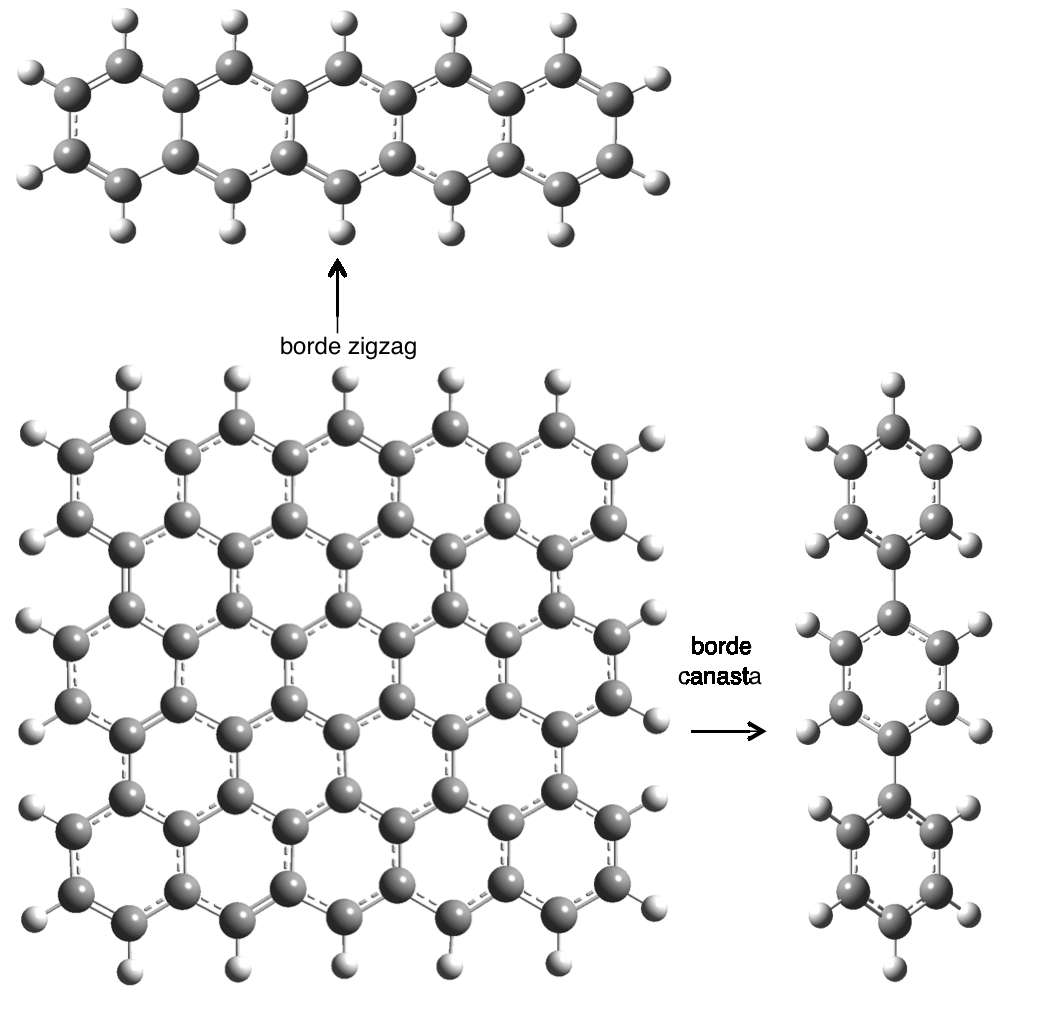} 
          \caption{Se muestran los diferentes bordes de una molcula aromtica simtrica}
          \label{bordes}
 \end{figure}

Buscando extender nuestros clculos a molculas aromticas ms grandes, especficamente a tiras de grafeno \cite{ribbons1}, de las cuales se hablar en la siguiente seccin, se encuentra que algunas propiedades fsicas de stas dependen del borde que tengan. La figura \ref{bordes} ilustra los  diferentes bordes en una molcula aromtica, uno es el borde \textit{zigzag} que es caracterstico para los poliacenos, que fueron estudiados en la seccin anterior. El otro borde se conoce como borde \textit{canasta} el cual es  caracterstico para los poli-para-fenilenos \cite{polyph1}. Estos poli-para-fenilenos han sido de gran inters para qumicos  debido a su simplicidad y simetra, tambin por su estabilidad trmica, conductividad elctrica y propiedades pticas \cite{polyph2}. Tienen amplia aplicacin en la industria del plstico \cite{plastic1, plastic2}.

\begin{figure}[htb!]
        \begin{subfigure}[b]{0.5\textwidth}
     \hspace*{0.3in}   \includegraphics[height=0.09\textheight]{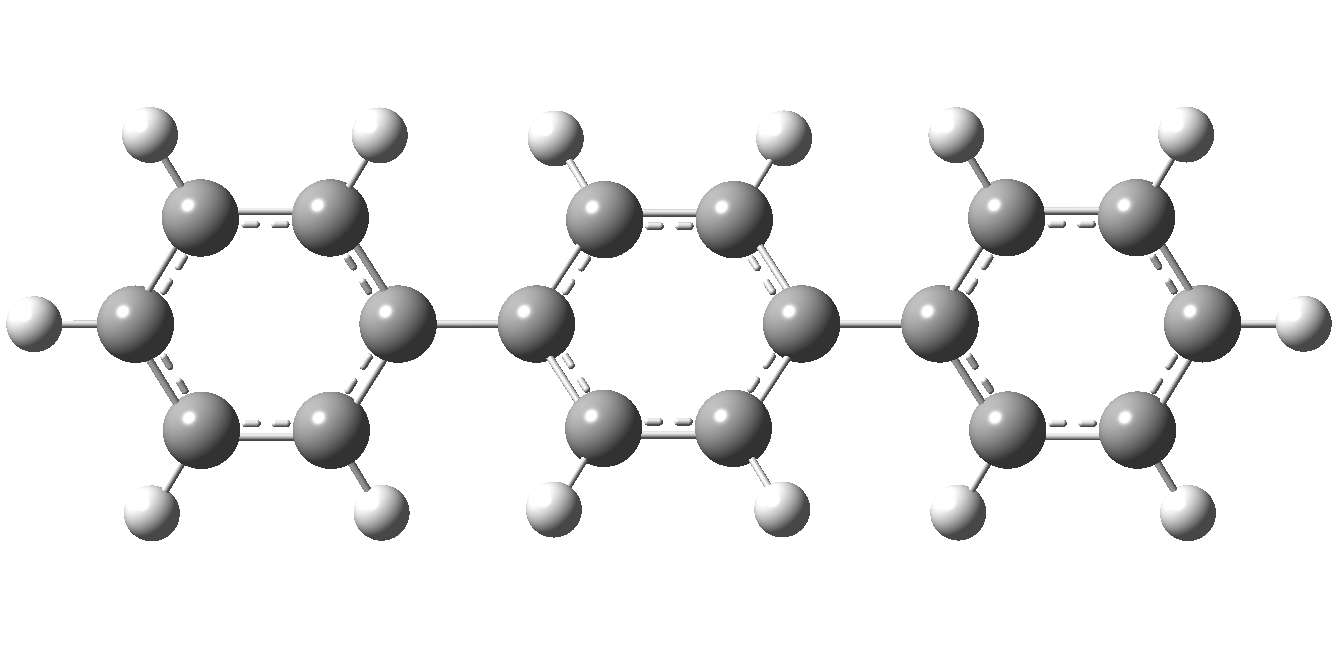} \caption{}
	\end{subfigure}
       	\begin{subfigure}[b]{0.5\textwidth}
           \hspace*{0.3in} \includegraphics[height=0.09\textheight]{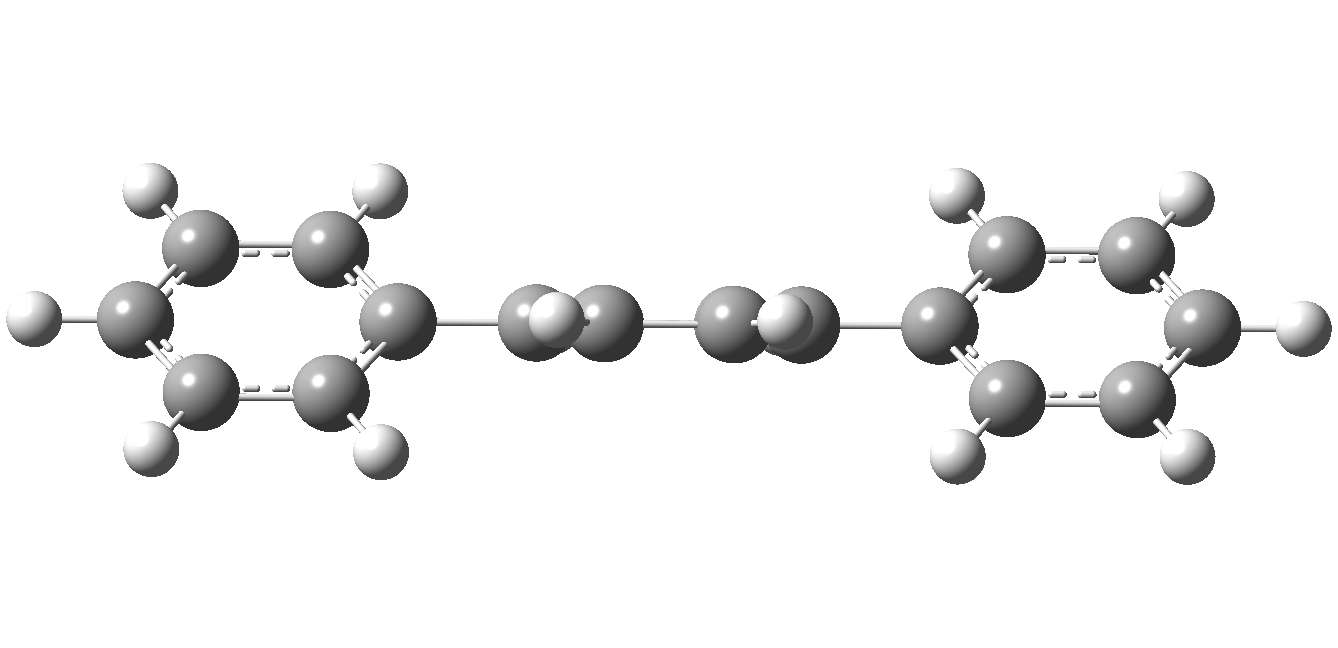} \caption{}
         \end{subfigure}\\
           \begin{subfigure}[b]{0.5\textwidth}
     \hspace*{0.1in}   \includegraphics[height=0.11\textheight]{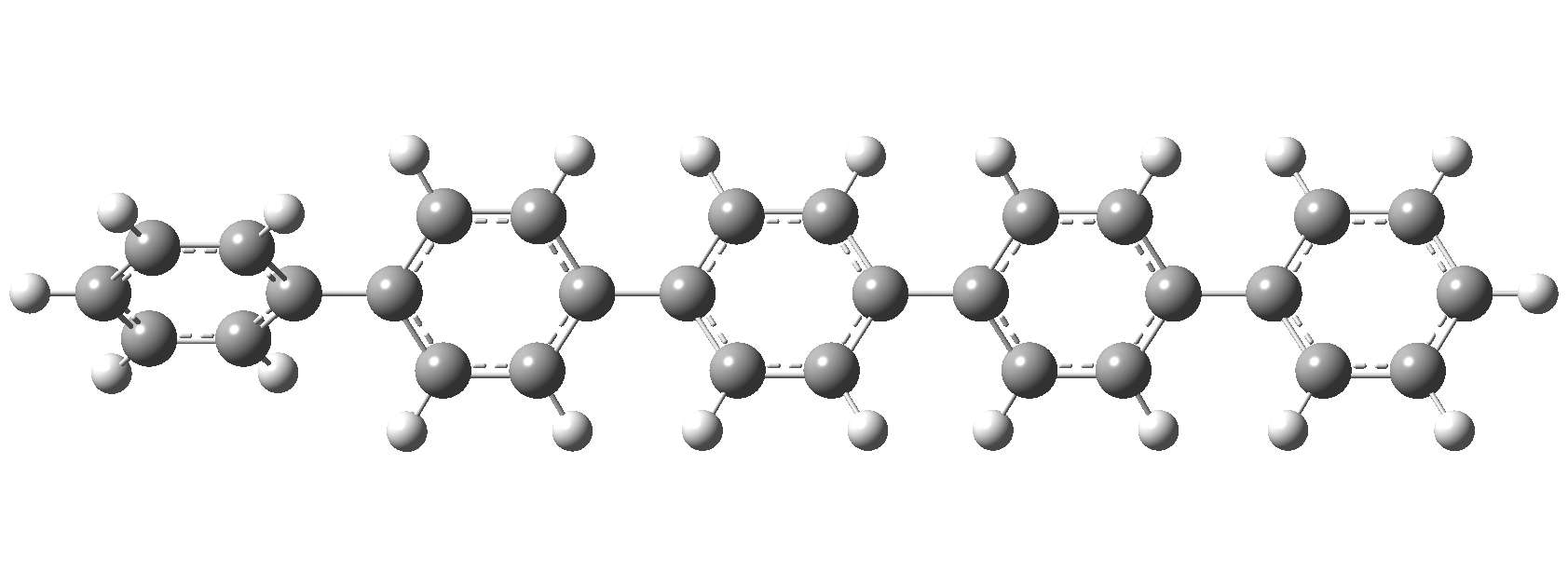} \caption{}
	\end{subfigure}
       	\begin{subfigure}[b]{0.5\textwidth}
         \includegraphics[height=0.11\textheight]{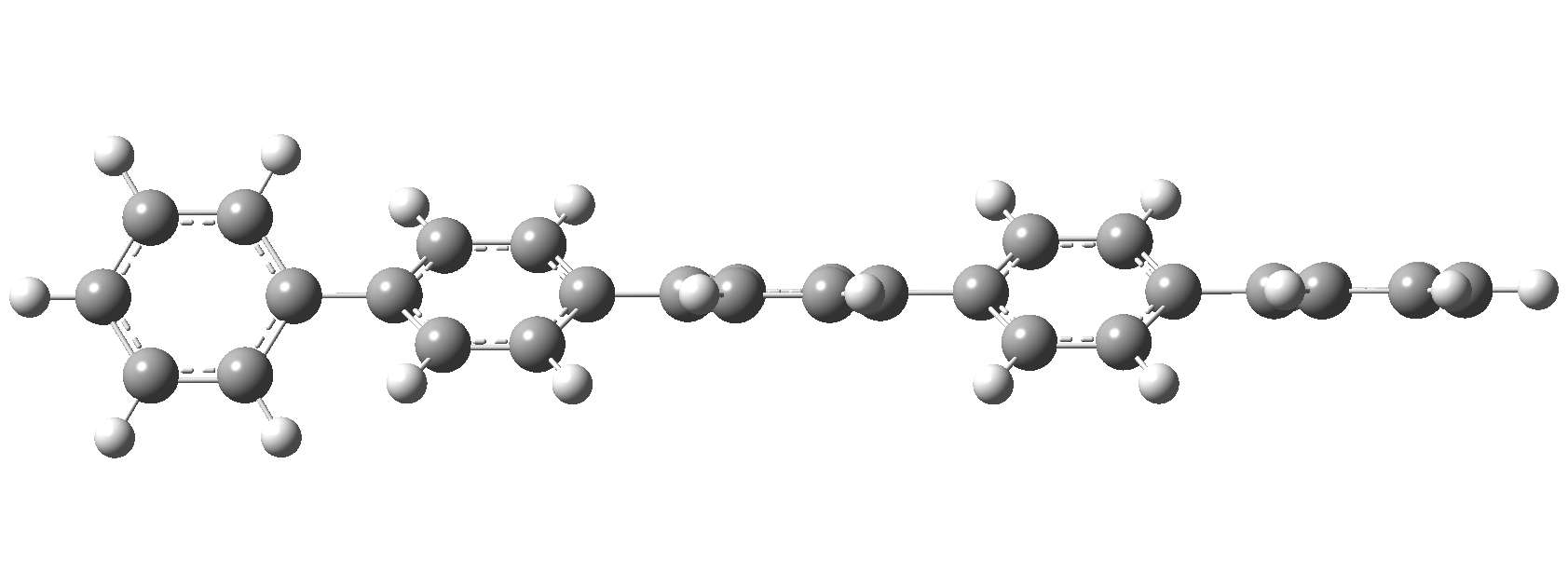}  \caption{}
         \end{subfigure}
           \caption{Se muestra la configuracin de mnima energa para el trifenileno  desde  ($a$) una vista frontal  y  ($b$) una vista lateral y el pentafenileno desde dos ngulos diferentes y el pentafenileno ($c$) y ($d$).}
          \label{polyph}\
 \end{figure}

La palabra para en el nombre poly-para-fenilenos (en ingls poli-para-phenylenes), hace referencia a la estructura, lineal, es decir, que no tiene bifurcaciones al unir los fenilenos.Los poli-para-fenilenos tienen la caracterstica, a diferencia de los poliacenos, que los anillos de benceno no se encuentran en el mismo plano como se ilustra en la figura \ref{polyph} los casos del trifenileno y pentafenileno.\\

\begin{figure}[htb!]
        \begin{subfigure}[b]{0.5\textwidth}
     \hspace*{0.1in}   \includegraphics[height=0.1\textheight]{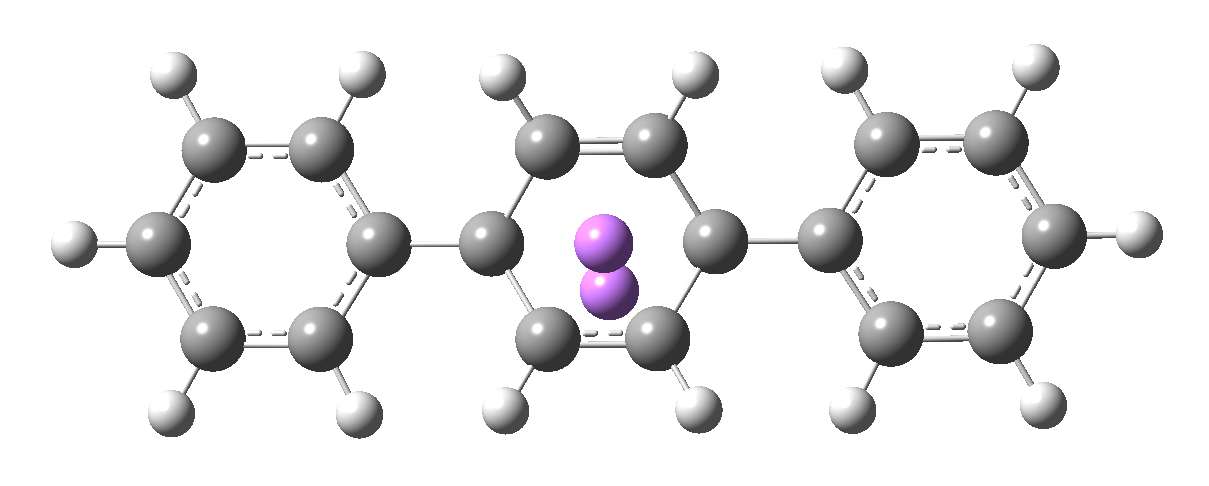} 
	\end{subfigure}
       	\begin{subfigure}[b]{0.5\textwidth}
         \includegraphics[height=0.1\textheight]{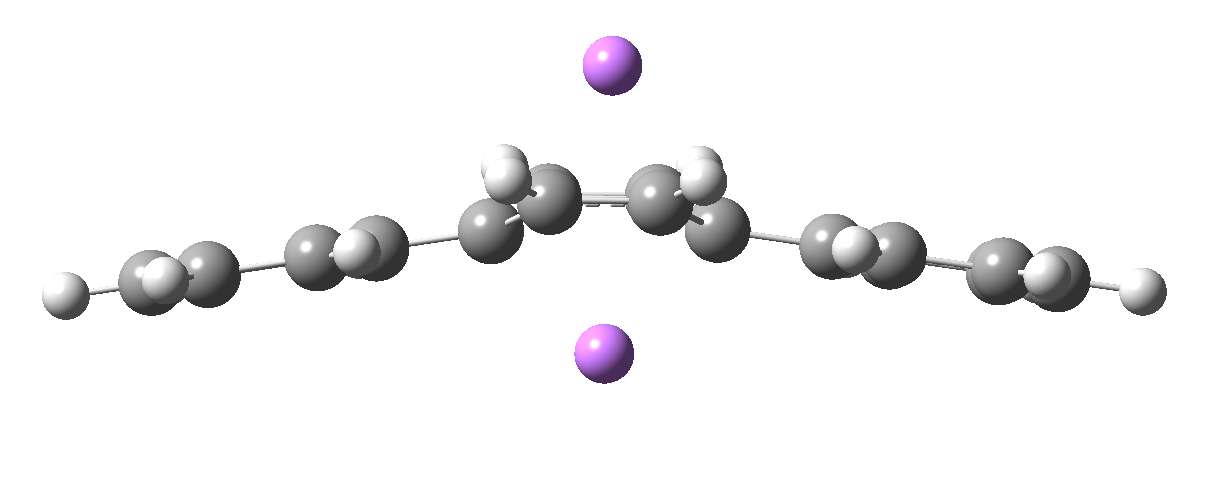}  
         \end{subfigure}
          \caption{Se muestra la configuracin de mnima energa para el trifenileno con adsorcin de dos Litios en el segundo anillo.}
          \label{triphenyl}
 \end{figure}

\begin{figure}[htb!]
        \begin{subfigure}[b]{0.5\textwidth}
     \includegraphics[height=0.12\textheight]{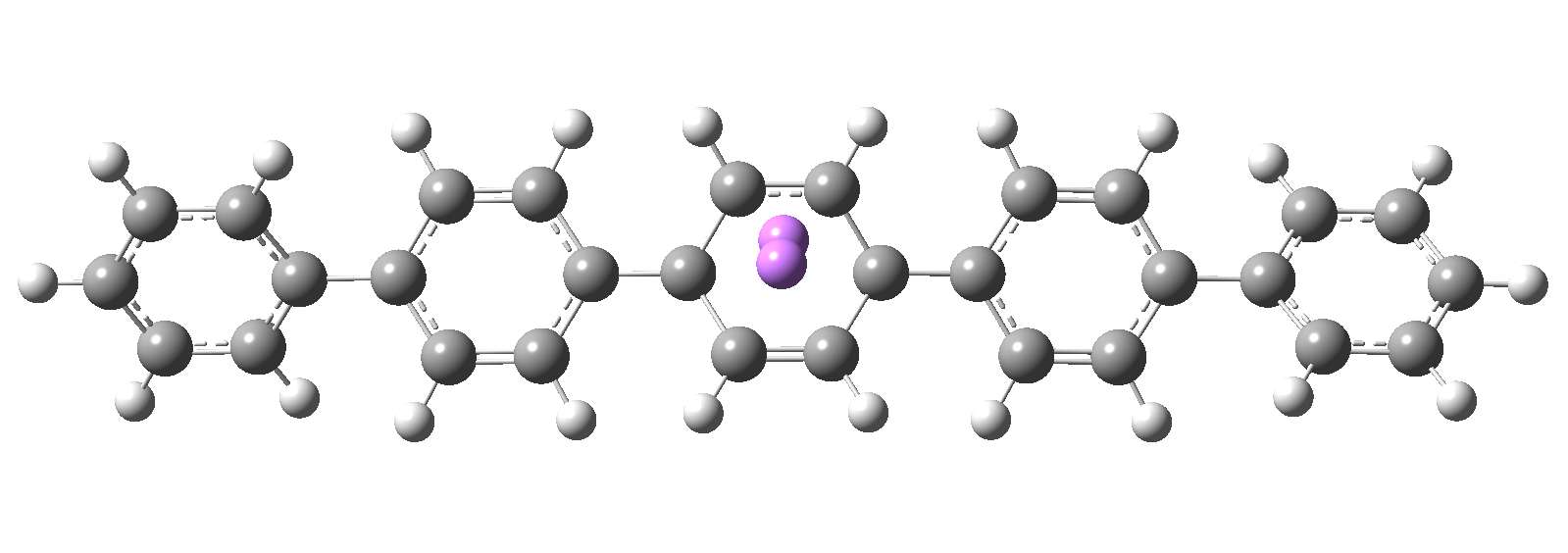} 
	\end{subfigure}
       	\begin{subfigure}[b]{0.5\textwidth}
         \includegraphics[height=0.12\textheight]{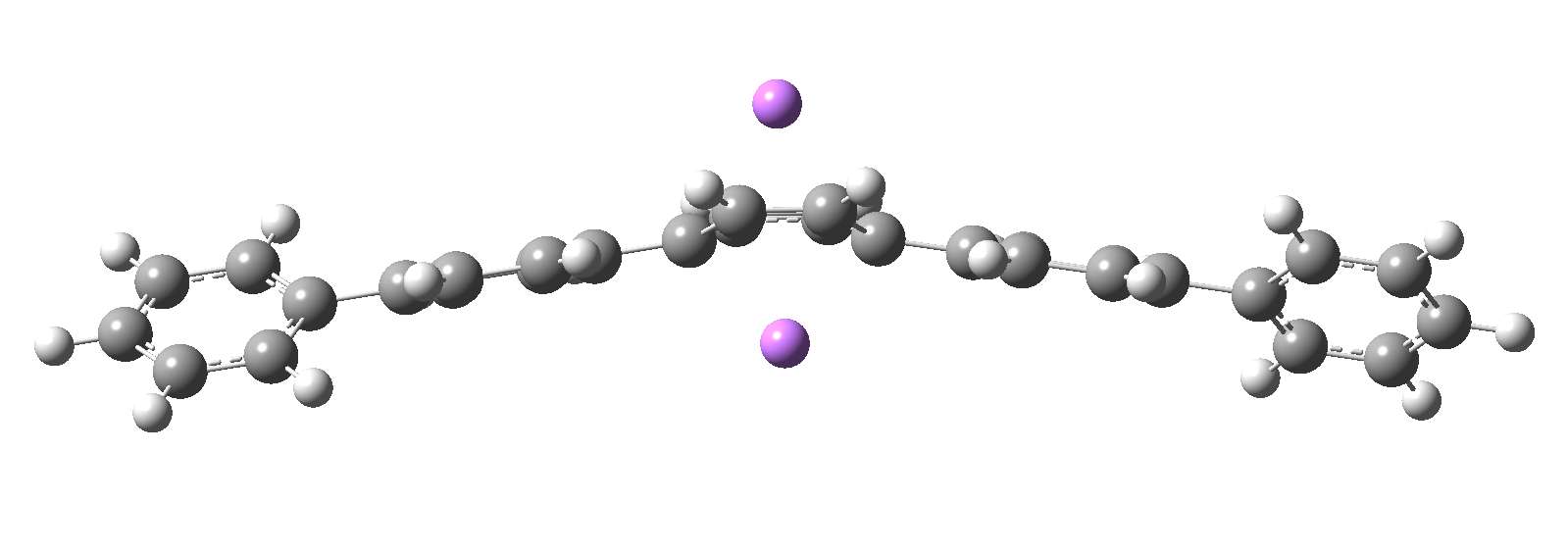}  
         \end{subfigure}
          \caption{Se muestra la configuracin de mnima energa para el pentafenileno con adsorcin de dos Litios en el tercer anillo.}
          \label{pentaphenyl}
 \end{figure}

Se realizaron clculos para la adsorcin de un par de Litios en el trifenileno y el pentafenileno, como se muestran sus configuraciones de mnima energa en las grficas \ref{triphenyl} y \ref{pentaphenyl} respectivamente. Lo primero que notamos en ambos casos es que los anillos mas cercanos al anillo donde se adsorbieron los dos Litios se alinean, es decir estos anillos rotaron.  Tambin presentan deformacin al adsorber el par de Litios

Tambin se hicieron clculos para la adsorcin de ms pares de Litios en cuyos casos no encontramos ningn comportamiento regular parecido al de los poliacenos.

\subsection{Tiras de Grafeno}

\begin{figure}[htb!]
   \hspace*{0.3in}  \includegraphics[height=0.5\textheight]{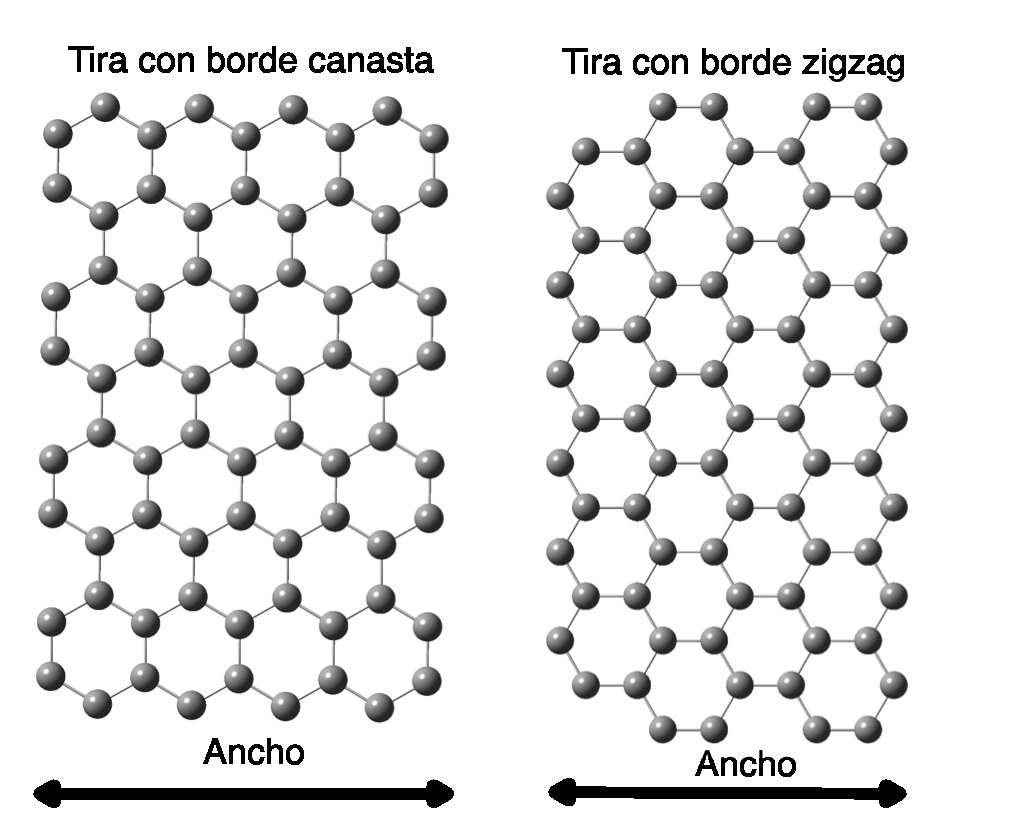} 
\caption{Estructura de tiras de grafeno con borde canasta y borde zigzag.}
          \label{ancho}
 \end{figure}

Las tiras de grafeno (tambin llamadas nano-cintas de grafeno o cintas nano-grafito, GNRs por sus siglas en ingls, \textit{Graphene Nano Ribbons}), son tiras de anillos de benceno con un ancho ultra-delgado ($< 20 nm$) como se ilustra en la figura \ref{ancho} . Las tiras  de grafeno se introdujeron originalmente como un modelo terico de Mitsutaka Fujita y co-autores para examinar efectos  de borde y de tamao nanomtrico en el grafito \cite{tiras1}-\cite{tiras3}.\\

\begin{figure}[htb!]
  \begin{subfigure}[b]{0.5\textwidth}
       \hspace*{0.3in} \includegraphics[height=0.22\textheight]{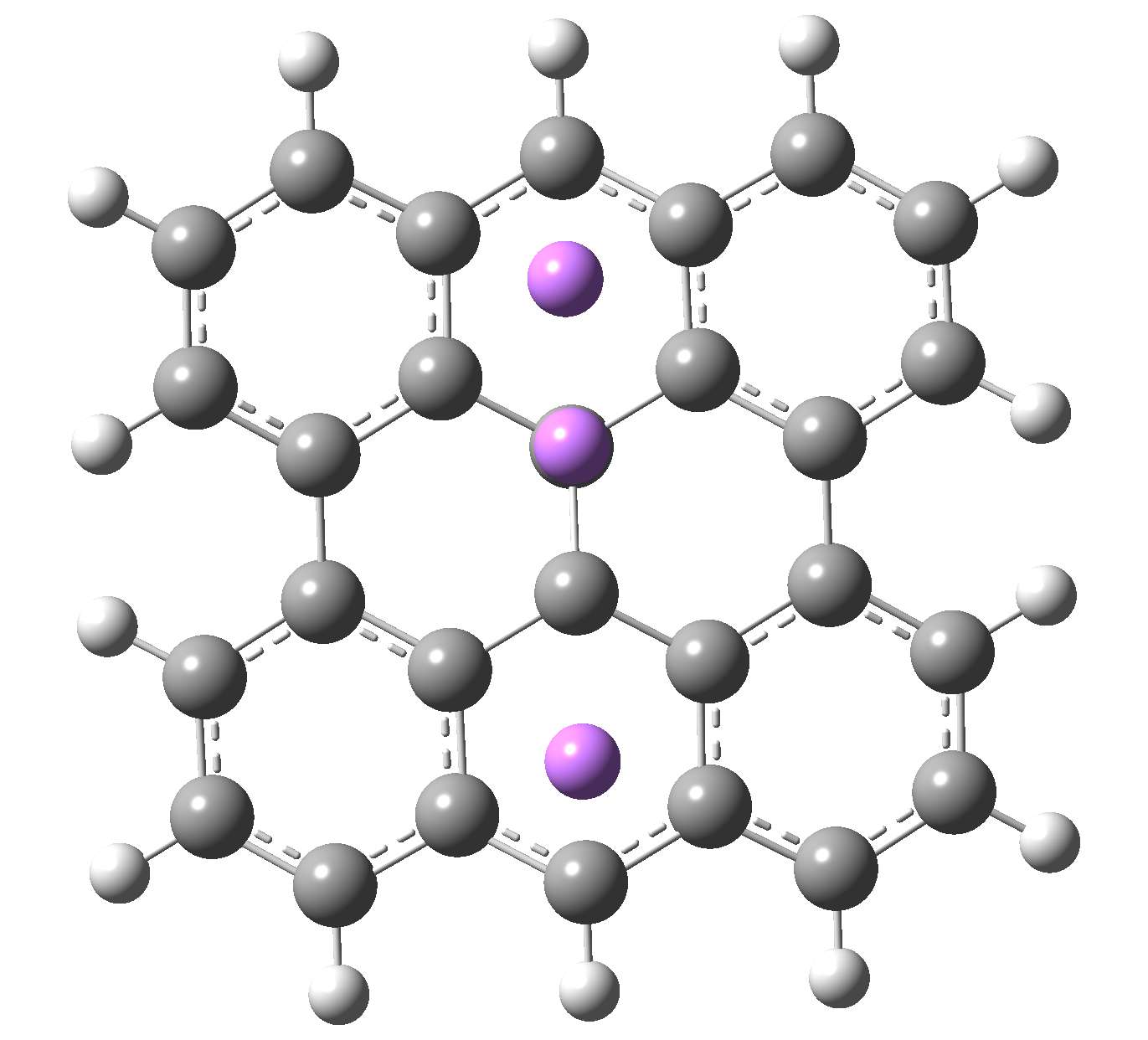}
        \caption{}
\end{subfigure}
	\begin{subfigure}[b]{0.5\textwidth}
       \hspace*{0.3in}\includegraphics[height=0.22\textheight]{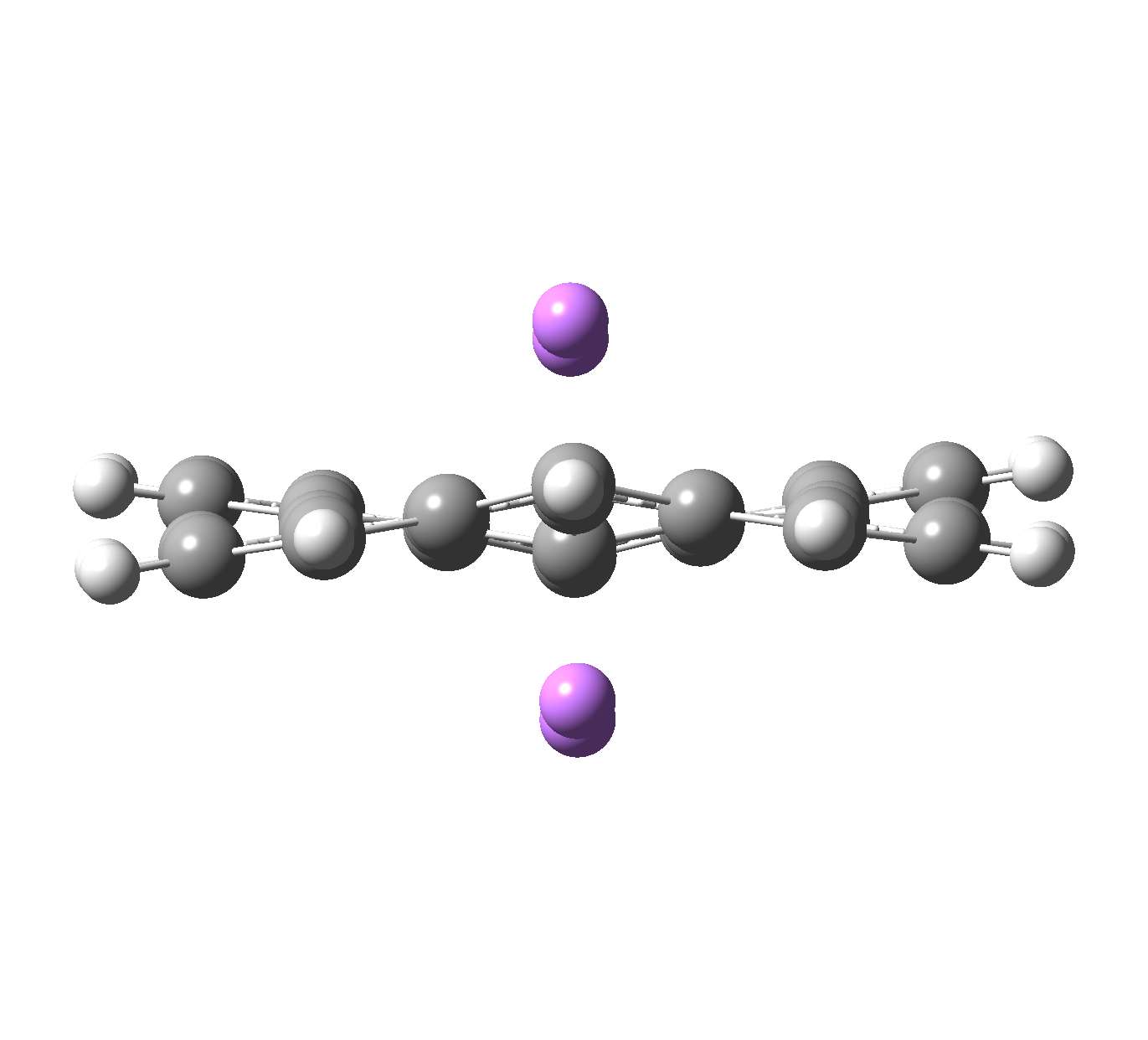}   
        \caption{}
          \end{subfigure}
          \caption{Adsorcin de dos pares de Litios en una tira de grafeno de 8 anillos ($a$) vista frontal y  ($b$) vista lateral de la configuracin optimizada usando el mtodo de DFT con el funcional B3LYP y el conjunto de bases 6-311g*}
          \label{8ring}
 \end{figure}

En este trabajo se extendieron los clculos a tiras de grafeno manteniendo la simetra translacional. En este caso usamos el mtodo de DFT con el funcional B3LYP y el conjunto de bases 6-311gd*. 
Empezamos con una hojuela de ocho anillos como se muestra en la figura \ref{8ring} en la cual se adsorben dos pares de Litios con los procedimientos anteriormente explicados, es decir explorando diferentes condiciones iniciales y distintos pasos en la adsorcin de Litios. En este caso obtuvimos que la configuracin de mnima energa muestra la deformacin obtenida en los poliacenos, en este caso, el del antraceno. Como un ejercicio de abstraccin, la vista frontal en la figura \ref{8ring}($a$) puede ser vista como la unin de dos antracenos, de acuerdo a esto vemos en la vista lateral  \ref{8ring}($b$) que el antraceno que est al frente se dobla  hacia abajo, mientras que el antraceno que est en el fondo se dobla hacia arriba. El ngulo de deformacin en los dos antracenos en sta estructura es de 158¼, una deformacin del mismo orden de magnitud a la obtenida para el caso del antraceno aislado que es de 150¼. Por otro lado no todos los Litios se adsorbieron en la mitad del anillo correspondiente, un par de ellos se adsorbe sobre un carbono como se ve en la figura \ref{8ring}($a$).\\

\begin{figure}[htb!]
  \begin{subfigure}[b]{0.5\textwidth}
        \hspace*{0.3in}\includegraphics[height=0.22\textheight]{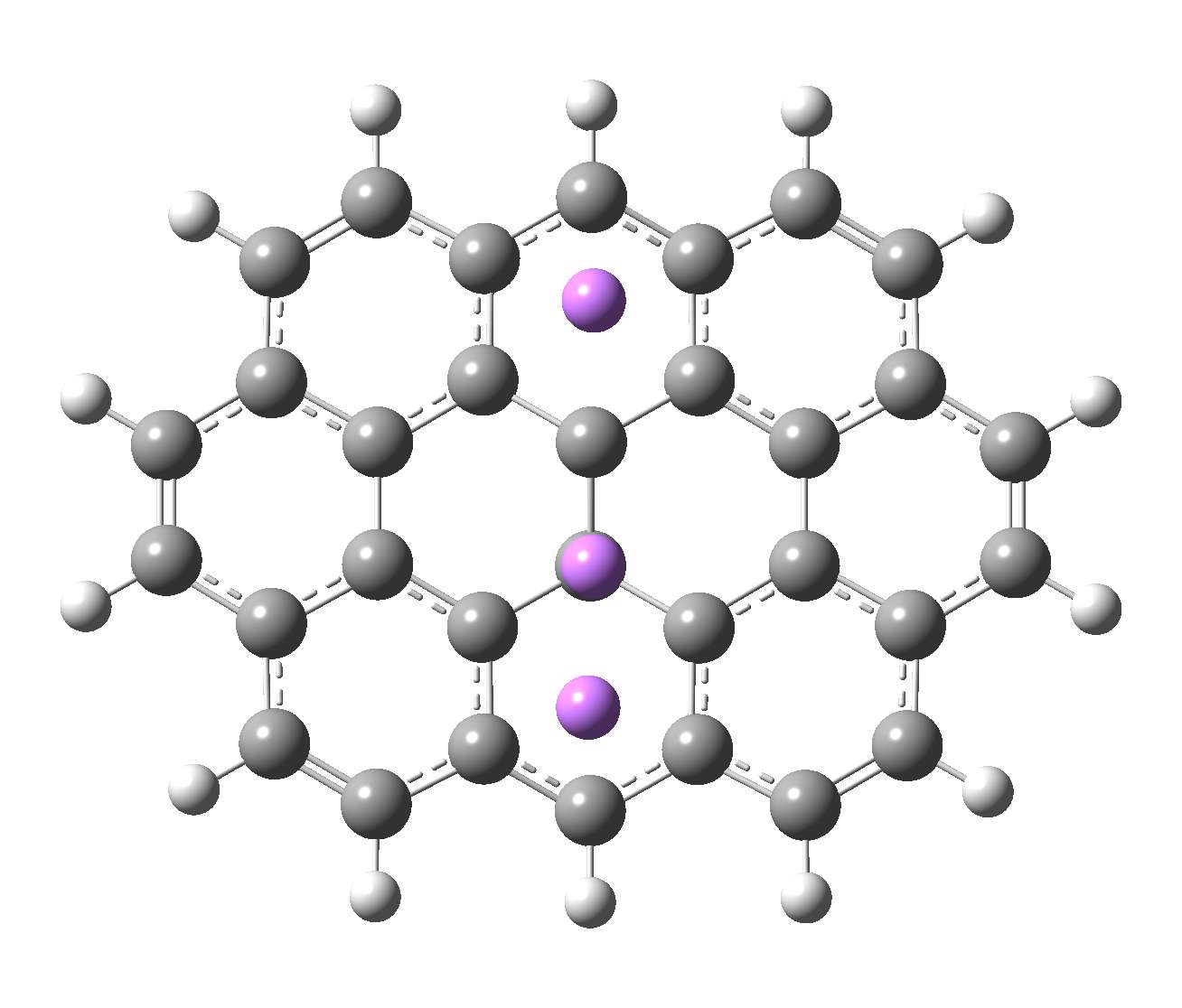}
        \caption{}
\end{subfigure}
	\begin{subfigure}[b]{0.5\textwidth}
      \hspace*{0.3in} \includegraphics[height=0.22\textheight]{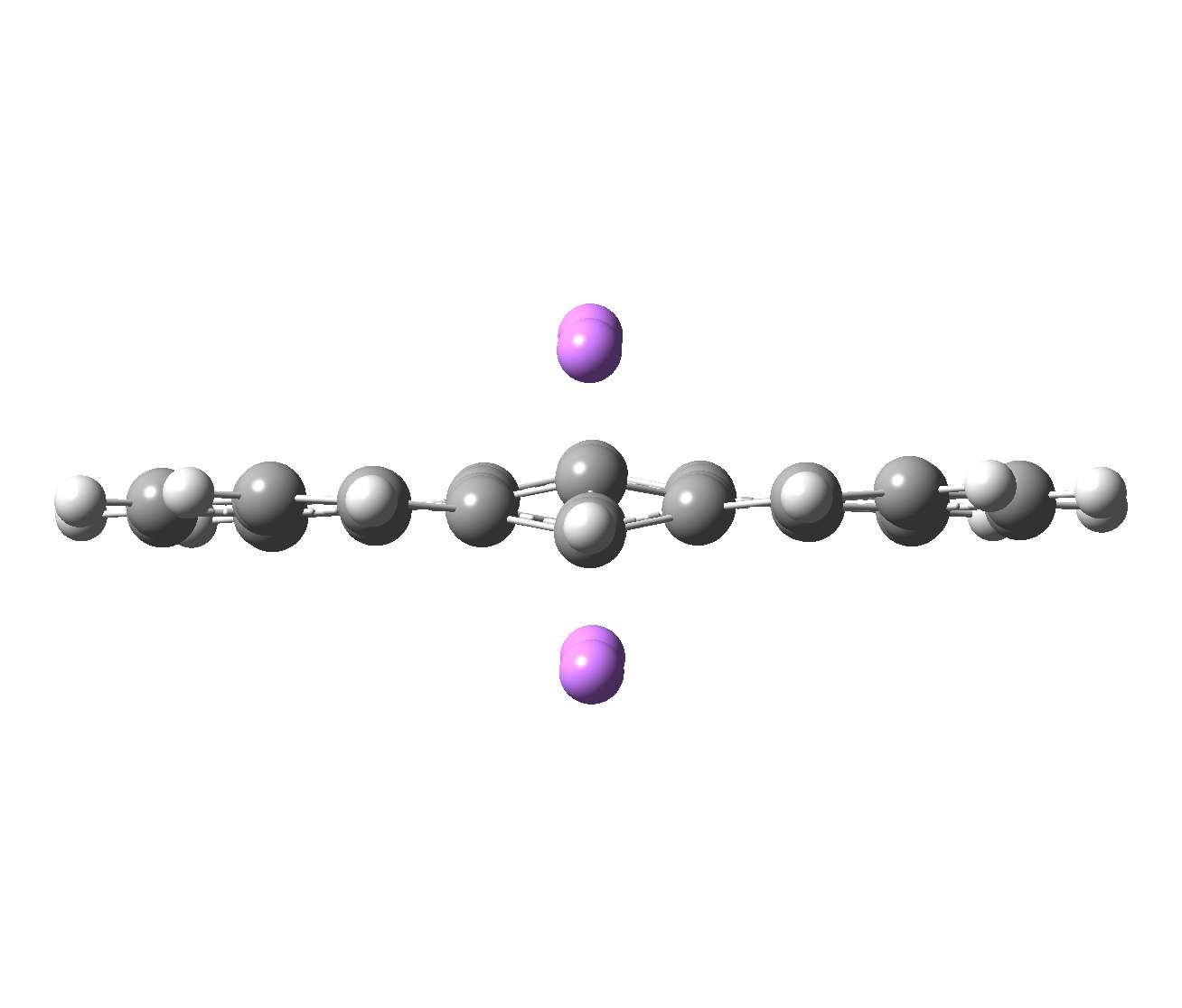}   
        \caption{}
          \end{subfigure}
          \caption{Adsorcin de dos pares de Litios en una tira de grafeno de 10 anillos ($a$) vista frontal y  ($b$) vista lateral de la configuracin optimizada usando el mtodo de DFT con el funcional B3LYP y el conjunto de bases 6-311g*}
          \label{10ring}
 \end{figure}

Al aumentar el nmero de anillos de la molcula y conservando la simetra se optimiz la tira de 10 anillos nuevamente con dos pares de Litios como se muestra en la figura \ref{10ring} usando el mtodo anteriormente mencionado se obtiene un comportamiento similar al descrito anteriormente con la deformacin de los antracenos, slo que en ste caso la deformacin se reduce nuevamente, es decir, el ngulo de deformacin para cada antraceno es de 163¼.\\
 
\begin{figure}[htb!]
  \begin{subfigure}[b]{0.5\textwidth}
     \hspace*{0.3in}   \includegraphics[height=0.33\textheight]{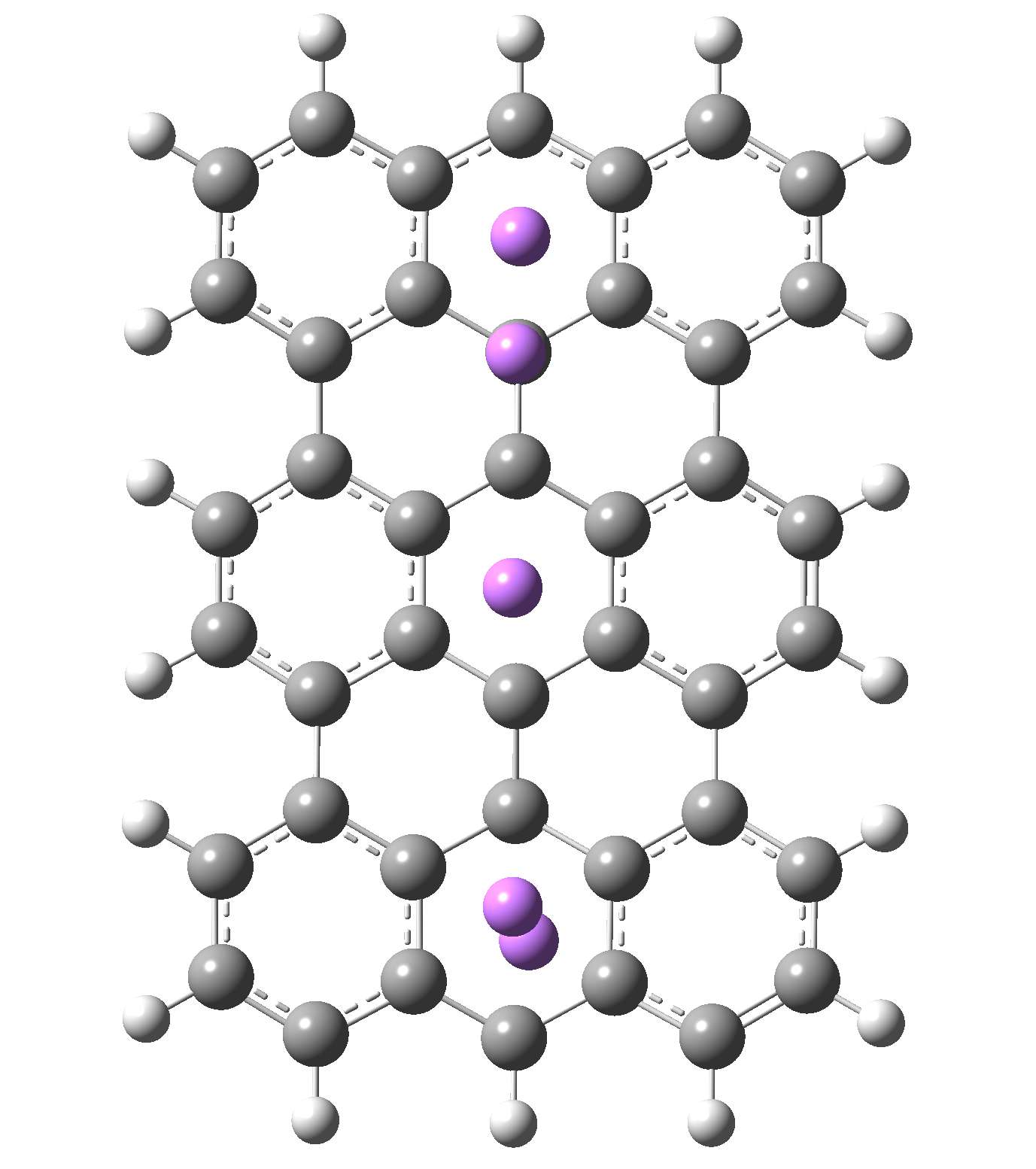}
        \caption{}
\end{subfigure}
	\begin{subfigure}[b]{0.5\textwidth}
       \hspace*{0.3in}\includegraphics[height=0.33\textheight]{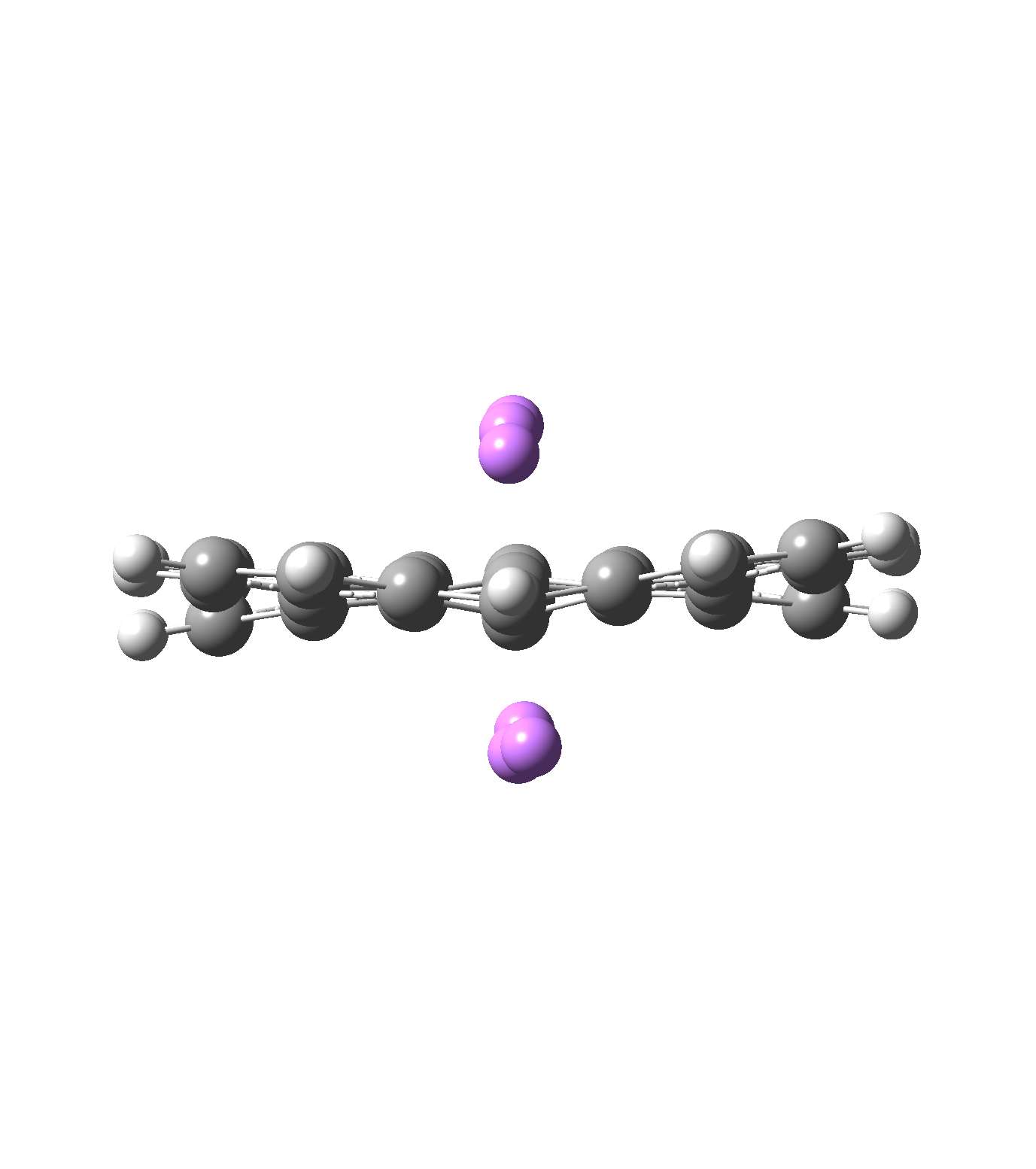}   
        \caption{}
          \end{subfigure}
          \caption{Adsorcin de tres pares de Litios en una tira de grafeno de 13 anillos ($a$) vista frontal y  ($b$) vista lateral de la configuracin optimizada usando el mtodo de DFT con el funcional B3LYP y el conjunto de bases 6-311g*.}
          \label{13ring}
 \end{figure}

El siguiente paso fue extender la molcula  en direccin vertical, conservando la simetra, y adsorber tres pares de Litios como se ve en la figura \ref{13ring}. Haciendo el mismo ejercicio de abstraccin que en el caso anterior se puede notar de la figura \ref{13ring}($a$) que la molcula esta compuesta por tres antracenos unidos, la optimizacin de dicha molcula con los tres pares de Litios dispuestos en el anillo central de cada antraceno nos da como resultado nuevamente la deformacin de los tres antracenos. Los antracenos externos tienen un ngulo de deformacin de 161¼ mientras que el del antraceno interno es de 171¼, es decir, que presenta una deformacin menor respecto a los antracenos externos. En cuanto a la direccin, se puede observar en la figura \ref{13ring}($b$) nuevamente que es alternante, es decir, los antracenos externos se doblan en una direccin y el interno en la direccin opuesta a stos. En este punto podemos concluir cualitativamente que la deformacin debido a la adsorcin de cierto nmero de Litios disminuye a medida que aumentamos el nmero de anillos.\\

\begin{figure}[htb!]
        \begin{subfigure}[b]{0.5\textwidth}
        \includegraphics[height=0.2\textheight]{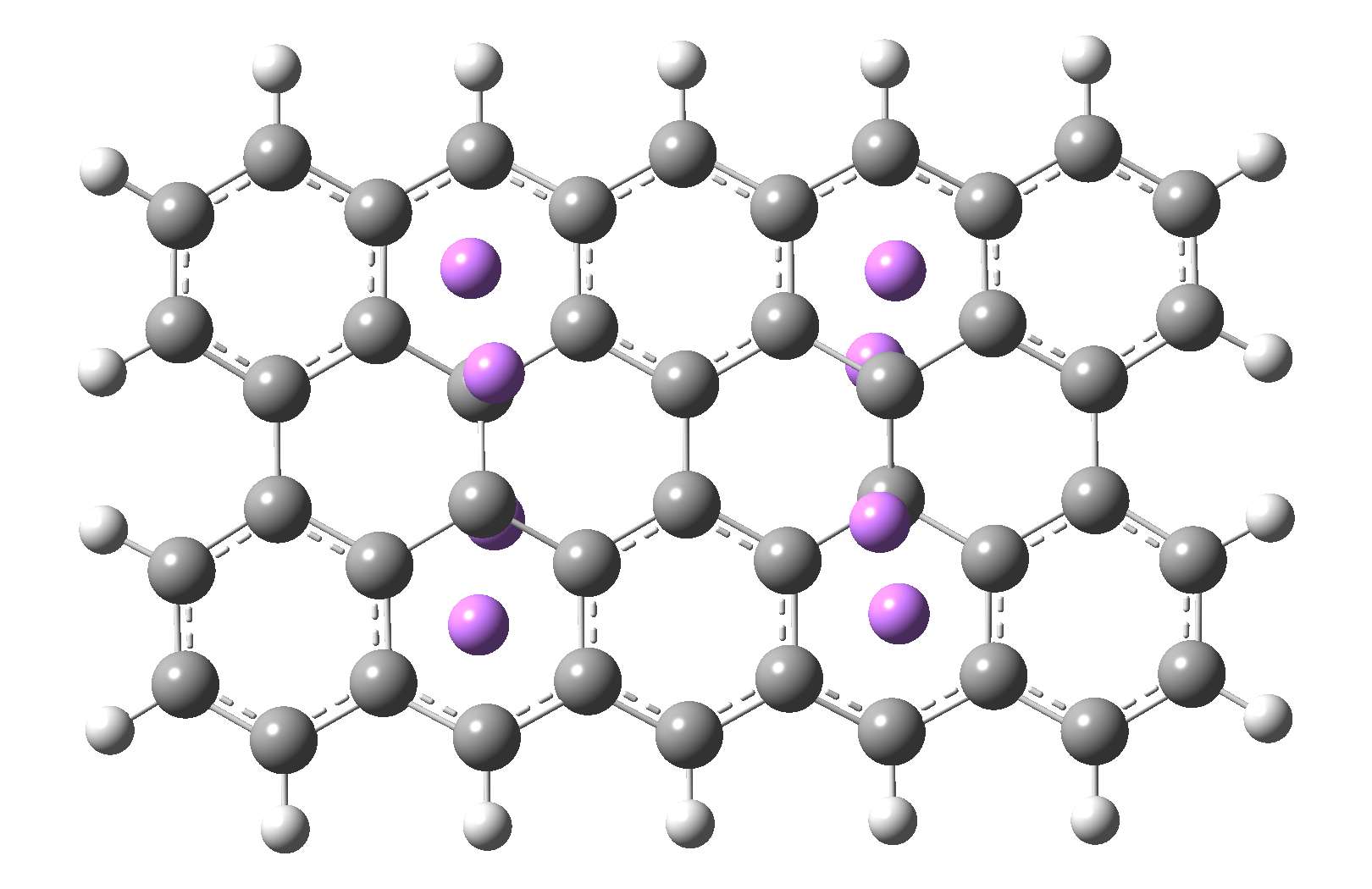}
            \caption{}
	\end{subfigure}
	\begin{subfigure}[b]{0.5\textwidth}
        \hspace*{0.3in}\includegraphics[height=0.2\textheight]{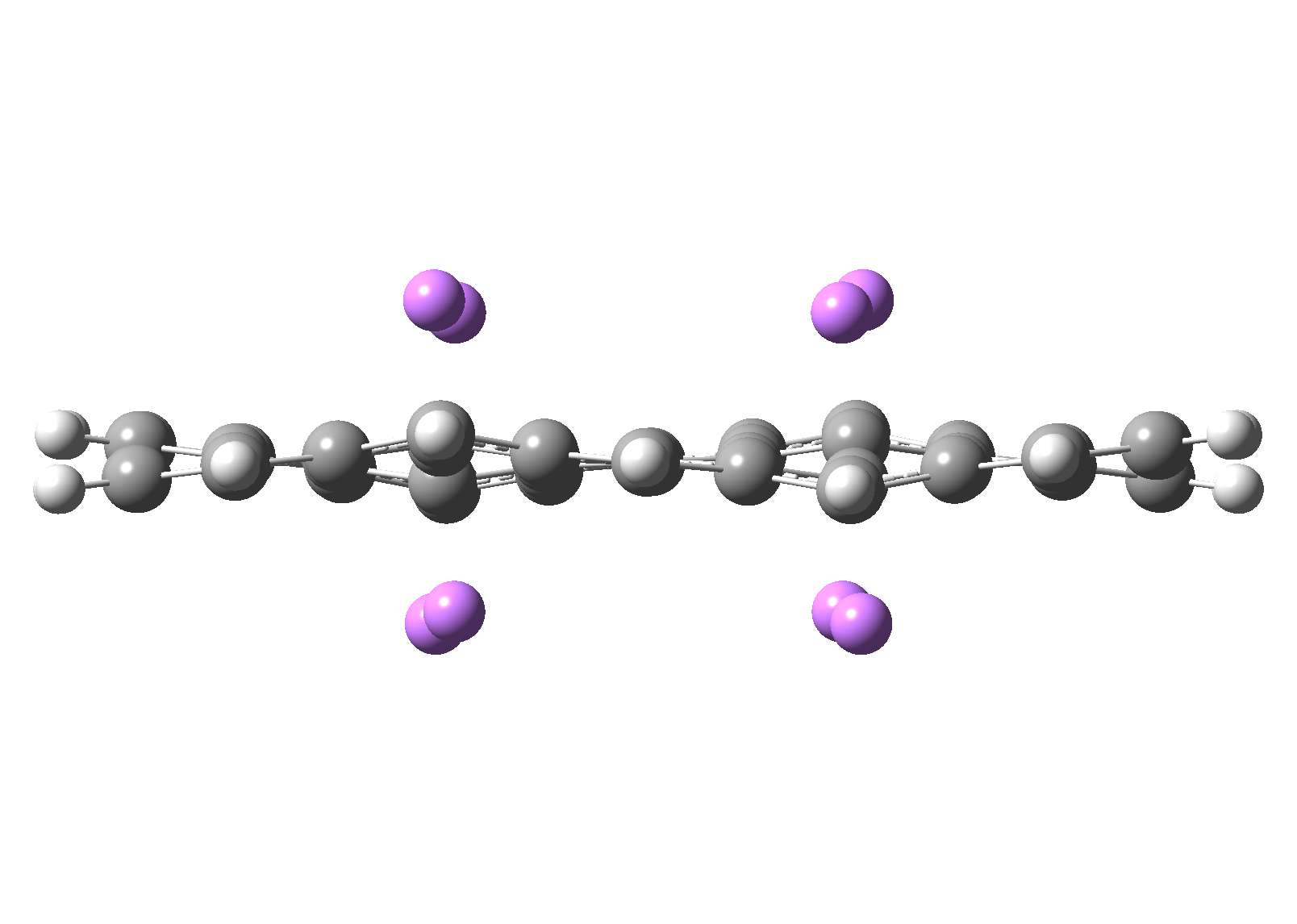}   
            \caption{}
         \end{subfigure}
\caption{Configuracin de un mnimo local de la energa de la adsorcin de 4 pares de Litios en una tira de grafeno de 14 anillos cuya estructura corresponde al zigzag. ($a$) Vista frontal y  ($b$) vista lateral de la configuracin optimizada usando el mtodo de DFT con el funcional B3LYP y el conjunto de bases 6-311g*.}
          \label{14ring}
 \end{figure}

\begin{figure}[htb!]
        \begin{subfigure}[b]{0.5\textwidth}
        \hspace*{-0.3in} \includegraphics[height=0.2\textheight]{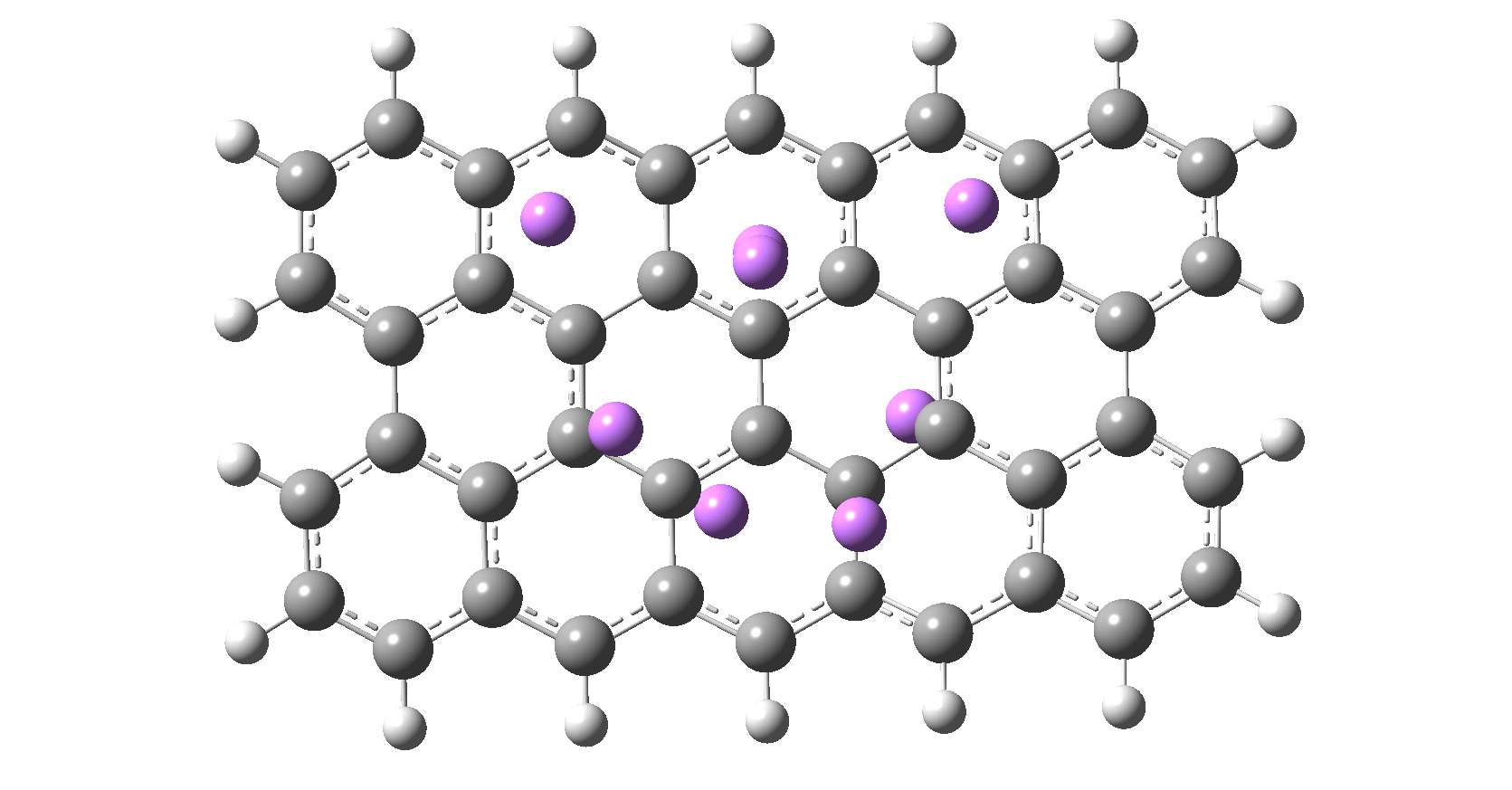}   
      \caption{}
	\end{subfigure}
	\begin{subfigure}[b]{0.5\textwidth}
        \hspace*{-0.3in}  \includegraphics[height=0.2\textheight]{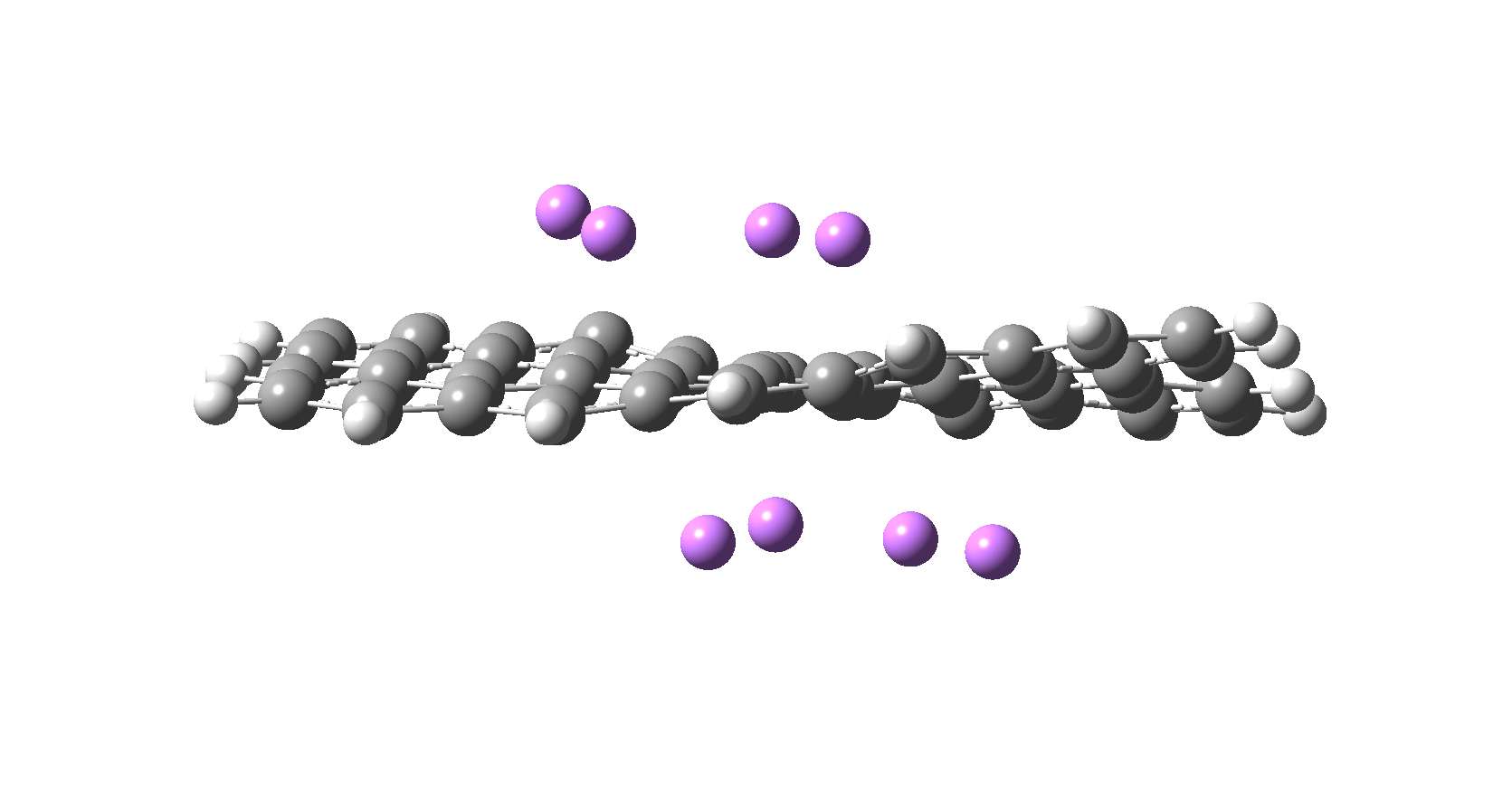}   
           \caption{}
         \end{subfigure}
\caption{Configuracin de un mnimo de energa de la adsorcin de 4 pares de Litios en una tira de grafeno de 14 anillos, cuya energa es ms baja que para el caso zigzag. ($a$) Vista frontal y  ($b$) vista lateral de la configuracin optimizada usando el mtodo de DFT con el funcional B3LYP y el conjunto de bases 6-311g*.}
          \label{14ring_min}
 \end{figure}

De la misma forma extendemos en direccin horizontal nuestra molcula conservando la simetra translacional en dicha direccin, trabajando as con una tira de 14 anillos y adsorbiendo 4 pares de Litios como se ve en la figura \ref{14ring}. Abstrayendo de la figura \ref{14ring}($a$), podemos decir que la molcula est compuesta de dos pentacenos unidos cada uno con la adsorcin de dos pares de Litios. Obtuvimos un mnimo local de energa, cuya energa de adsorcin es de -198 kcal/mol. Dicha configuracin se muestra en la figura  \ref{14ring}. En este caso se reproduce el comportamiento de la adsorcin de dos pares alternados de Litios en un pentaceno que da como resultado el zigzag. La orientacin del zigzag de cada uno de los pentacenos es opuesta respecto al otro como se vio en el antraceno.  En la figura \ref{14ring_min} se muestra la configuracin de un mnimo de energa ms bajo que el caso anterior cuya energa de adsorcin es -219.1kcal/mol. En la figura se observa una torsin de la molcula aromtica y por la ubicacin de los Litios sobre sta podemos decir que el fenmeno de la distorsin de Peierls no aparece en estructuras ms grandes que las consideradas anteriormente y esto es fcil de entender pues sistemas ms grandes no pueden ser pensados como cuasi unidimensionales, como s pudimos modelarlo en el caso de los poliacenos. \\

\begin{figure}[htb!]
          \begin{subfigure}[b]{0.5\textwidth}
       \includegraphics[height=0.2\textheight]{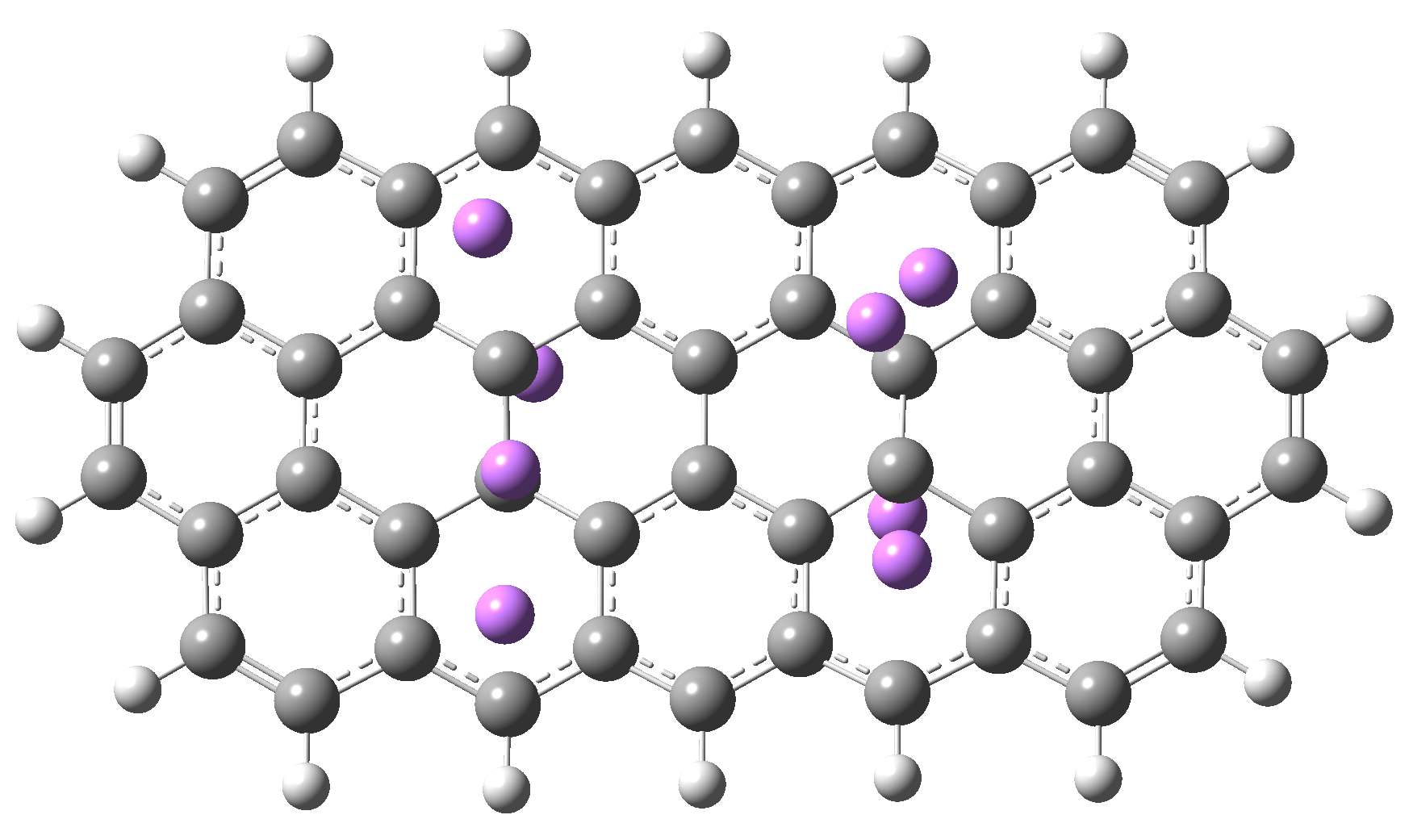}   
         \caption{}
	\end{subfigure}
	\begin{subfigure}[b]{0.5\textwidth}
        \includegraphics[height=0.2\textheight]{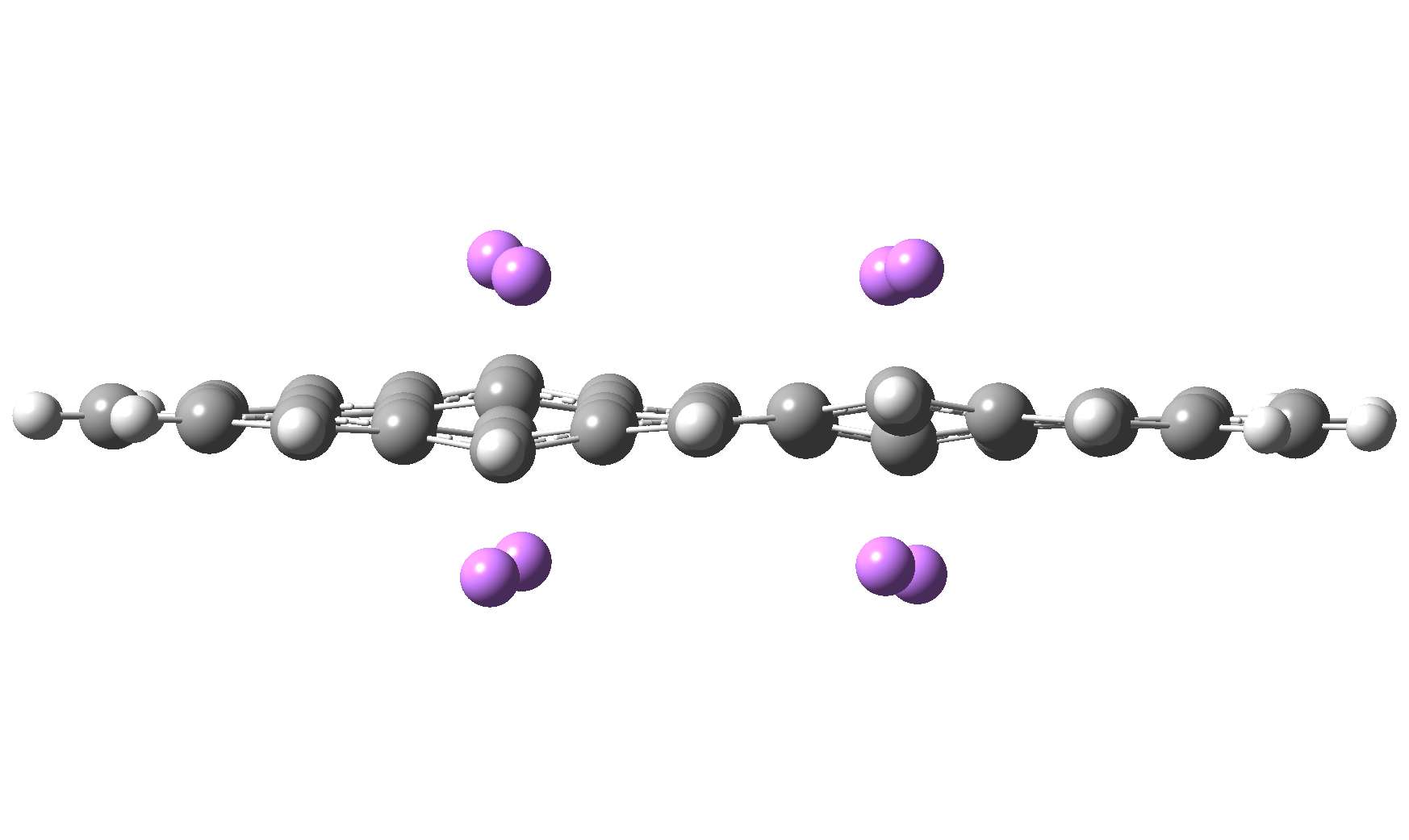}  
              \caption{}
         \end{subfigure}
         \caption{Configuracin de un mnimo local de energa de la adsorcin de 4 pares de Litios en una tira de grafeno de 16 anillos cuya estructura corresponde al zigzag. ($a$) Vista frontal y  ($b$) vista lateral de la configuracin optimizada usando el mtodo de DFT con el funcional B3LYP y el conjunto de bases 6-311g*.}
          \label{16ring}
 \end{figure}

\begin{figure}[htb!]
          \begin{subfigure}[b]{0.5\textwidth}
     \hspace*{-0.3in}  \includegraphics[height=0.19\textheight]{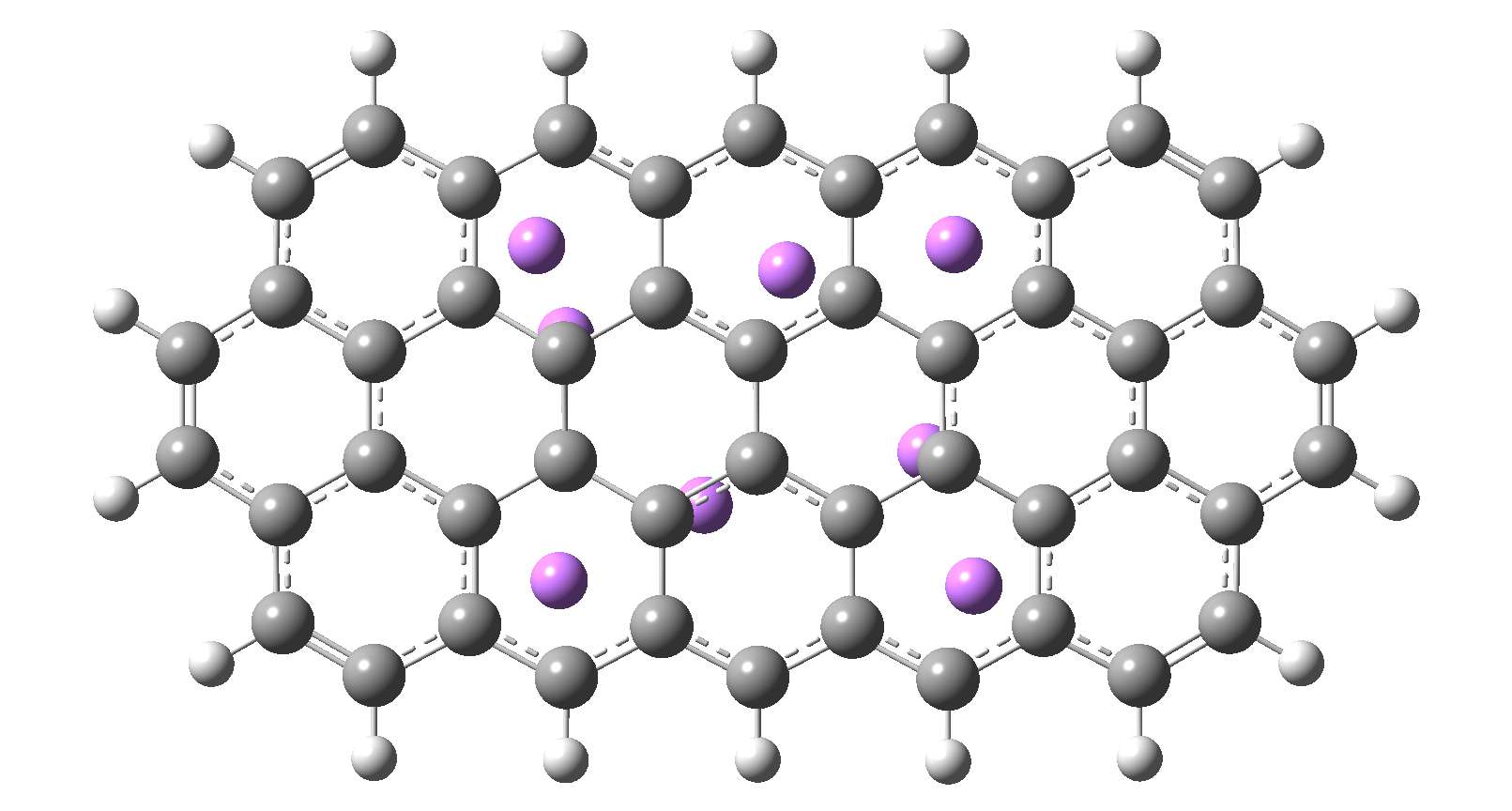}   
         \caption{}
	\end{subfigure}
	\begin{subfigure}[b]{0.5\textwidth}
       \hspace*{-0.2in} \includegraphics[height=0.19\textheight]{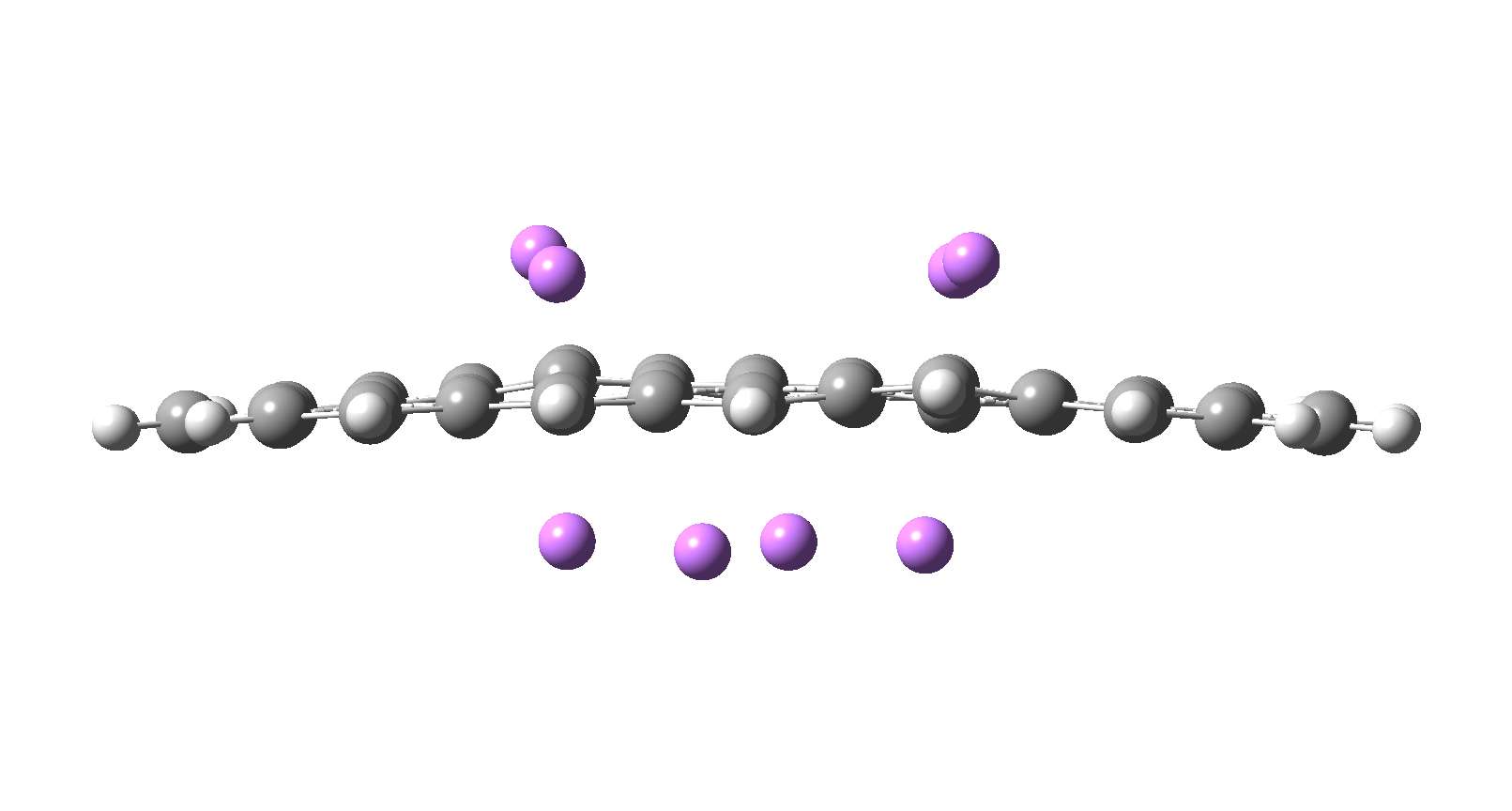}  
              \caption{}
         \end{subfigure}
         \caption{Configuracin de un mnimo de energa de la adsorcin de 4 pares de Litios en una tira de grafeno de 16 anillos, cuya energa es ms baja que para el caso zigzag.  ($a$) Vista frontal y  ($b$) vista lateral de la configuracin optimizada usando el mtodo de DFT con el funcional B3LYP y el conjunto de bases 6-311g*.}
          \label{16ring_min}
 \end{figure}

Adicionamos dos anillos a la molcula aromtica anterior y obtenemos nuevamente que la configuracin con el zigzag, como se ve en la figura \ref{16ring}, no corresponde al minimo absoluto de la energa. El valor de la energa de adsorcin es de -168.6 kcal/mol. En comparacin, se muestra en la figura \ref{16ring_min} la configuracin de un mnimo de energa ms bajo que el del zigzag, cuyo valor de la energa de adsorcin es de -176.2 kcal/mol.\\
 
\begin{figure}[htb!]
        \begin{subfigure}[b]{0.5\textwidth}
       \includegraphics[height=0.16\textheight]{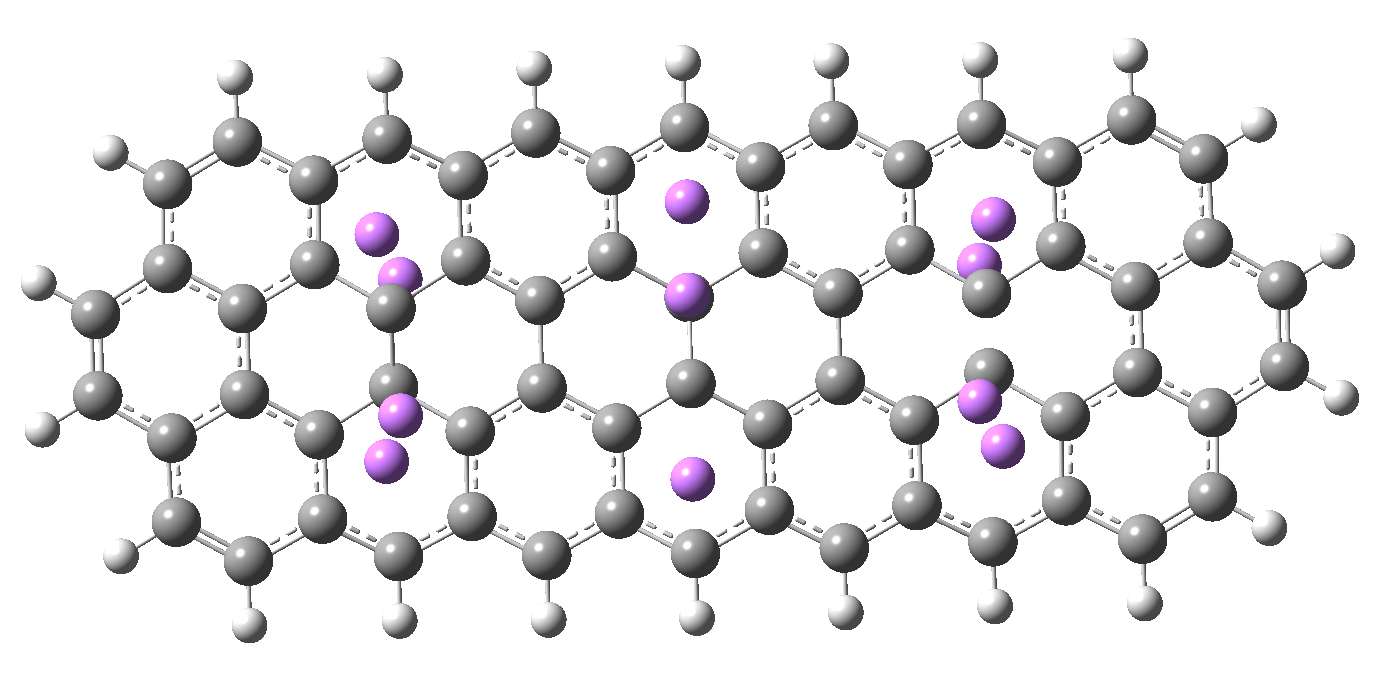}
        \caption{}
	\end{subfigure} 
	\begin{subfigure}[b]{0.5\textwidth}
       \hspace*{-0.1in} \includegraphics[height=0.16\textheight]{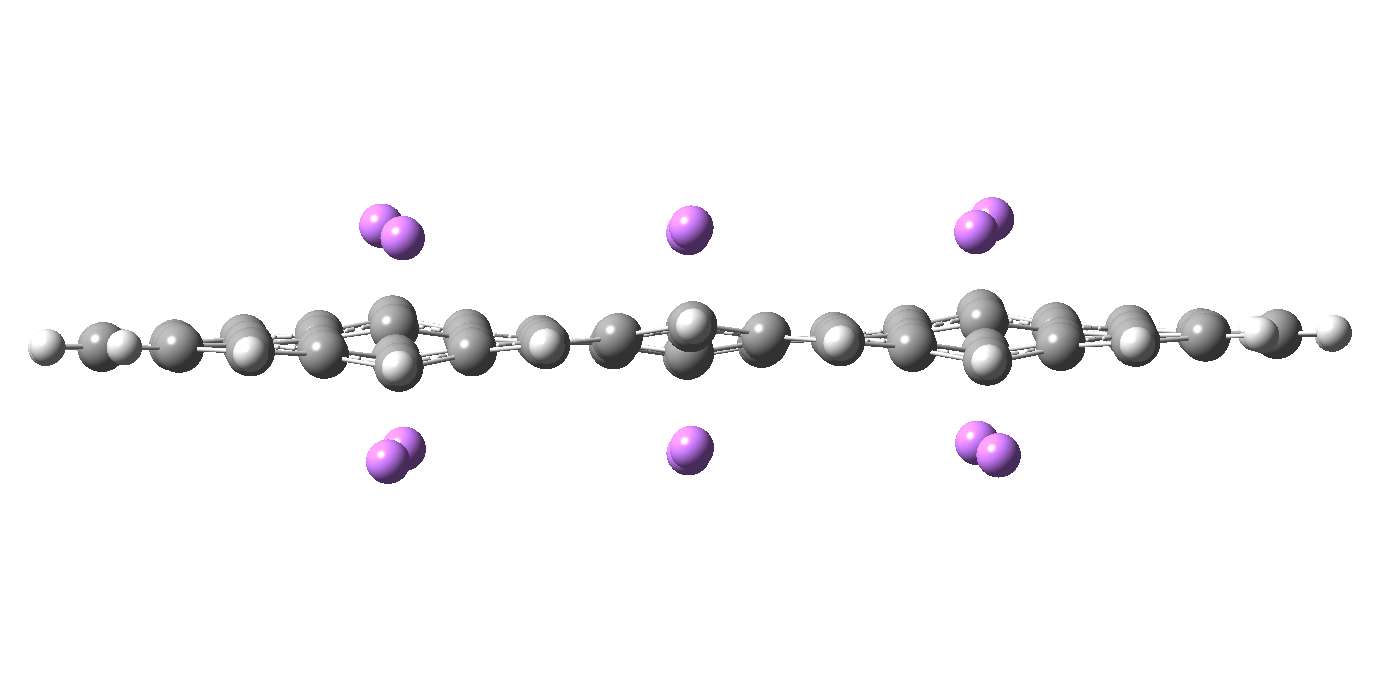}  
          \caption{}
         \end{subfigure}
         \caption{configuracin de un mnimo local de energa de la adsorcin de 6 pares de Litios en una tira de grafeno de 22 anillos cuya estructura corresponde al zigzag. ($a$) Vista frontal y  ($b$) vista lateral de la configuracin optimizada usando el mtodo de DFT con el funcional B3LYP y el conjunto de bases 6-311g*.}
          \label{22ring}
 \end{figure}

\begin{figure}[htb!]
        \begin{subfigure}[b]{0.5\textwidth}
       \hspace*{-0.33in}\includegraphics[height=0.165\textheight]{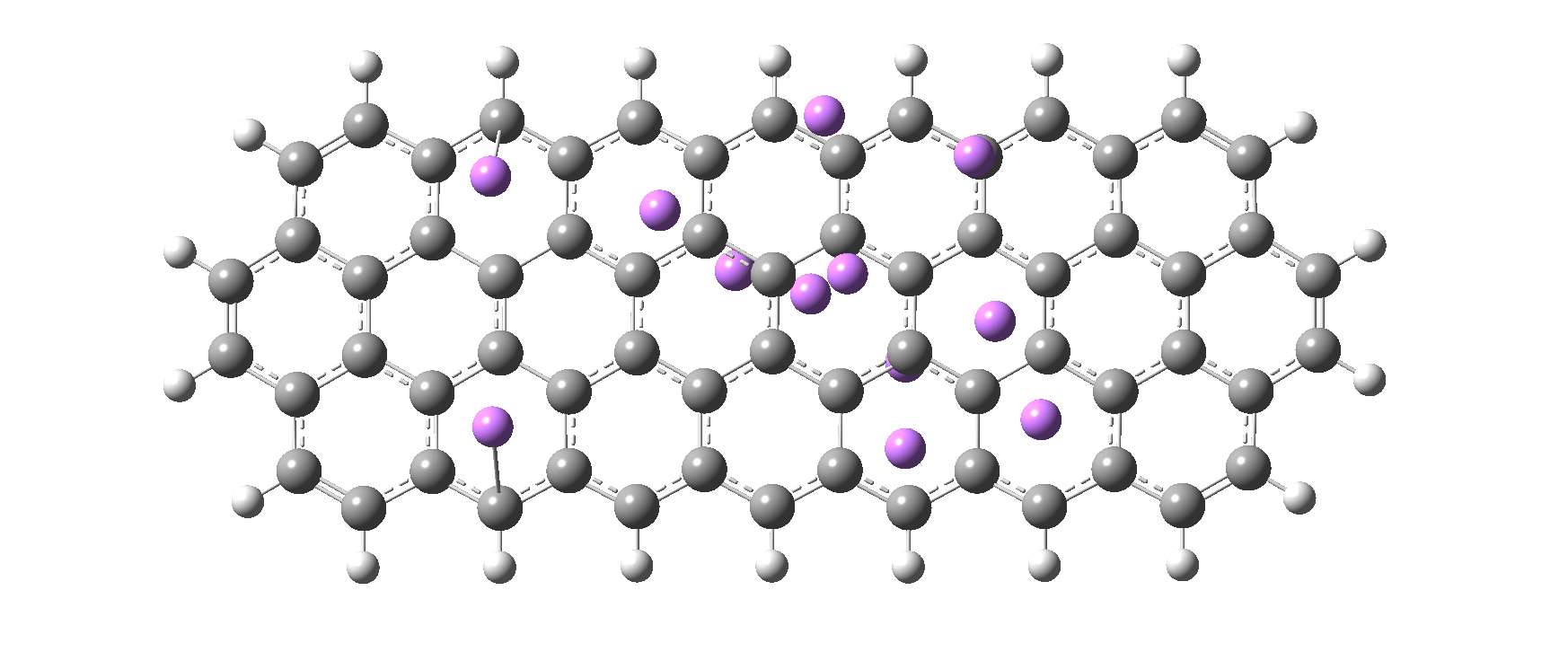}    
               \caption{}
	\end{subfigure}
	\begin{subfigure}[b]{0.5\textwidth}
        \hspace*{-0.25in} \includegraphics[height=0.165\textheight]{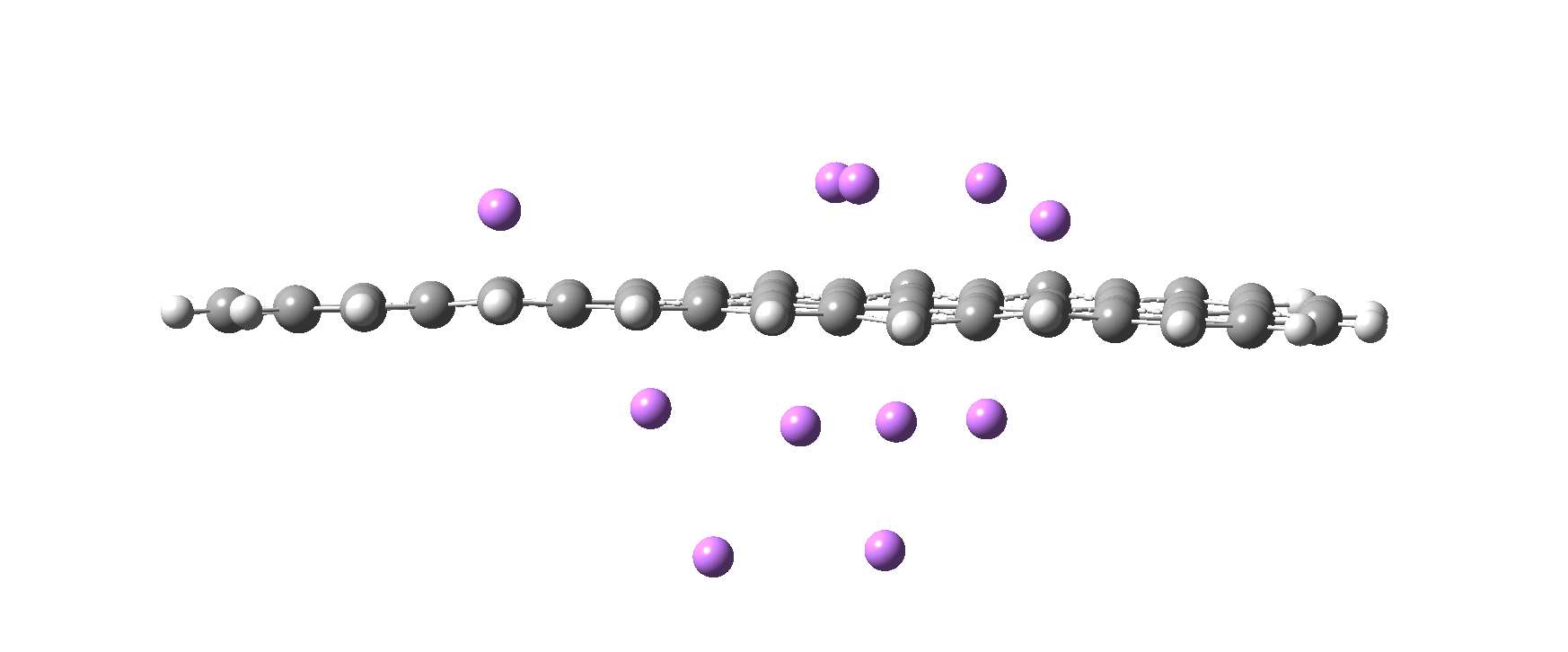} 
        \caption{}
         \end{subfigure}
         \caption{Configuracin de mnima energa de la adsorcin de 6 pares de Litios en una tira de grafeno de 22 anillos cuya energa es ms baja que para el caso zigzag. ($a$) Vista frontal y  ($b$) vista lateral de la configuracin optimizada usando el mtodo de DFT con el funcional B3LYP y el conjunto de bases 6-311g*.}
          \label{22ring_min}
 \end{figure}

Tambin realizamos el clculo extendiendo la molcula aromtica horizontalmente a 22 anillos, conservando la simetra, y adsorbiendo 6 pares de Litios. Nuevamente la estrutura optimizada resultante para el zigzag, como se muestra en la figura \ref{22ring}, corresponde a un mnimo local de energa, con una energa de adsorcin de -265.4 kcal/mol, mientras que, la configuracin mostrada en la figura \ref{22ring_min} corresponde a una energa ms baja, para la cual la energa de adsorcin es de -308.kcal/mol.\\

Finalmente, se realizaron clculos peridicos en los que se obtiene la configuracin zigzag, correspondiente a un mnimo local de energa de -169 kcal/mol, como se muestra en la figura \ref{period_fl}, mientras que la configuracin mostrada en la figura \ref{period_min} corresponde a una energa de -200 kcal/mol.\\

\begin{figure}[htb!]
 \begin{subfigure}[b]{0.5\textwidth}
     \hspace*{0in}\includegraphics[height=0.2\textheight]{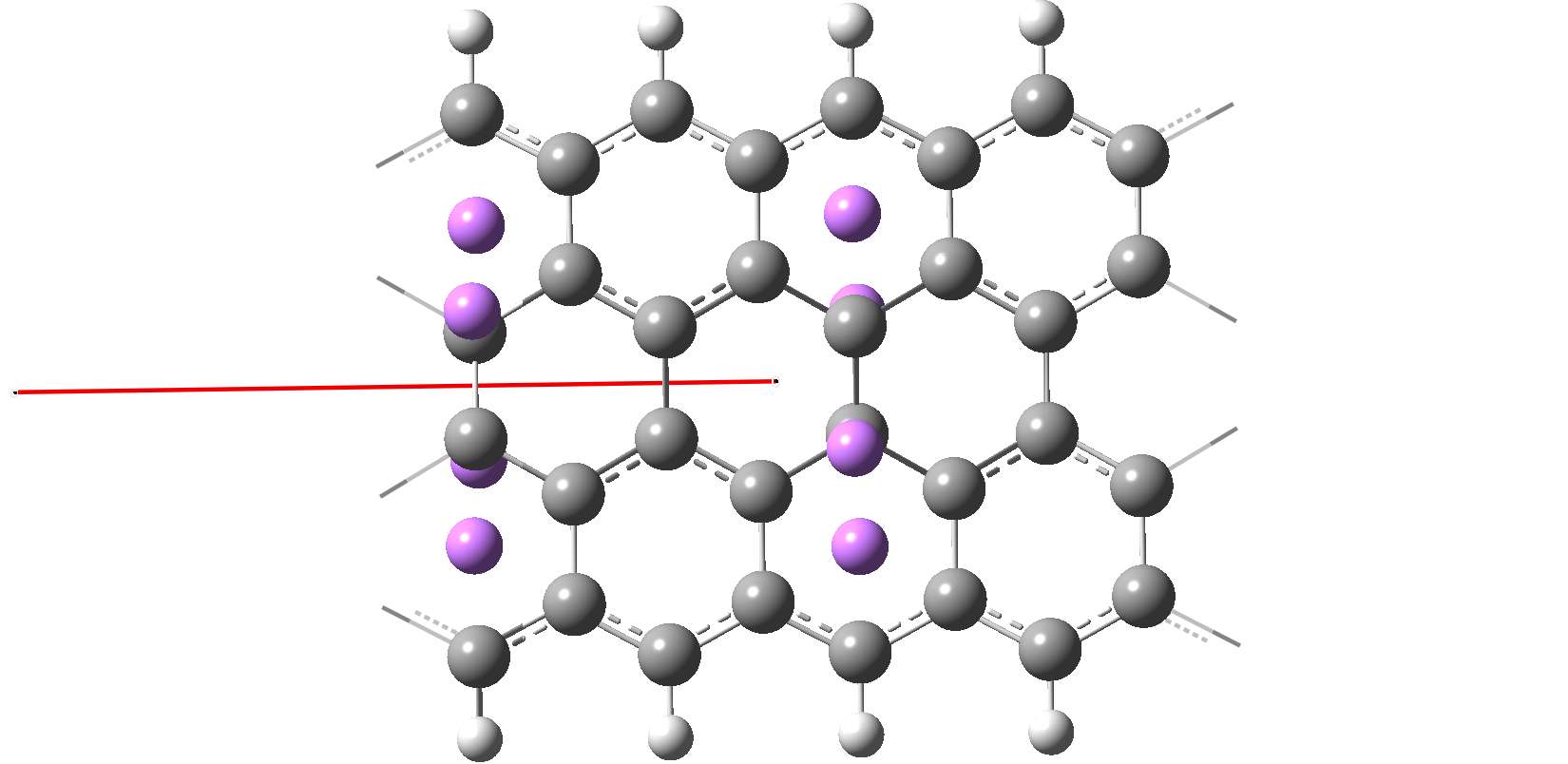}	\caption{}
         \end{subfigure}
	 \begin{subfigure}[b]{0.5\textwidth}
    \hspace*{0in} \includegraphics[height=0.2\textheight]{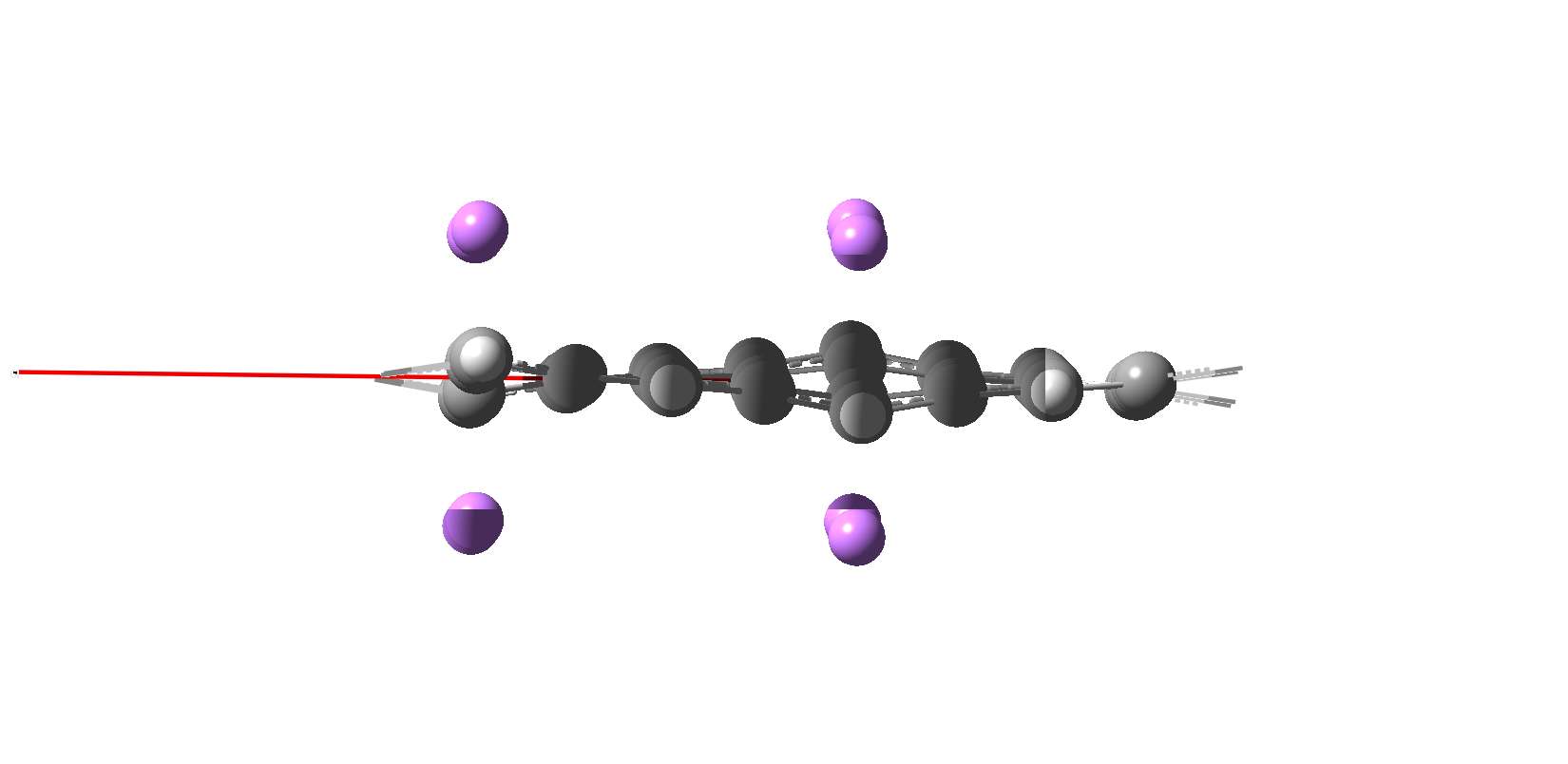}  	\caption{}
	\end{subfigure}\\

         \caption{ Se muestra la configuracin de un mnimo local de energa de la celda unitaria, con 8 pares de Litios, obtenida en el clculo peridico con el mtodo de DFT, el funcional B3LYP y el conjunto de bases 6-31g y cuya estructura corresponde al zigzag. ($a$) Vista frontal y  ($b$) vista lateral.}
         \label{cu_period_local}
\end{figure}

 \begin{figure}[htb!]
$\begin{array}{cc}
\includegraphics[height=0.23\textheight]{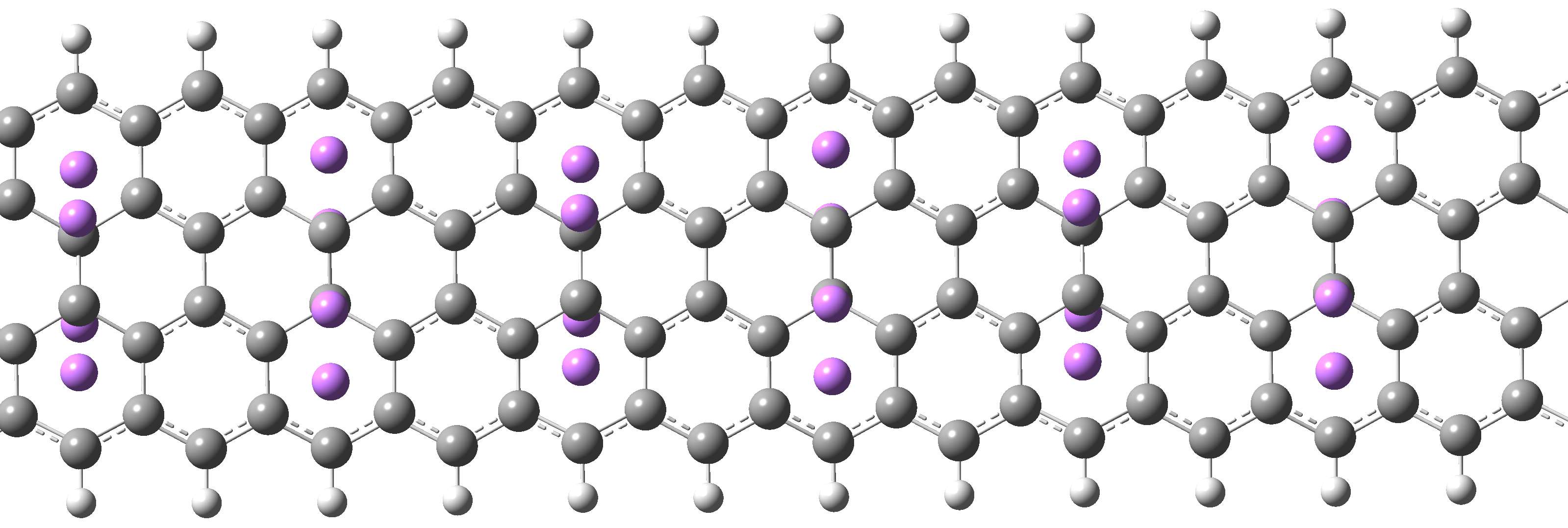}\\
\includegraphics[height=0.23\textheight]{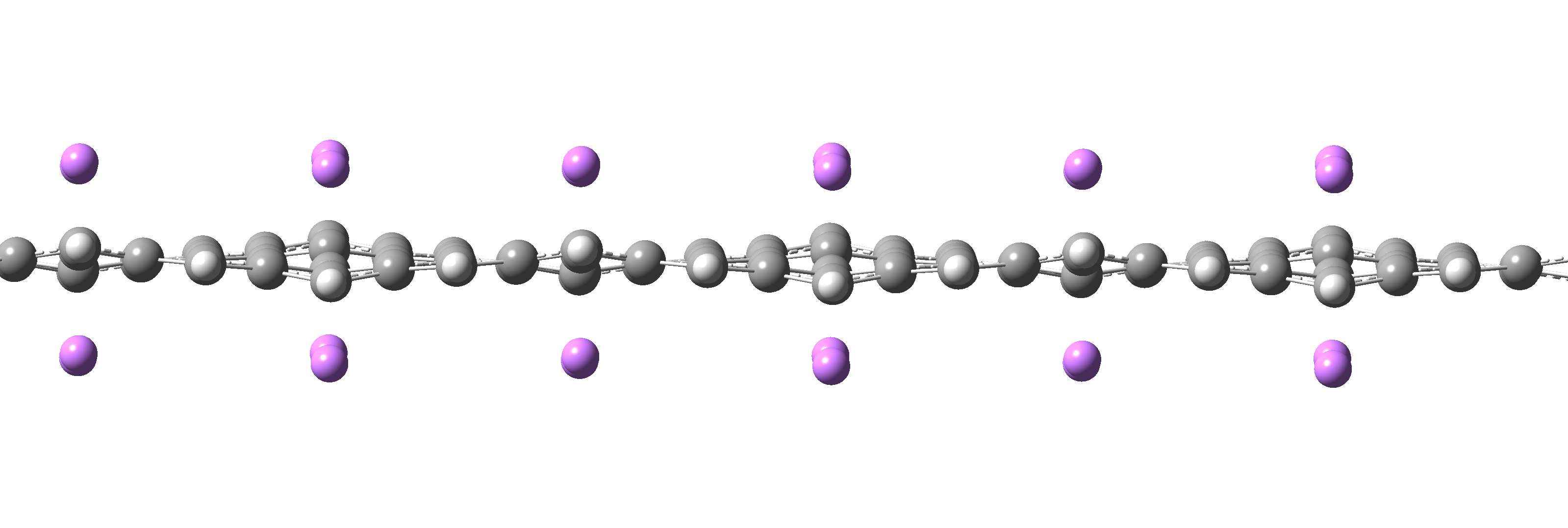}\\
\end{array}$
\caption{Se muestra la celda unitaria con ocho Litios, mostrada en la figura \ref{cu_period_local}, repetida tres veces. ($a$) Vista frontal y  ($b$) vista lateral.}
          \label{period_fl}
 \end{figure}

\begin{figure}[htb!]
 \begin{subfigure}[b]{0.5\textwidth}
     \hspace*{0in}\includegraphics[height=0.2\textheight]{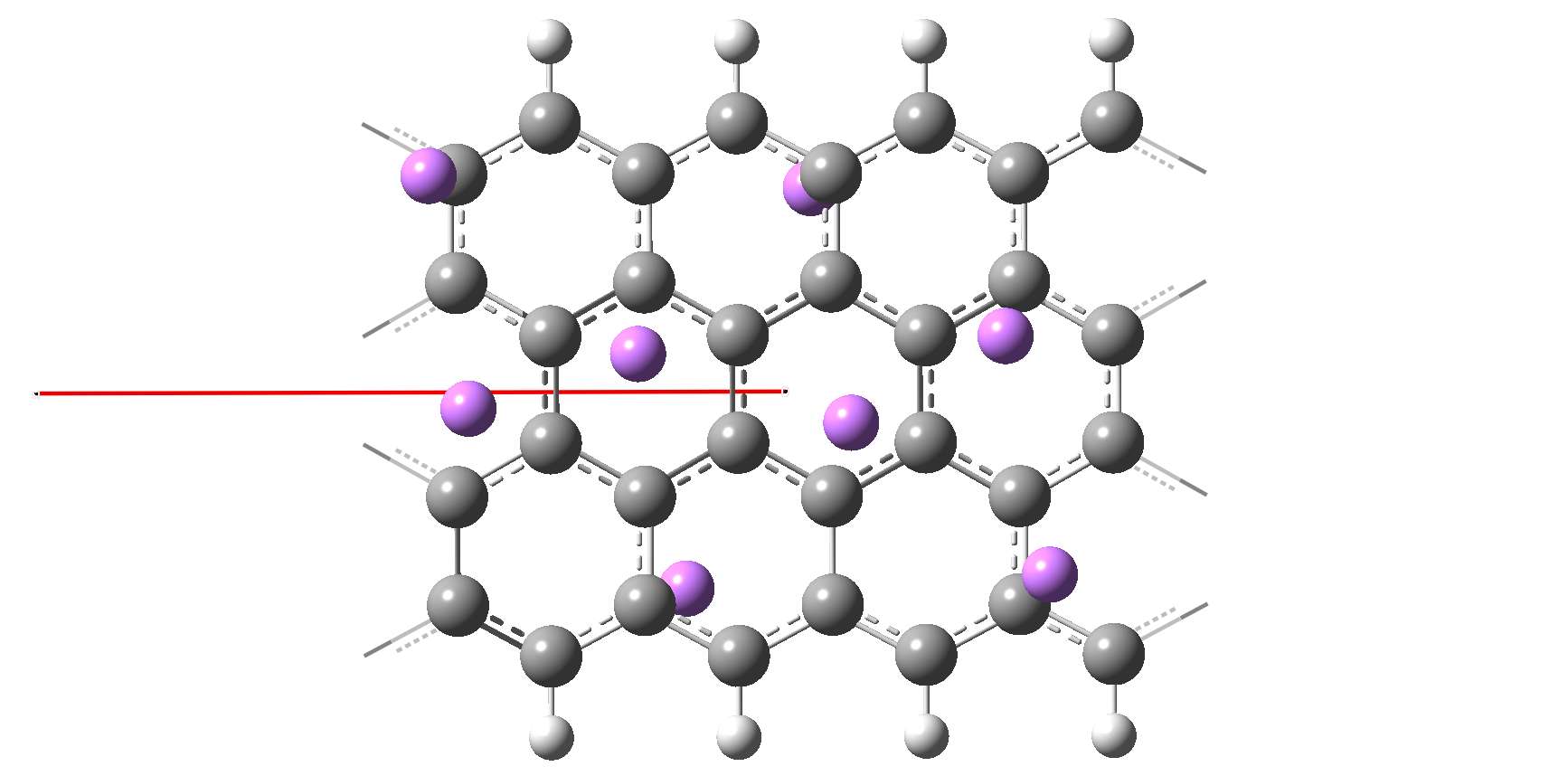}	\caption{}
         \end{subfigure}
	 \begin{subfigure}[b]{0.5\textwidth}
    \hspace*{0in} \includegraphics[height=0.2\textheight]{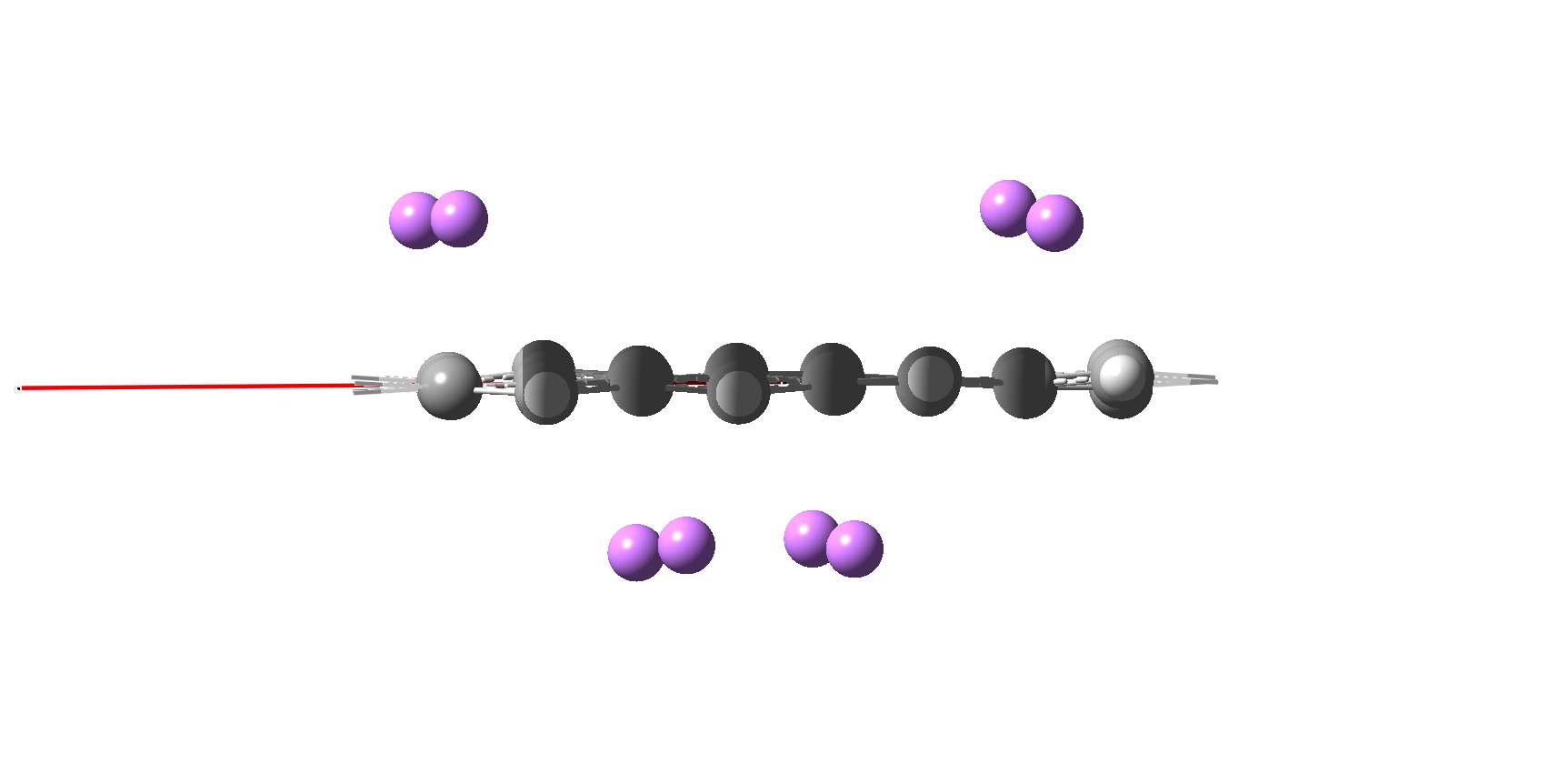}  	\caption{}
	\end{subfigure}\\

         \caption{ Se muestra la configuracin del mnimo absoluto de energa de la celda unitaria, con 8 pares de Litios, obtenida en el clculo peridico con el mtodo de DFT, el funcional B3LYP y el conjunto de bases 6-31g y cuya energa es ms baja que para el caso zigzag. ($a$) Vista frontal y  ($b$) vista lateral.}
         \label{cu_period_min}
\end{figure}

 \begin{figure}[t]
$\begin{array}{c}
 \hspace*{-0.1in}\includegraphics[height=0.21\textheight]{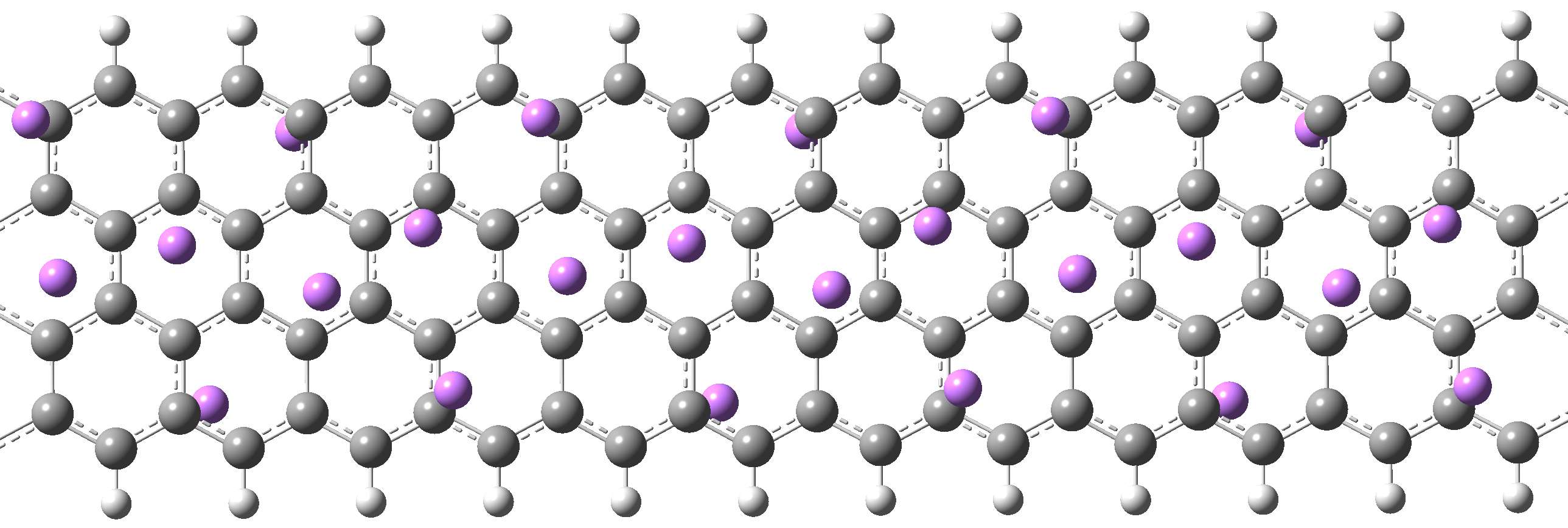}\\
 \hspace*{-0.1in}\includegraphics[height=0.21\textheight]{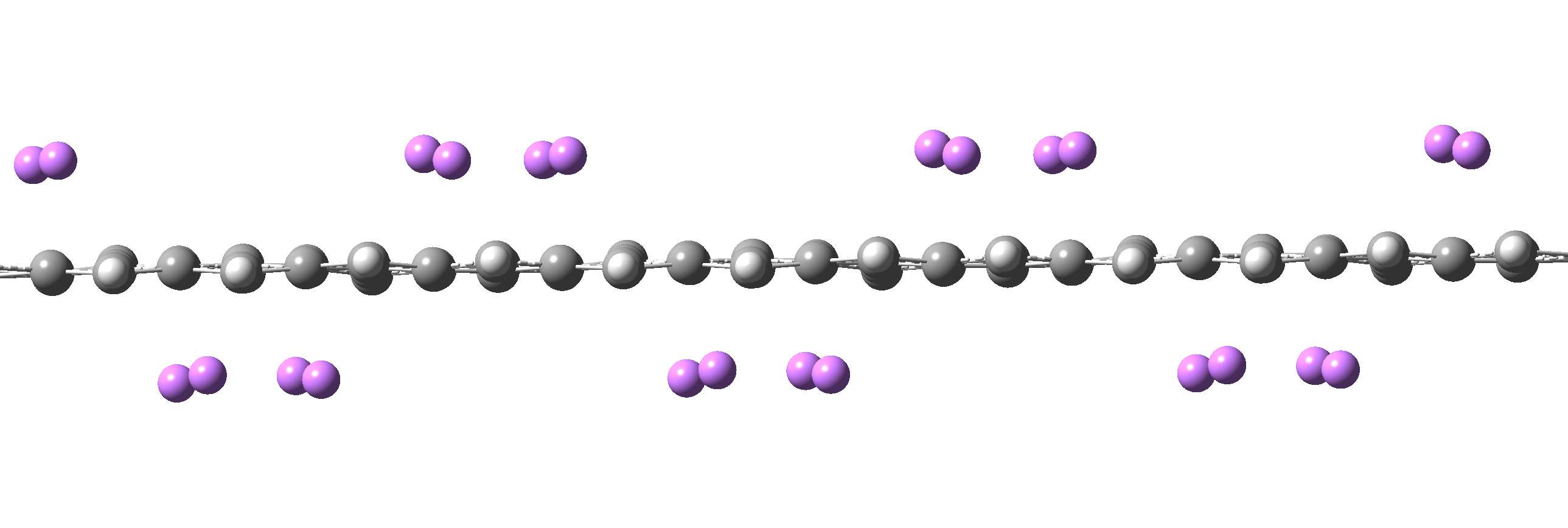}
\end{array}$
\caption{Se muestra la celda unitaria con ocho Litios, mostrada en la figura \ref{cu_period_min}, repetida tres veces. ($a$) Vista frontal y  ($b$) vista lateral.}
 \label{period_min}        
 \end{figure}

 \clearpage

\section{Reacciones}

\subsection{Efectos Catalticos}

 Los hidrocarburos clorados son compuestos de tomos de Hidrgeno y Carbono a los que se reemplazan algunos de sus Hidrgenos con tomos de Cloro. Se ha encontrado que la descomposicin de hidrocarburos clorados con y sin la presencia de agua, se facilita mediante el uso de radicales altamente reactivos que se pueden obtener a partir de perxido  de hidrgeno (H$_2$O$_2$) o el ozono (O$_3$) \cite{13_12dicl}-\cite{18_12dicl}. Es importante lograr un mejor entendimiento de estas reacciones debido a la abundancia de los hidrocarburos clorados en el ambiente y su impacto negativo en organismos vivos \cite{react4}. En particular, los metanoles (CH$_3$OH)  y formaldehdos (CH$_2$ el ms simple de ellos) clorinados, son usados y estudiados en diferentes reas de la qumica \cite{area1,area2}, dentro de las cuales est la qumica de la atmosfera \cite{atmosfera}.

Los efectos del confinamiento en nanotubos han sido recientemente un tema de inters en descomposicin de reacciones  \cite{menshut}, catlisis \cite{catal}, entre otros.  En el primer caso se sabe que la presencia de los nanotubos puede afectar la geometra de los reactantes, las barreras de energas, as como la energa resultante de las reacciones \cite{react4}. Incluso se ha encontrado a nivel  experimental que los nanotubos se pueden usar como recipientes de reaccin para la polimerizacin de C$_{60}$O \cite{nano}. El estudio hecho por Halls y sus colaboradores  revela el impacto de la presencia de nanotubos en reacciones como Menshutkin SN2 \cite{menshut}, as como otros mecanismos relacionados \cite{nano1}-\cite{nano4}.\\

De los resultados mencionados anteriormente surge el interrogante de si un efecto similar es producido al usar hojuelas de grafeno, teniendo en cuenta el trabajo previo \cite{js1} en el que se mostr que la adsorcin de metales a superficies de grafeno puede incrementar el potencial de las molculas adsorbidas. La comprensin de estos mecanismos podra mejorar el impacto negativo de este tipo de productos para los organismos vivos al contribuir a su reduccin, es decir, sera importante entender la manera en que las superficies (molculas aromticas grandes) pueden ser utilizados para mejorar la naturaleza qumica de las reacciones.\\

Para resolver el interrogante planteado y como aplicacin de los mtodos utilizados en este trabajo de tesis, se han explorado las energas de reaccin usando hojuelas de grafeno con la adsorcin de un tomo de Litio para llevar a cabo la descomposicin qumica del clorometanol (CH$_2$(OH)Cl), el diclorometanol (CH(OH)Cl$_2$) y el formaldehdo de cloro (ClCHO) \cite{react1}-\cite{react3}, que son unas, de muchas reacciones que se producen en la atmsfera \cite{atmos} y que fueron estudiadas en nanotubos \cite{react4}, lo cual nos permite comparar resultados. \\

El clorometanol se ha observado experimentalmente a bajas temperaturas en matrices \cite{matr} y en fase gaseosa \cite{pgas1,pgas2}, y en ambos ambientes se descompone en formaldehdos y en cloruro de Hidrgeno (HCl), ste ltimo es un gas txico. El dicloromeano (CH(OH)Cl$_2$) tambin decae en formaldehido y HCl \cite{pgas2}. La descomposicin del formaldehdo de Cloro (ClCHO) en solucin acuosa y en fase gas termina en monxido de Carbono (CO) y HCl \cite{formald,formaldg}, el primero de los cuales es un gas txico.\\

El primer par de reacciones a considerar representan la descomposicin del clorometanol seguida de la hidratacin del formaldehdo  \cite{react3}:

\beq \text{CH}_2 \text{(OH)Cl} \to \text{HCHO + HCl}\label{react1}\eeq

\beq \text{HCHO + H}_2\text{O} \to \text{CH}_2	\text{(OH)}_2\label{react2}\eeq

El siguiente par de ecuaciones representan la descomposicin del diclorometanol a formaldehdo de Cloro y cloruro de Hidrgeno, seguido de la descomposicin del formaldehdo de Cloro en monxido de Carbono y cloruro de Hidrgeno  \cite{react3}:

\beq \text{CH (OH) Cl}_2 \to \text{ClCHO + HCl}\label{react3}\eeq

\beq \text{ClCHO}\to \text{CO + HCl} \label{react4}\eeq

Las siguientes reacciones representan la descomposicin del doclorometanol en formaldehdo de Cloro, cloruro de Hidrgeno y agua, seguido de la hidratacin del formaldehdo de cloro y finalmente la descomposicin de  CH(Cl)(OH)$_2$ \cite{react3}:

\beq \text{CH(OH) Cl}_2 +\text{ H}_2\text{O}\to \text{ClCHO} + \text{H}_2\text{O + HCl}\label{react5}\eeq

\beq \text{ClCHO + H}_2\text{O}\to \text{CH(Cl)(OH)}_2\label{react6}\eeq

\beq \text{CH(Cl)(OH)}_2 + \text{H}_2\text{O} \to \text{HCOOH + HCl + H}_2\text{O}\label{react7}\eeq

Estas reacciones fueron optimizadas con clculos ab-initio usando los mtodos de M¿llerÐPlesset a segundo orden (MP2) \cite{moller} y DFT/B3LYP y obtuvieron los mismos resultados en ambos casos \cite{react3}.\\

Para nuestro estudio usamos una hojuela de siete anillos para el anlisis, por su simetra y reducido tamao. Se realizaron clculos de optimizacin de la geometra usando las bases STO-3G para la optimizacin y las bases 6-311++g** para clculos de energa. En la tabla \ref{react_ener} se muestran las energas de las reacciones \ref{react1}-\ref{react7}, calculadas en general como la diferencia entre la energa de los productos y la energa de los reactantes.  $\Delta E_I$ representa la energa de la reacciones  \ref{react1}-\ref{react7} aisladas,  $\Delta E_{II}$ representa la energa de la reacciones  \ref{react1}-\ref{react7} sobre la hojuela de grafeno, $\Delta E_{III}$ representa la energa de la reacciones  \ref{react1}-\ref{react7} sobre la hojuela de grafeno con la adsorcin de Li. Finalmente, con el fin de comparar los resultados las energas $\Delta E_{IV}$ que corresponden a las energas de las reacciones obtenidas con el mtodo MP2, que es una mejora del mtodo de Hartree-Fock pues aade efectos de correlacin electrnica a travs de la teora de perturbaciones. Las energas $E_{IV}$ son tomadas del trabajo de Trzaskowski y colaboradores \cite{react4} usando nanotubos de carbono para el confinamiento de las reacciones con MP2 y bases 6-31+g**.\\

 \begin{table*}\centering
\begin{tabular}{ccccc}
\hline
Reaccin No. &\textbf{$E_{I}$} &  \textbf{$E_{II}$}& \textbf{$E_{III}$} &  \textbf{$E_{IV}$}\\
 \hline
1&	6.75	&13.56&	14.88&	-15.51\\
2	&-8.76	&-13.48	&-17.25&	12.95\\
3&	-5.06&	-6.33&	-6.76	&0.00\\
4&	4.59	&8.46	&8.79	&-8.81\\
6&	-5.05 &	4.10	&1.13	&44.43\\
7&	-1.76&	1.02	&-4.74&	-24.77\\\hline
\end{tabular}
\caption{Energas relativas (calculadas con el B3LYP/6-311++g**//B3LYP/STO-3g) de las diferentes reacciones qumicas \ref{react1}-\ref{react7}  en kcal/mol donde $\Delta E_{I}$: corresponde a las reacciones qumicas aisladas; $\Delta E_{II}$: corresponde a las reacciones qumicas en la superficie de la hojuela de grafeno; $\Delta E_{III}$: corresponde a las reacciones en el complejo Li+ hojuela y $\Delta E_{IV}$: corresponde a las energas de las reacciones en el  nanotubo, realizadas con MP2/6-31+g* \cite{react4} para un modelo de las descomposiciones en un espacio confinado. }
\end{table*}

\begin{figure}[htb!]
 \hspace*{-0.3in}\includegraphics[height=1\textheight]{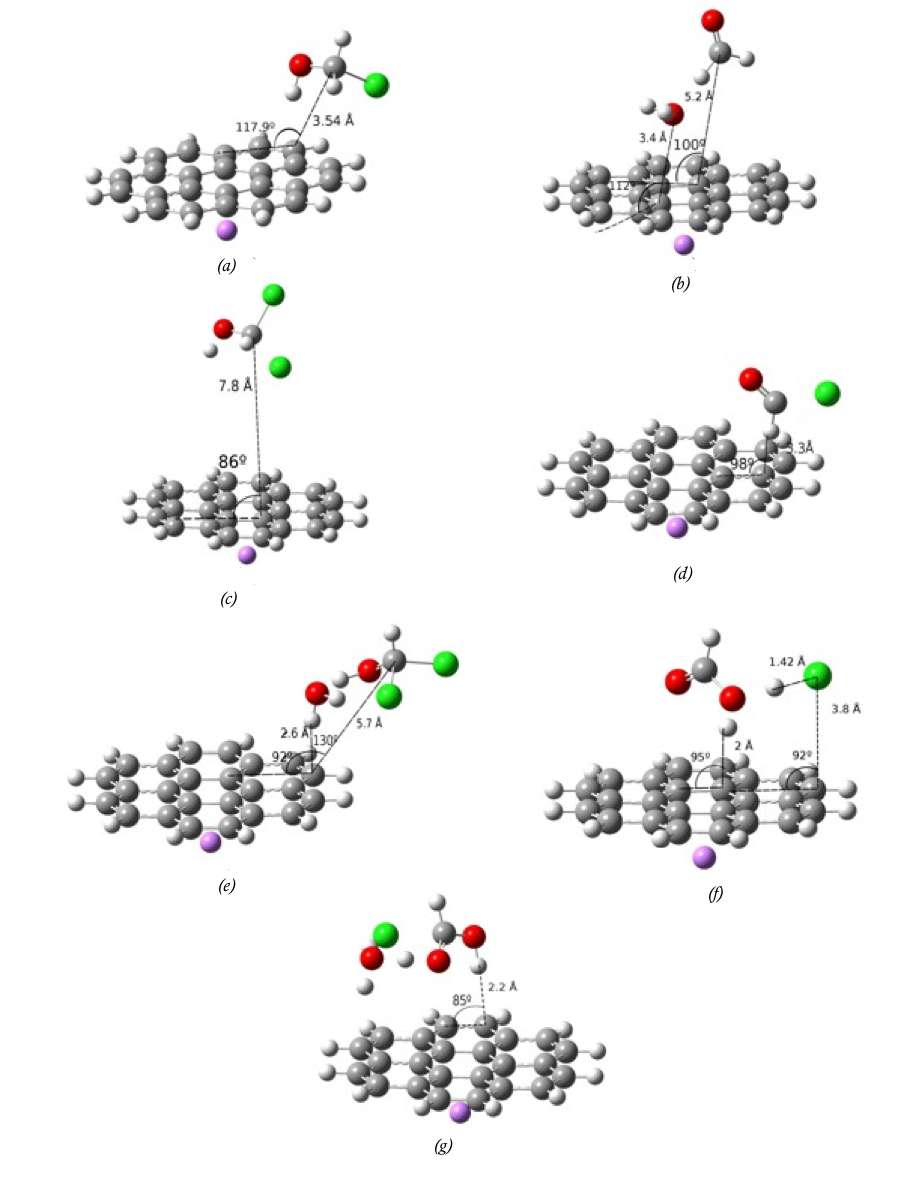}
\caption{Representaciones grficas de los reactivos optimizados (a nivel B3LYP/STO-3G) en la superficie de la hojuela + Li, donde las longitudes de enlace estn en angstroms (\AA) y los ngulos se expresan en grados (¡), correspondientes a  las reacciones $(a)$ \ref{react1}, $(b)$ \ref{react2}, $(c)$ \ref{react3}, $(d)$ \ref{react4}, $(e)$ \ref{react5}, $(f)$ \ref{react6},$(g)$ \ref{react7}}
          \label{reactants}
 \end{figure}

\begin{figure}[htb!]
 \hspace*{-0.3in}\includegraphics[height=1\textheight]{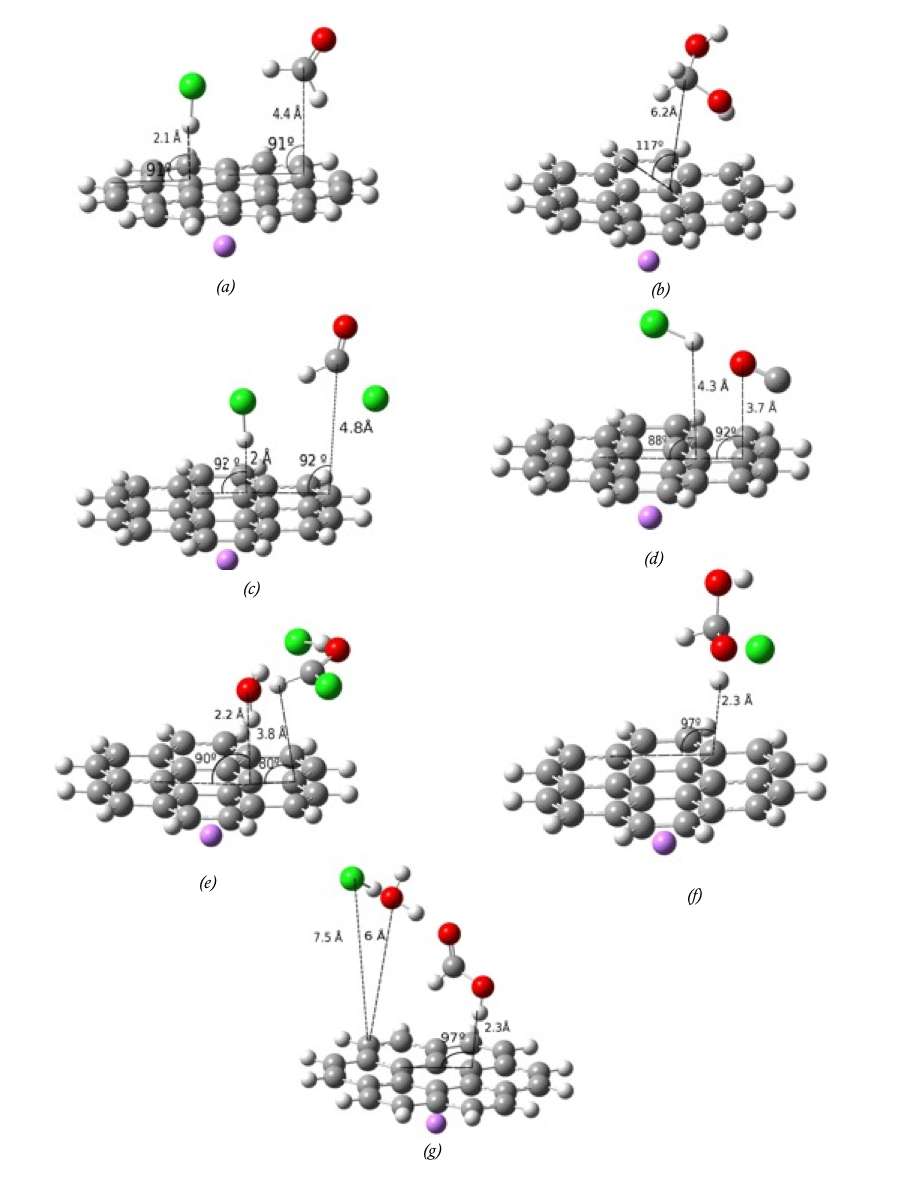}
\caption{Representaciones grficas de los productos optimizados (a nivel B3LYP/STO-3G) en la superficie de la hojuela + Li, donde las longitudes de enlace estn en angstroms (\AA) y los ngulos se expresan en grados (¡), correspondientes a  las reacciones $(a)$ \ref{react1}, $(b)$ \ref{react2}, $(c)$ \ref{react3}, $(d)$ \ref{react4}, $(e)$ \ref{react5}, $(f)$ \ref{react6},$(g)$ \ref{react7}}
          \label{products}
 \end{figure}

 Las estructuras moleculares correspondientes a los reactivos  en el Li + hojuela aromtica  de cada una de las reacciones se optimizaron en el nivel B3LYP con bases STO-3G y  se muestran en la figura \ref{reactants}.  Los productos se muestran en la figura \ref{products}. El tomo de Li distorsiona ligeramente el centro de la superficie del grafeno, una tendencia que se observa con frecuencia en las estructuras de fulereno \cite{fule}.\\

Los resultados de los clculos para cada reaccin son variados:\\

\textit{Reaccin \ref{react1}}: $ \text{CH}_2 \text{(OH)Cl} \to \text{HCHO + HCl}$\\

De la tabla \ref{react_ener} se puede observar que la energa de reaccin es significativamente ms alta en el Li + superficie de grafeno que en la reaccin aislada y tambin ms grande que la reaccin sin el metal . Si comparamos los resultados anteriores utilizando confinamiento de nanotubos \cite{react4}, los clculos sugieren que en presencia de tales condiciones, sta descomposicin es ms reactiva. Teniendo en cuenta que  para muchas de estas reacciones qumicas se prefiere una barrera de reactividad o energa ms alta, ya que minimiza la produccin de radicales libres nocivos, nuestro resultado es ms favorable.\\

En cuanto a la geometra de las estructuras qumicas, la  posicin y la orientacin de CH$_2$(OH)Cl con respecto a la superficie, es similar al caso en el que no se adsorbe Li a la superficie. La medida de los ngulos entre los tomos de carbono colineales de la hojuela se obtiene una pequea deformacin de alrededor de 3 ¼.\\

En la estructura de los productos se obtienen resultados diferentes respecto al caso del la hojuela sin Li. Las molculas estn casi fuera de la hojuela con una orientacin muy diferente. La distancia entre el tomo de Litio y la hojuela con reacciones no cambia comparado con el caso de la hojuela y el tomo de Litio slamente.\\

 \textit{Reaccin \ref{react2}}: HCHO + H$_2$O $\to$ CH$_2$	(OH)$_2$\\
Se observa que  la energa utilizando el nanotubo est en el orden de 13 kcal/mol lo que aumenta la barrera en comparacin con la reaccin aislada. Observamos que la reaccin en la superficie de hojuela tiene una energa de reaccin de -13,48 kcal / mol. Cuando el metal se adsorbe a la hojuela se obtiene una energa de reaccin de -17,25 kcal / mol. Mientras que estos resultados son mejores para el caso de nanotubos, en la primera iniciacin del mecanismo00
, para nuestro caso se limitarn potencialmente la reaccin qumica de proceder a estos productos finales.\\

Estructuralmente, los reactivos y los productos tienen formaciones qumicas interesantes. Para las reacciones en comparacin con o sin el Li las estructuras difieren, pues las molculas estn casi fuera de la superficie de la hojuela. Para los productos la posicin y orientacin de la molcula CH$_2$(OH)$_2$ con respecto a la superficie de la hojuela son similares al caso que no hay tomo de Litio.\\

  \textit{Reaccin \ref{react3}}:  CH (OH) Cl$_2 \to$ ClCHO + HCl\\
Nuestros resultados usando adsorcin de metal no son tan notables como las de la superficie del nanotubo. Sin embargo, ya que es una reaccin en cadena de la ecuacin \ref{react1}, la cual en nuestro caso tiene una barrera  relativamente alta para evitar la formacin de tales molculas.
La posicin y la orientacin de la sustancia reaccionante CH (OH) Cl$_2$ con respecto a la superficie de la hojuela es similar en ambos casos, con y sin Li. Los ngulos entre los tomos de carbono colineales de la hojuela tienen una pequea deformacin de 3¼. Podemos considerar cualitativamente que la molcula se encuentra en el centro de la hojuela con adsorcin de Li. Para los productos la posicin y orientacin de las molculas con respecto a la hojuela son similares en el caso sin Li.\\

 \textit{Reaccin \ref{react4}}: ClCHO$\to$ CO + HCl \\
Para esta reaccin las energas con la hojuela son ms bien constantes (con y sin Li) y mayor que en el caso aislado. Sin embargo, si se compara la eficacia de la inhibicin de la reaccin en la superficie en comparacin con el nanotubo se observa una gran diferencia. El valor de la reaccin aislado es 4,59 kcal / mol en comparacin con 8,79 kcal / mol y -8,81 kcal/mol para el Li+grafeno y nanotubos respectivamente.\\

La molcula de la sustancia reaccionante en hojuela sin Li est cualitativamente en el centro de la superficie de la hojuela y se separa de sta 4.1  \AA. Es  de resaltar que la molcula se desplaza desde el centro de la hojuela a la esquina de la misma para el caso con adsorcin de Li. En el caso de los productos en comparacin con el caso sin Li hay un cambio y las molculas estn casi fuera de la superficie de la hojuela. \\

  \textit{Reaccin \ref{react5}}: $CH(OH) Cl_2 + H_2O\to ClCHO + H_2O + HCl$\\
 Nuestros resultados son bastante mejores en relacin con el caso de nanotubos, como podemos ver que lleva a un potencial de 14,51 kcal / mol. Esto se traduce en la dificultad de que ocurra la reaccin al adsorber el metal en la hojuela. Cuando no utilizamos un metal las energas de reaccin son ms bajas y hacen bastante factible de que esta descomposicin ocurra lo que conduce a un impacto ecolgico negativo. Estructuralmente, los sistemas no sufren cambios drsticos respecto al uso o no de Li y permanecen cerca de la hojuela. La distancia del tomo de Li a la superficie de la hojuela es de 1,9 \AA{} y el Li permanece en el centro del anillo, posicin que mantiene en las reacciones anteriores tambin.\\

  \textit{Reaccin \ref{react6}}: ClCHO + H$_2$O$\to$ CH (Cl) (OH)$_2$\\
 Se observa que la energa de reaccin en el interior del nanotubo es ms grande que en nuestro caso,  sin embargo de la reaccin \ref{react5} vimos que la diferencia en nuestro mecanismo tendra poca probabilidad de ocurrir. Desde un punto de vista estructural, la distancia del tomo de Li a la hojuela es 1,90 \AA{}  y permanece en el centro del anillo.
Con respecto a los productos,  las molculas para este caso sin tomo de Li estn cualitativamente en el centro de la superficie de la hojuela y con la presencia del metal se desplazan hacia afuera de la misma.\\

 \textit{Reaccin \ref{react7}}: CH(Cl)(OH)$_2$ + H$_2$O $\to$ HCOOH + HCl + H$_2$O\\
Para la ltima reaccin de nuestros resultados en la superficie de la hojuela de grafeno son menos favorables que en el interior del nanotubo. Esto se traduce en el hecho que estas reacciones sobre la hojuela sern menos probable que tenga lugar que dentro del nanotubo o en el caso aislado. 
Estructuralmente la tendencia general observada para los reactivos y productos es que se necesita el metal para que algunas molculas permanezcan centradas en la superficie de la hojuela.\\

Este estudio debe servir como un marco bsico para el futuro de la experimentacin en el campo de los inhibidores de descomposicin atmosfrica utilizando metal de adsorcin en las superficies de grafeno para reducir la razn a la que las reacciones qumicas especficas tienen lugar.\\

Es importante hacer notar que se han mejorado las barreras de energa para los mecanismos de descomposicin de muchas de estas reacciones qumicas con respecto a los nanotubos . Esto puede sugerir una capacidad de ciertas hojuelas de grafeno para actuar como amortiguadores ambientales en las reacciones qumicas perjudiciales. Con el uso de la adsorcin de metal a las superficies de grafeno fue posible minimizar las energas de reaccin en situaciones en las que se quiere suprimir las reacciones qumicas perjudiciales.\\

Estos resultados apoyan el hecho de que las hojuelas de grafeno pueden ser fcilmente susceptibles a reacciones qumicas si estas inducen un potencial externo \cite{sucept}. Aqu se ha inducido tal potencial con el uso de la adsorcin de Li a la superficie de una hojuela de grafeno por lo tanto se puede decir que a travs de sta adsorcin es posible manipular y controlar ciertas reacciones qumicas. A este estudio sigue la pregunta  experimental interesante si dicha aplicacin, con el fin de controlar reacciones qumicas en la atmsfera, se puede realizar. Creemos, a partir de los resultados de los clculos, que mientras que en ciertas situaciones el confinamiento en el interior de los nanotubos conduce a mejores resultados que el modelo planteado aqu,  la adsorcin de Li a la  hojuela de grafeno acta como supresor importante sobre la primera reaccin que conduce a la reduccin de produccin de productos intermedios perjudiciales. Adicionalmente,  se han implementado estrategias de adsorcin en grafeno \cite{ad_exp1}-\cite{ad_exp3}  mientras que an no est claro cmo controlar las reacciones dentro de los nanotubos. En una nota lateral, cabe mencionar que el clculo tambin se realiz mediante doble adsorcin de metal sobre el lado apuesto a la reaccin  \ref{react7}  en  la hojuela de grafeno. Para este sistema las energas de reaccin en realidad se vuelven ms exotrmica ( -7,51 kcal / mol a la B3LYP/6-311 + + G ** nivel de la teora). Como una aplicacin potencialmente til podemos utilizar la relacin de concentracin de tomos de Li para controlar el resultado de mecanismos qumicos especficos.\\


\chapter{Conclusiones}

\begin{itemize}

\item Se estudi la adsorcin de Litio a hojuelas de grafeno y se observ que cuando se adsorben dos litios en el mismo anillo de la hojuela y del mismo lado, sta se deforma a pesar de que el estado en el que la hojuela permanece plana es posible, por lo tanto sta deformacin corresponde a un rompimiento espontneo de simetra. Tambin se observ que a medida que se aumenta el tamao de la hojuela el efecto de deformacin es menor. Se observaron deformaciones  tambin en hojuelas de Boro-Nitrgeno.

\item Con el fin de  simplificar el problema y obtener un mejor entendimiento del rompimiento de simetra, se redujo el sistema de hojuelas de grafeno a poliacenos. Inicialmente comparamos los resultados obtenidos de los clculos realizados para la optimizacin de doble adsorcin de Litio, de lados opuestos del poliaceno, empleando los mtodos de HF y DFT y obtenemos que el rompimiento de la simetra es un efecto de campo medio, que en realidad es un tanto atenuado por correlaciones. 

\item En clculos de DFT para la mltiple adsorcin de pares de Litios en anillos cercanos, no adyacentes de poliacenos, se encontr que el estado de mnima energa corresponde a un zigzag, incluso para el clculo peridico. Considerando que los poliacenos se puedan considerar un sistema cuasi-unidimensional, encontramos que ste resultado coincide con la distorsin de Peierls, efecto que no se puede generalizar a todos los alcalinos, pues la mltiple adsorcin de pares de los dems alcalinos no muestran tal efecto.

\item Encontramos que la distorsin de Peieles es observable en hojuelas pequeas, con la simetra apropiada, pero no se puede generalizar a hojuelas de tamao mayor y menos a tiras de grafeno, ya que los estados de mnima energa para stos casos no corresponden sta.

\item Se propone un mecanismo por medio del cual las reacciones de descomposicin de el clorometano, el diclorometano y el formaldehdo de cloro ocurren sobre una hojuela aromtica con el fin de reducir la posibilidad de produccin de ciertos radicales dainos para los humanos. 

\end{itemize}


\chapter{Perspectivas}

\begin{figure}[htb!]
$\begin{array}{cc}
\hspace*{0.3in}\includegraphics[height=0.13\textheight]{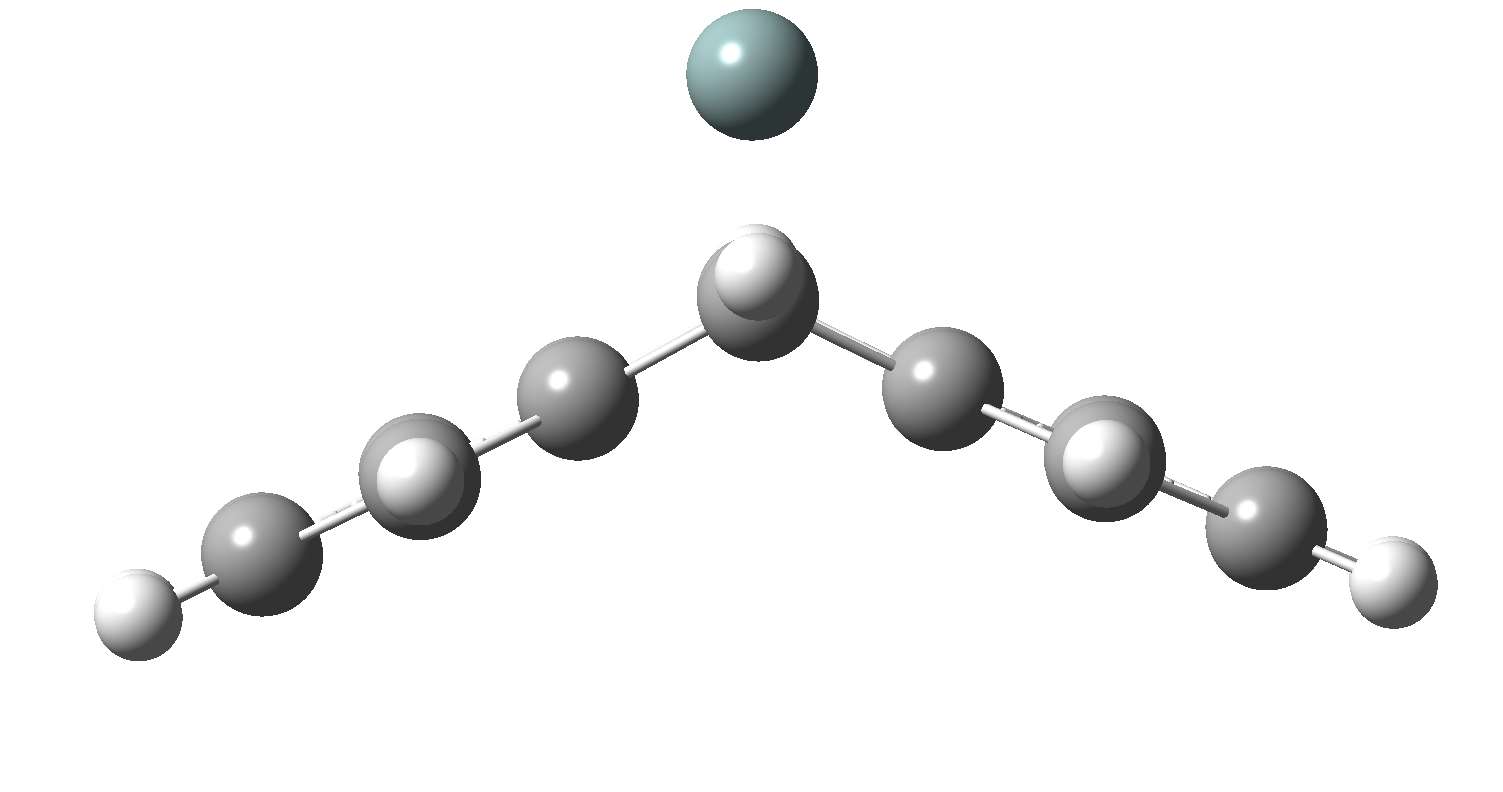}&
\hspace*{0.3in} \includegraphics[height=0.15\textheight]{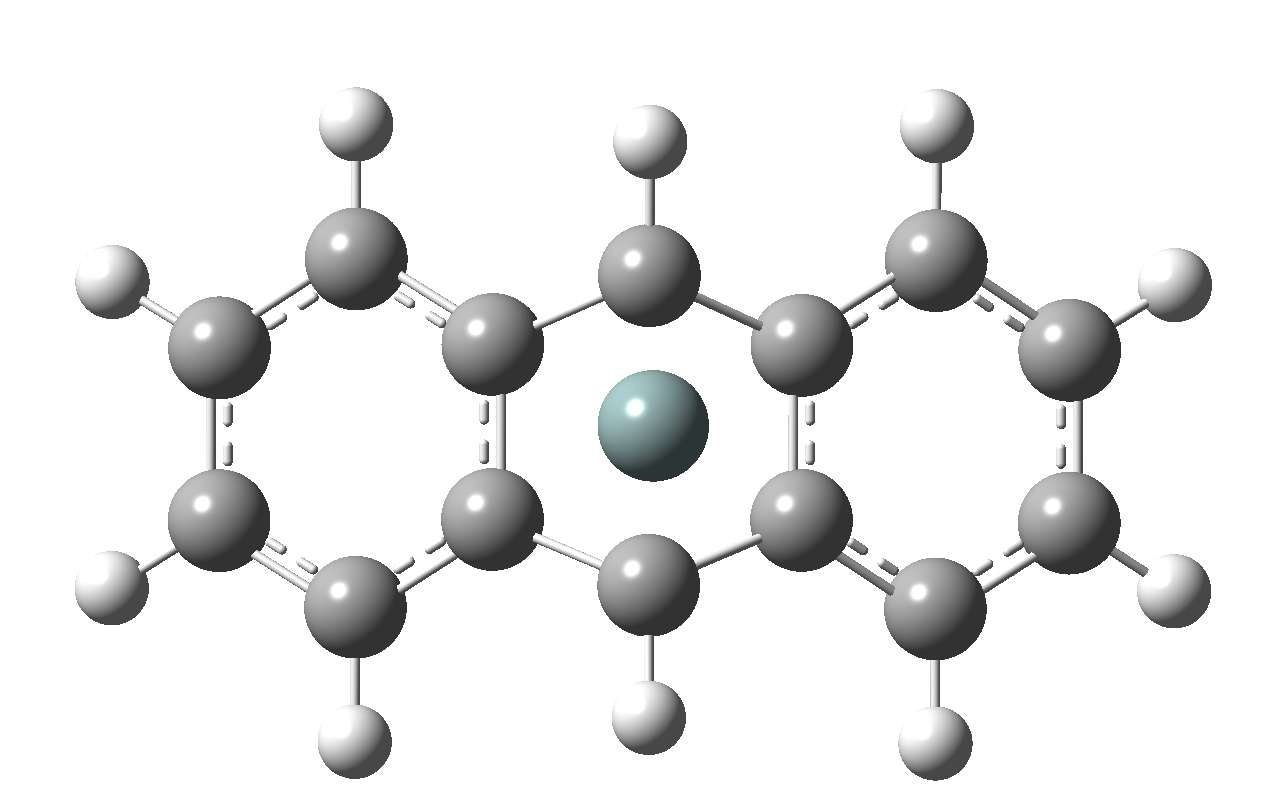}
\end{array}$
\caption{Estructura optimizada de la adsorcin de un tomo de Silicio al antraceno con el mtodo de  DFT/B3LYP y el conjunto de bases 6-311g*. }  
\label{III_one_si}      
 \end{figure}

Estos resultados abren un amplio campo de investigacin con el que pretendemos continuar. En la aplicacin a reacciones, la adsorcin del metal est restringida a un slo lado de la hojuela de grafeno, debido a que los tomos de Li adsorbidos del mismo lado en el que se encuentran los reactantes y productos llevara a una participacin directa de estos tomos en la reaccin. Por sta misma razn no pudimos estudiar el efecto de deformaciones fuertes debidas al Litio ya que no se pudo usar la doble adsorcin de Litio que, como vimos, produce una mayor deformacin. Con esto en mente, hemos iniciado la exploracin de adsorcin de otros tomos que produzcan mayor deformacin con slo adsorber un tomo en poliacenos. Hasta el momento hemos encontrado que el Silicio, el Germanio y el Selenio lo hacen. Como ejemplo se muestra el caso del Silicio. Para ello se realizaron clculos con DFT/B3LYP y el conjunto de bases 6-311g* para la adsorcin de un Si en antraceno como muestra la figura \ref{III_one_si}. La energa de adsorcin del Si es  -64.37 kcal/mol y para el Li es -21.58 kcal/mol mientras que las transferencias de carga son similares con 0.610945 $e$ y 0.650351 $ e$ respectivamente. \\

\begin{figure}[htb!]
    \centering

\includegraphics[height=0.2\textheight]{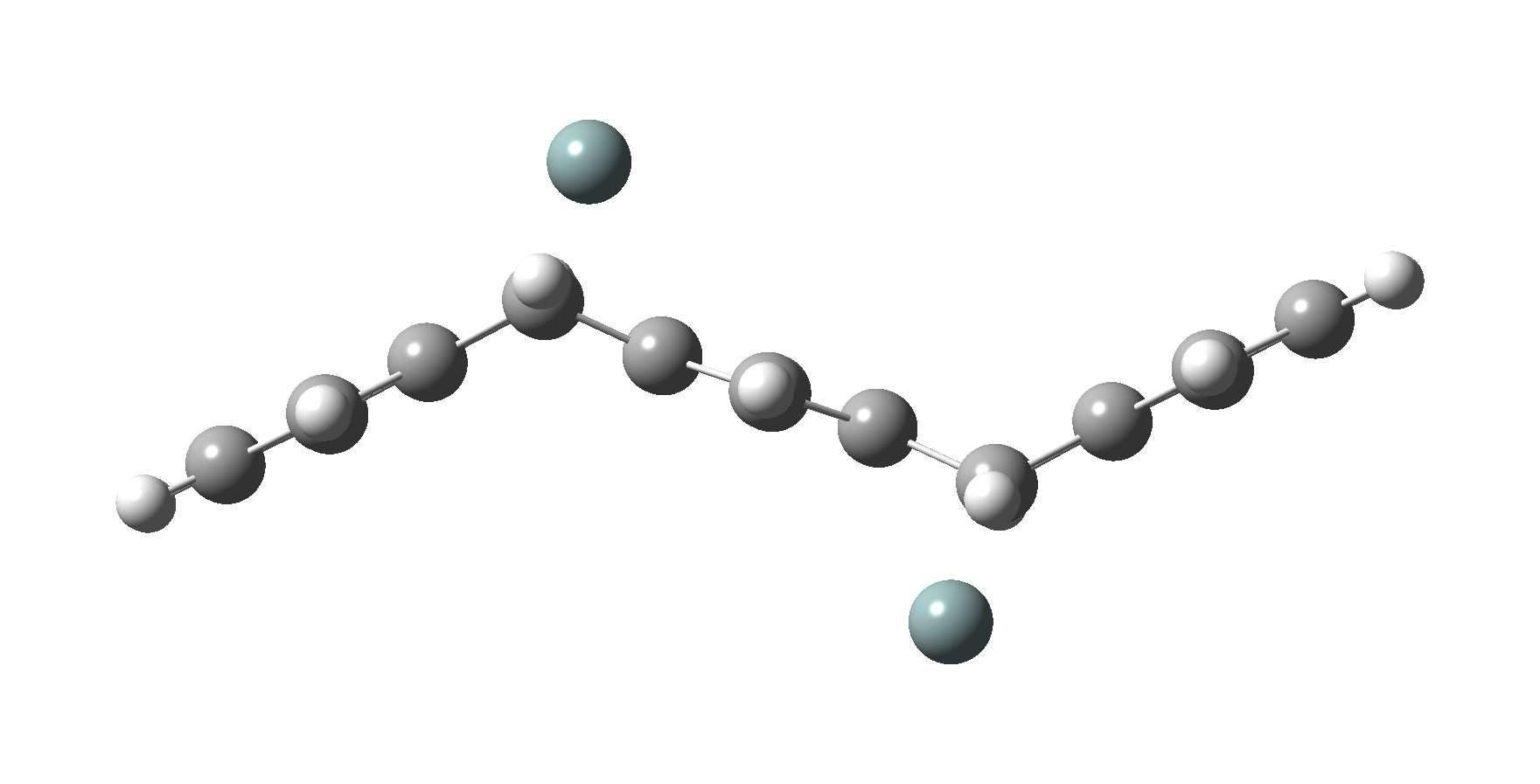}
\caption{Estructura optimizada de la adsorcin de dos tomos de Silicio de lados opuestos en el segundo y cuarto anillo del pentaceno con el mtodo de  DFT/B3LYP y el conjunto de bases 6-311g*.}
          \label{V_two_si_zig}
 \end{figure}

\begin{figure}[htb!]
  \centering
\includegraphics[height=0.2\textheight]{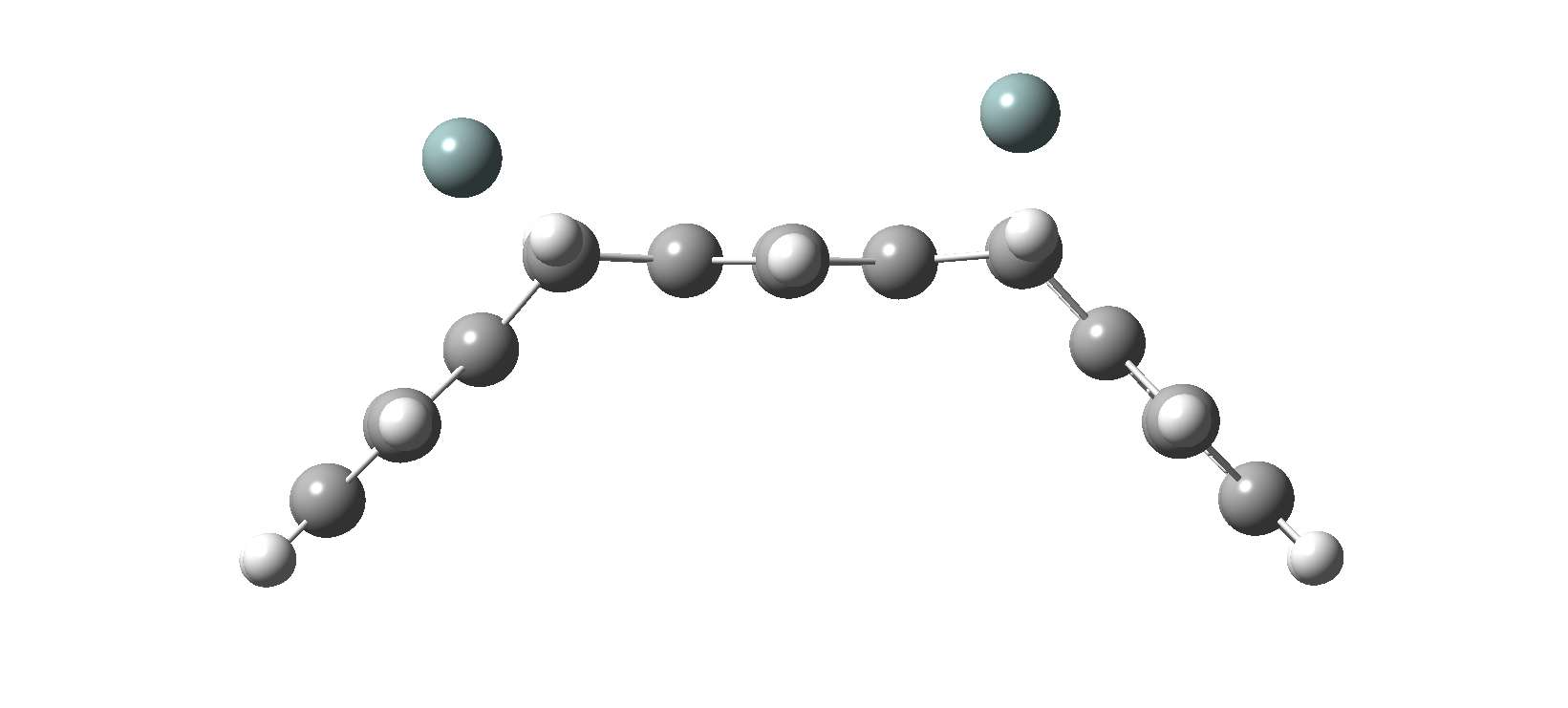}
\caption{Estructura optimizada de la adsorcin de dos tomos de Silicio del mismo lado en el segundo y cuarto anillo del pentaceno con el mtodo de  DFT/B3LYP y el conjunto de bases 6-311g*.}
          \label{arco}
 \end{figure}

Partiendo de pentaceno se ha encontrado la estructura del zigzag con slo adsorber dos tomos de Si en lados opuestos como se muestra en la figura \ref{V_two_si_zig} pero si se inicia con ambos del mismo lado se obtiene con casi la misma energa un arco como se ve en la figura \ref{arco} (-894753.45 kcal/mol la energa zigzag y -894757.04 kcal/mol la energa de arco).\\

Estos resultados preliminares abren varias perspectivas: Primero vemos que los efectos sobre la geometra no son meramente efectos de transferencia de carga porque en el caso de antraceno con transferencias parecidas, los ngulos son muy distintos, mientras que con transferencias diferentes (doble adsorcin de Li versus simple de Si) los ngulos son parecidos.
De otra parte el ejemplo nos da esperanzas de producir hojuelas y hojas con fuerte deformacin y transferencia de carga similar a la de la doble adsorcin de Li. Esto tambin nos deja pensar en obtener tiras de grafeno en forma de canales y otras formas, que pueden tener efecto sobre las reacciones por su greometra.



\begin{thebibliography}{23}

 
\bibitem{peierls}
R.E. Peierls, \textit{Quantum theory of Solids},
\textsl{Clarendon Press, Oxford}, (1955).

\bibitem{js1} 
Jalbout, Abraham F.; Seligman, Thomas H.,
\textsl{ Journal of Computational and Theoretical Nanoscience}, Volume 6, Number 3, pp. 541-544(4) (2009)

\bibitem{geim}
A. K. Geim, 
\textsl{ Science,} 324: 1530 (2009).

\bibitem{fullerenes}
 Kroto, H.W.; Heath, J.R.; O'Brien, S.C.; Curl, R.F.; Smalley, R.E., \textit{C60: Buckminsterfullerene},
 \textsl{ Nature,}  (318):14, pp 162-163 (1985).
 
  \bibitem{nanotubes}
  Monthioux, Marc; Kuznetsov, V. ,
  \textsl{Carbon}, 44 (9): 1621 (2006)
  
   \bibitem{ribbons1}
 Nakada K., Fujita M., Dresselhaus G. and Dresselhaus M.S. \textit{Edge state in graphene ribbons: Nanometer size effect and edge shape dependence}. 
 \textsl{Physical Review B}, 54 (24): 17954  (1996).
 
  \bibitem{aromatic}
 H. E. Armstrong,
 \textsl{Proceedings of the Chemical Society (London)}, 6 (85): 95Ð106 (1890)
 
 \bibitem{aromaticity}
 Schleyer, Paul von Ragu, \textit{Introduction:  Aromaticity}. 
 \textsl{Chemical Reviews} 101 (5): 1115Ð8 (2001).

\bibitem{capas}
Partoens, B. and Peeters, 
\textsl{F.M. Phys. Rev. B}, 74:075404 (2006).

 \bibitem{1947}
 Wallace, P. R., 
 \textsl{Physical Review}, 71 (9): 622 (1947).

\bibitem{novo}
K. S. Novoselov, A. K. Geim, S. V. Morozov, D. Jiang, M. I. Katsnelson, I. V. Grigorieva, S. V. Dubonos and A. A. Firsov.
\textsl{Nature}, 438:197-200 (2005).

\bibitem{gr_exp1}
A. Bermudez, M. A. Martin-Delgado, and E. Solano,
\textsl{Phys. Rev. Lett.,} 99, 123602 (2007).

\bibitem{gr_exp2}
Michael S. Fuhrer,
\textsl{Nature}, 459 (7250): 1037 (2009).

\bibitem{gr_exp3}
L. Lamata, J. Len, T. Schtz, and E. Solano, 
\textsl{Phys. Rev. Lett., }98, 253005 (2007).

\bibitem{gr_exp4}
S. Longhi,
\textsl{ Opt. Lett.,} 35, 1302 (2010).

\bibitem{gr_exp5}
J. A. Franco-Villafae, E. Sadurn, S. Barkhofen, U. Kuhl, F. Mortessagne, and T. H. Seligman,
\textsl{Phys. Rev. Lett.,} 111, 170405 (2013).

\bibitem{mobil}
K.I. Bolotina, K.J. Sikesb, Z. Jianga, M. Klimac, G. Fudenberga, J. Honec, P. Kima, H.L. Stormer,
\textsl{Solid State Communications}, 146:351Ð355 (2008). 

\bibitem{dft}
Hohenberg, Pierre; Walter Kohn, \textit{Inhomogeneous electron gas},
\textsl{Physical Review} 136 (3B): B864-B871 (1964).

\bibitem{thomas}
L. H. Thomas, 
\textsl{Prog. Cambrige Phil. Soc.,} 23, 542 (1927).

\bibitem{fermi}
E. Fermi,
\textsl{Rend. Acad. Naz. Lincei,} 6, 602 (1927).

\bibitem{ks}
W. Kohn and L. J. Sham,
\textsl{Phys. Rev. A.}, 140:1133 (1965).

\bibitem{relativity}
On the Electrodynamics of Moving Bodies. 
\textsl{Annalen der Physik} 17: 891-921 (1905).

\bibitem{ssb}
Edited by Henryk Arodz, Jacek Dziarmaga, Wojciech Hubert Zurek, \textit{Patterns of Symmetry Breaking},
\textsl{Originally published by Kluwer Academic Publishers},  Series II: Mathematics, Physics and Chemistry - Vol. 127 (2003).

 \bibitem{article_flakes}
 A.F. Jalbout, Y.P. Ortiz, T.H. Seligman.
 \textsl{Chem. Phys. Lett., } 564 p.  69-72 (2013).

\bibitem{jahn-teller}
H. Jahn and E. Teller.,
\textsl{Proc. Roy. Soc.}, A \textbf{161}(905): pp 220-235 (1937).

\bibitem{group}
Hamermesh M.,  \textit{Group Theory and Its Application to Physical Problems}. 
\textsl{New York: Dover}, (1989).

\bibitem{acenes}
Clar, E., \textit{Polycyclic Hydrocarbons},
\textsl{Academic Press: London,}, Vol. 1, pp 4?5 (1964). 

\bibitem{poly1}
K. Hummer, C. Ambrosch-Draxl,  
\textsl{Phys. Rev. B} 71:81202Ð81205 (2005). 
 
  
\bibitem{poly2}
J.E. Anthony, 
\textsl{Chem. Rev. },106:5028Ð5048 (2006). 

 
\bibitem{poly3}
J.L. Brdas, J.P. Calbert, D.A. da Silva Filho, J. Cornil, 
\textsl{Proc. Natl. Acad. Sci. USA}, 99:5804Ð5809 (2002). 
 
  
\bibitem{poly4}
M. Bendikov, F. Wudl, D.F. Perepichka, 
\textsl{Chem. Rev.}, 104 :4891Ð4945 (2004). 

 \bibitem{poly5}
  Anthony, J. E., 
  \textsl{Angewandte Chemie International} 47 pp 452-483 (2007).
 

\bibitem{ofet}
Anthony, J. E.,
\textsl{Angew. Chem.,} 47, 452 (2008).


\bibitem{oled}
Wolak, M. A., Jang, B.-B., Palilis, L. C. and Kafafi, Z. H.,
\textsl{ J. Phys. Chem. B,}Ê108, 5492 (2004).

\bibitem{opv}
Rand, B. P., Genoe, J., Heremans, P. and Poortmans, J. 
\textsl{Prog. Photovoltaics,}Ê 15, 659  (2007).

\bibitem{pentacene} 
Gross, L.; Mohn, F; Moll, N; Liljeroth, P; Meyer, G 
\textsl{Science} 325 (5944): 1110-4 (2009).

\bibitem{hartree} 
D. R. Hartree,
\textsl{Proc. Camb. Phil. Soc.}, 24, 111(1928).

 \bibitem{fock}
 V. Fock, 
\textsl{Zeits. f. Physik.},  61:126 (1930).

 \bibitem{articulo}
 Y.P. Ortiz, T. H. Seligman,
 \textsl{AIP Proceedings,} vol 1323, pp 257-264 (2010).

 \bibitem{tiras1}
 Fujita M., Wakabayashi K., Nakada K. and Kusakabe K,
 \textsl{Journal of the Physics Society Japan,} 65 (7): 1920  (1996).
 
 \bibitem{polyph1}
 James G. Speight, Peter Kovacic, Fred W. Koch,	 
 \textsl{ J. MACROMOL. XI.-REVS. MACROMOL. CHEM.}, C5(2), 295-386 (1971).

 \bibitem{tiras2}
Nakada K., Fujita M., Dresselhaus G. and Dresselhaus M.S, 
\textsl{Physical Review B,} 54 (24): 17954 (1996).

\bibitem{tiras3}
Wakabayashi K., Fujita M., Ajiki H. and Sigrist M, 
\textsl{Physical Review B,} 59 (12): 8271 (1999). 

\bibitem{atmos}
F. Keppler, D. B. Harper, T. Rockmann, R. M. Moore and J. T. G. Hamilton,
\textsl{Atmos. Chem. Phys., }5:2403Ð2411 (2005).

 \bibitem{art_atm}
Y.P. Ortiz,  A.F. Jalbout,
 \textsl{Chem. Phys. Lett., } 564 p.  73-77 (2013).
 



\bibitem{ap-b-o}
Max Born; J. Robert Oppenheimer, 
\textsl{Annalen der Physik}, 389 (20): 457Ð484 (1927).


\bibitem{sakurai}
J. Sakurai, \textit{Modern Quantum Mechanics},
\textsl{Addison-Wesley}, (1994).

\bibitem{demtro}
Wolfgang Demtrder:  \textit{Molecular Physics} ,
\textsl{WILEY-VCH Verlag GmbH \& Co. KGaA}, (2005).

\bibitem{messiah}
Albert Messiah, \textit{Quantum Mechanics},
\textsl{Dover publications INC}, (1999).

\bibitem{solids}
Karlheinz Schwarz
\textsl{Journal of Solid State Chemistry,}176, Issue 2, pp 319Ð328 (2003).

\bibitem{parr}
Robert G. Parr, Weitao Yang, \textit{Density Functional Theory of Atoms and Molecules,}
\textsl{Oxford University Press.}, (1989)

\bibitem{finn}
Marcelo Alonso, Edawrd J. Finn
\textsl{Fsica vol I, Mecnica},  Fondo Educativo Interamericano, S.A. (1971).

\bibitem{fowler}
Michael Fowler, \textit{Graduate Quantum Mechanics Notes},
\textsl{University of Virginia}, (2013).

\bibitem{kittel}
Charles Kittel, \textit{Introduction to Solid State Physics},
\textsl{John Wiley \& Sons, Inc.,} eight edition, (2005).

\bibitem{onedim-conduct}
C. K. Chiang, M. J. Cohen, A. F. Garito, A. J. Heeger, C. M. Mikulski and A. G. MacDiarmid,
\textsl{Solid State Commun.,} 18, 1451 (1976).

\bibitem{teller-landau}
 E. Teller, \textit{An historical note}, on a page inserted before the preface in R. Englman, \textit{The Jahn-Teller Effect in Molecules and Crystals},\\
 \textsl{Wiley-Interscience, London, (1972).}; \\ 
 E. Teller, \textit{The Jahn-  Teller Effect - Its history and applicability},
 \textsl{ Physica A} 114:14-18.(1982).
 


\bibitem{sto}
Slater, J.C., \textit{Atomic Shielding Constants},
\textsl{Phys. Rev.}  v 36: 57 (1930).

\bibitem{fboys}
Boys, S. F.,  
\textsl{Proc. R. Soc. [London]}, A 200 (1063): 542Ð554 (1950).

 \bibitem{pople}
Pople, J. A.; D. Beveridge, \textit{Approximate Molecular Orbital Theory}.
\textsl{McGraw-Hill.} (1970).

\bibitem{lectstand}
Standard, Jean M.,
\textsl{Illinois State University}, Lecture Handouts of Computation of Molecular Properties (2013).

\bibitem{teogauss}
Emili Besal and Ramon Carb-Dorca,
\textsl{J. Math Chem.,}  49:1769Ð1784 (2011).

 \bibitem{hfslater}
 J.C. Slater,
 \textsl{Physical Review}, 81:385-390 (1951).
 
  \bibitem{beck1}
A. D. Becke
\textsl{J. Chem. Phys.}, 85:7184 (1986).

 \bibitem{beck2}
A. D. Becke
\textsl{J. Chem. Phys.}, 84:4524 (1986).

 \bibitem{DK}
A.E. De Pristo and J.R. Kress,
\textsl{J. Chem. Phys.}, 86:1425 (1987).

 \bibitem{beck3}
A. D. Becke,
\textsl{Phys. Rev. A}, 38:3098 (1988).

 \bibitem{perdew}
J.P. Perdew and Y. Wang,
\textsl{Phys. Rev. B}, 33:8800 (1986).

 \bibitem{fila}
 M. Filatov y W. Thiel, 
 \textsl{Phys. Rev. A}, 57:189 (1998).

 \bibitem{ceperley}
 D. M. Ceperley and B. J. Alder, \textit{Ground State of the Electron Gas by a Stochastic Method}, 
 \textsl{Phys. Rev. Lett.} 45:566 (1980).
 
 \bibitem{vosko}
S. H. Vosko, L. Wilk and M. Nusair, 
\textsl{Can. J. Phys.}, 58:1200 (1980).

\bibitem{coef1}
 Y. Wangand and J.P. Perdew,
 \textsl{Phys. Rev. B.}, 43:8911 (1991).
 
 \bibitem{coef2}
J.P. Perdew and Y. Wangand, 
 \textsl{Phys. Rev. B.}, 43:13244 (1992).
 
  \bibitem{perdewc86}
J.P. Perdew,
 \textsl{Phys. Rev. B}, 33:8822 (1986).

 \bibitem{lyp88}
C. Lee, W. Yang and R.G. Parr
 \textsl{Phys. Rev. B}, 37:785 (1988).
 
  \bibitem{coef3}
F. Abu-Awwad and P. Politzer,
 \textsl{J. Comput. Chem.}, 21:227 (2000).
 
 \bibitem{docto}
 Juan Carlos Sancho Garca, Tesis de doctorado,
 \textsl{Universidad de Alicante} (2001).
 
 \bibitem{pople70}
 W. J. Hehre, W. A. Lathan, R. Ditchfield, M. D. Newton, and J. A. Pople, \textit{Gaussian 70}, 
 \textsl{Quantum Chemistry Program Exchange,}Program No. 237 (1970).
 
 \bibitem{moller}
M¿ller, Christian; Plesset, Milton S., 
\textsl{ Phys. Rev.} 46 (7): 618Ð622 (1934).
 
  \bibitem{gaussian}
M. J. Frisch et. al., GAUSSIAN09,
\textsl{Revision A.02}, Gaussian Inc., (2009).

\bibitem{nwchem}
 M. Valiev, E.J. Bylaska, N. Govind, K. Kowalski, T.P. Straatsma, H.J.J. van Dam, D. Wang, J. Nieplocha, E. Apra, T.L. Windus, W.A. de Jong, ``NWChem: a comprehensive and scalable open-source solution for large scale molecular simulations" 
 \textsl{Comput. Phys. Commun.} 181, 1477 (2010).
 
\bibitem{cp2k} 
J. VandeVondele, M. Krack, F. Mohamed, M. Parrinello, T. Chassaing and J. Hutter, "Quickstep: fast and accurate density functional calculations using a mixed Gaussian and plane waves approach",
\textsl{Comp. Phys. Comm.} 167, 103 (2005).
 
 



 
 
 
 \bibitem{polyexp1}
 Payne M. M., Parkin S. R., Anthony J. E. 
 \textsl{Journal of the American Chemical Society}, 127 (22): 8028Ð9 (2005).
   
 \bibitem{polyexp2}
Rajib Mondal, Bipin K. Shah, and Douglas C. Neckers 
\textsl{J. Am. Chem. Soc.},  128(30) pp 9612 - 9613 (2006).

\bibitem{polyexp3}
Chen, Kew-Yu; Hsieh, HH; Wu, CC; Hwang, JJ; Chow, TJ . 
\textsl{Chemical Communications} (10): 1065Ð7 (2007).

\bibitem{hidro}
L. Jeloaica, V. Sidis,
\textsl{Chemical Physics Letters,} 300:157Ð162 (1999).
 
 \bibitem{respaldo1}
 Alan Hinchliffe and Humberto J. Soscun Machado,
 \textsl{Int. J. Mol. Sci.}, 1, 8-16. (2000)
 
 \bibitem{respaldo2} 
Rohoullah Firouzi, Mansour Zahedi,
 \textsl{ Journal of Molecular Structure: THEOCHEM}, 862:7Ð15 (2008). 
 
   \bibitem{Ab_error}
   Abraham. F. Jalbout,
   \textsl{Journal of Theoretical and Computational Chemistry} Vol. 6, No. 2, 269Ð279 (2007).

\bibitem{gth}
S. Goedecker, M. Teter, and J. Hutter, \textit{Separable dual-space Gaussian pseudopotentials}, 
\textsl{Phys. Rev. B}, 54, 1703-1710 (1996).
 
 \bibitem{molopt}
Joost VandeVondele and Juerg Hutter,
\textsl{ J. Chem. Phys.,} 127, 114105 (2007).

 
 \bibitem{polyph2}
 Tour, J. 
 \textsl{Advanced Materials}  6:190 (1994).
 
 \bibitem{plastic1}
 David Parker, Jan Bussink, Hendrik T. van de Grampel, Gary W. Wheatley, Ernst-Ulrich Dorf, Edgar Ostlinning, Klaus Reinking, "Polymers, High-Temperature" 
 \textsl{Ullmann's Encyclopedia of Industrial Chemistry, Wiley-VCH: Weinheim.,} (2002).
 
 \bibitem{plastic2}
 Uyama, Hiroshi; Ikeda, Ryohei; Yaguchi, Shigeru; Kobayashi, Shiro;
 \textsl{ Polymers from Renewable Resources. ACS Symposium Series} 764. p. 113 (2001).
 
 \bibitem{13_12dicl}
Buxton, G. V.; Greenstock, C.-L.; Helman, W. P.; Ross, A. B.,
\textsl{J. Phys. Chem. Ref. Data}, 17, 513-886 (1988).

\bibitem{14_12dicl}
 Neta, P.; Huie, R. E.; Ross, A. B.,
 \textsl{J. Phys. Chem. Ref. Data}, 17, 1027-1284 (1988).
 
\bibitem{16_12dicl}
  Asmus, K.-D.,
 \textsl{Methods Enzymol.}, 105, 167-178 (1984).
  
\bibitem{18_12dicl}
Bothe, E.; Janata, E.
\textsl{ Radiat. Phys. Chem.}, 44, 455-458 (1994).

 
  \bibitem{react4}
Trzaskowski, B, L. Adamowicz, L. ,
\textsl{ Theor Chem Accts}, 124: 95 (2009).

\bibitem{area1}
Wallington, T. J.; Schneider, W. F.; Barnes, I.; Becker, K. H.; Sehested, J.; Nielsen, O. J.,
\textsl{Chem. Phys. Lett.}, 322, 97-102 (2000).

\bibitem{area2}
Neta, P.; Huie, R. E.; Ross, A. B.,
\textsl{J. Phys. Chem. Ref. Data} 17, 1027-1284 (1988).

\bibitem{atmosfera}
Wayne, R. P. \textit{Chemistry of Atmospheres}, 
\textsl{Oxford University Press: Oxford, United Kingdom, 3rd ed.}, (2000).

\bibitem{menshut}
Halls MD, Schlegel HB,
\textsl{J. Phys. Chem. B} 106:1921 (2002). 
 
 \bibitem{catal}
 Xiulian Pan and Xinhe Bao
 \textsl{Acc. Chem. Res.}, 44 (8), pp 553Ð562 (2011).

\bibitem{nano}
Britz DA, Khlobystov AN, Porfyrakis K, Ardavan A, Briggs  GAD,
\textsl{ Chem. Commun.,} 37 (2005).

\bibitem{nano1}
Lu T, Goldfield EM, Gray SK,
\textsl{J. Phys. Chem. C,} 112:2654 (2008).

\bibitem{nano2}
Lu T, Goldfield EM, Gray SK,
\textsl{J. Phys. Chem. C, } 112:15260 (2008).


\bibitem{nano3}
Yuan S, Wang X, Li P, Li F, Yuan S,
\textsl{J. Appl. Phys.} 104:054310 (2008).

\bibitem{nano4}
Barajas-Barraza RE, Ramirez-Ruiz JA, Guirado-Lopez RA,
\textsl{J. Comput. Chem. Nanosci.} 5:2255, (2008).


\bibitem{react1}
Tyndall GS, Wallington TJ, Hurley MD, Schneider WF,
\textsl{J. Phys. Chem., } 97:1576-1582 (1993).

\bibitem{react2}
Wallington TJ, Schneider WF, Barnes I, Becker KH, Sehested J, Nielsen OJ,
\textsl{Chem. Phys., }322:97-102 (2000).

\bibitem{react3}
 Phillips DL, Zhao C, Wang D, 
\textsl{J. Phys. Chem. A, }109:9653 (2005).

\bibitem{matr}
Knuttu, H.; Dahlqvist, M.; Murto, J.; Rsnen, M.,
\textsl{J. Phys. Chem.}, 92, 1495-1502 (1988).

\bibitem{pgas1}
Tyndall, G. S.; Wallington, T. J.; Hurley, M. D.; Schneider, W. F.,
\textsl{ J. Phys. Chem.}, 97, 1576-1582 (1993).

\bibitem{pgas2}
Wallington, T. J.; Schneider, W. F.; Barnes, I.; Becker, K. H.; Sehested, J.; Nielsen, O.,
\textsl{J. Chem. Phys. Lett.}, 322, 97-102 (2000).

\bibitem{formald}
Dowidiet, P.; Mertens, R.; von Sontag, C.,
\textsl{J. Am. Chem. Soc.,} 118, 11288-11292 (1996).

\bibitem{formaldg}
Hisatsune, C.; Heicklen, J. Can. J.,
\textsl{Spectrosc.}, 18, 77-81 (1973).

\bibitem{fule}
Pavanello, M., Jalbout, A.F., Trzaskowski, B., Adamowicz, L.,
\textsl{  Chem. Phys. Letts.,} 442:339 (2007). 

\bibitem{sucept}
Qi, Y., Mazur, Y., Hipps, K.W., 
\textsl{RSC Adv, }2:10579 (2012). 
 
\bibitem{ad_exp1}
Geunsik Lee, Bongki Lee, Jiyoung Kim and Kyeongjae Cho, 
\textsl{J. Phys. Chem. C}, 113 (32), pp 14225Ð14229 (2009).

 \bibitem{ad_exp2}
F. Schedin, A. K. Geim, S. V. Morozov, E. W. Hill, P. Blake, M. I. Katsnelson and K. S. Novoselov,
\textsl{Nature Materials}, 6, 652 - 655 (2007). 

\bibitem{ad_exp3}
Eric C. Mattson, Kanupriya Pande, Miriam Unger, Shumao Cui , Ganhua Lu , M. Gajdardziska-Josifovska, Michael Weinert, Junhong Chen, and Carol J. Hirschmugl,
\textsl{ J. Phys. Chem. C}, 117 (20), pp 10698Ð1070 (2013).
 
\end{thebibliography}
\end{document}